\documentclass[11pt]{article}
\pdfoutput=1

\usepackage{jheppub} 

\usepackage{amsmath,amssymb,epsfig,amsfonts}
\usepackage{graphicx}
\usepackage{epsf}
\usepackage{makeidx}
\usepackage{multirow}
\usepackage[all]{xy}
\usepackage{hyperref}
\usepackage{verbatim} 
\usepackage{rotating}
\usepackage{xcolor}
\usepackage{multirow}
\usepackage{bm}
\usepackage{cleveref}
\usepackage{slashed}
\usepackage[normalem]{ulem}
\usepackage{extarrows}

\usepackage{graphicx}
\usepackage{latexsym,amsmath,amsfonts,amssymb}
\usepackage{mathrsfs}
\usepackage{bbm}
\usepackage{bm}
\usepackage{subfig}
\usepackage{paralist}

\usepackage{tikz}
\usepackage{diagbox}
\usetikzlibrary{positioning}
\usetikzlibrary{calc}
\usetikzlibrary{decorations.pathreplacing,calligraphy}

\usepackage{xstring}
\usetikzlibrary{decorations.pathmorphing} 
\usetikzlibrary{decorations.markings} 
\usetikzlibrary{arrows} 
\usetikzlibrary{shapes} 
\usetikzlibrary{matrix} 
\usetikzlibrary{positioning} 
\usepackage[english]{babel} 
\usepackage[autostyle]{csquotes}
\usepackage{subfig}

\tikzstyle{brane}=[draw]
\tikzset{D7/.style={circle, draw=black, inner sep=0pt, fill=white, minimum size=3mm}}
\tikzset{hasse/.style={circle, fill,inner sep=2pt}}
\tikzset{flavor/.style={regular polygon,fill=white,regular polygon sides=4,inner sep=2.5pt, draw}}
\tikzset{gauge/.style={circle, draw,inner sep=2.5pt}}
\tikzset{gaugeb/.style={circle, draw,fill=black,inner sep=2.5pt}}
\tikzset{gauger/.style={circle, draw,fill=cyan,inner sep=2.5pt}}
\tikzset{gaugeg/.style={circle, draw,fill=red,inner sep=2.5pt}}
\tikzset{bd/.style={circle, draw=black, inner sep=0pt, fill=black, minimum size=2mm}}
\tikzset{wd/.style={circle, draw=black, inner sep=0pt, fill=white, minimum size=2mm}}
\tikzset{Dynkin/.style={circle, draw=black, inner sep=0pt, fill=white, minimum size=2mm}}
\tikzstyle{ligne}=[draw, thick] 
\tikzset{doublearrow/.style={ draw=black!75, color=black!75, thick, double distance=3pt, }}

\numberwithin{equation}{section}  

\graphicspath{ {figs/} }

\newtheorem{Theorem}{Theorem}[section]

\newcommand{\be}{\begin{equation}}
\newcommand{\ee}{\end{equation}}
\newcommand{\ba}{\begin{aligned}}
\newcommand{\ea}{\end{aligned}}

\def\ket#1{{|{#1}\rangle}}

\def\unit{{1\kern-.65ex {\rm l}}}
\def\1{{1\kern-.65ex {\rm l}}}

\def\Hom{\textrm{Hom}}

\newcount\hour \newcount\minute
\hour=\time \divide \hour by 60
\minute=\time
\def\now{%
\ifnum \hour<13
  \ifnum \hour=0 \advance \hour by 12 \number\hour:\else \number\hour:\fi%
     \ifnum \minute<10 0\fi%
     \number\minute%
\ A.M.%
\else \advance \hour by -12 \number\hour:%
  \ifnum \minute<10 0\fi%
  \number\minute%
  \ P.M.%
\fi%
}

\makeatother

\def\mb{\mathbb}

\def\mc{\mathcal}

\def\bp{\begin{pmatrix}}
\def\ep{\end{pmatrix}}

\newcommand{\bea}{\begin{equation} \begin{aligned}}
 \newcommand{\eea}{\end{aligned} \end{equation}}
\newcommand{\bit}{\begin{itemize}} 
\newcommand{\eit}{\end{itemize}}

\renewcommand{\d}{\partial }

\newcommand{\h}{\widehat}

\newcommand{\MG}{{\mathbf X}} 

\newcommand{\CY}{{\MG}} 

\newcommand{\FT}{{\mathcal{T}_\MG^{\rm 5d}}} 

\tikzstyle{brane}=[draw]
\tikzset{D7/.style={circle, draw=black, inner sep=0pt, fill=white, minimum size=3mm}}
\tikzset{hasse/.style={circle, fill,inner sep=2pt}}
\tikzset{flavor/.style={regular polygon,fill=white,regular polygon sides=4,inner sep=2.5pt, draw}}
\tikzset{gauge/.style={circle, draw,inner sep=2.5pt}}
\tikzset{gaugeb/.style={circle, draw,fill=black,inner sep=2.5pt}}
\tikzset{gauger/.style={circle, draw,fill=cyan,inner sep=2.5pt}}
\tikzset{gaugeg/.style={circle, draw,fill=red,inner sep=2.5pt}}
\tikzset{SUd/.style={circle, draw=black, inner sep=0pt, fill=yellow, minimum size=2mm}}
\tikzset{bd/.style={circle, draw=black, inner sep=0pt, fill=black, minimum size=2mm}}
\tikzset{wd/.style={circle, draw=black, inner sep=0pt, fill=white, minimum size=2mm}}
\tikzset{Dynkin/.style={circle, draw=black, inner sep=0pt, fill=white, minimum size=2mm}}
\tikzstyle{ligne}=[draw, thick] 
\tikzset{doublearrow/.style={ draw=black!75, color=black!75, thick, double distance=3pt, }}

\usepackage{booktabs}
\usepackage{makecell}

\newcommand{\dd}{\mathrm{d}}
\renewcommand{\d}{\partial }

\newcommand{\op}[1]{\operatorname{#1}}

\tikzset{global scale/.style={
    scale=#1,
    every node/.append style={scale=#1}
  }
}

\begin{document}

\baselineskip=18pt  
\numberwithin{equation}{section}  
\allowdisplaybreaks  


%
%


\thispagestyle{empty}

\vspace*{0.8cm} 
\begin{center}
{{\Huge  A note on the quantization of M2-branes on local Calabi-Yau manifolds
}}

\vspace*{1.5cm}

 {\large Yutong Huang$^{1}$, Qinjian Lou$^1$, Jiahua Tian$^2$, Yi-Nan Wang$^{1,3}$}

\vspace*{1.0cm}


\smallskip
{\it $^1$ School of Physics,\\
Peking University, Beijing 100871, China\\ }

\smallskip
{\it $^2$ School of Physics,\\
East China Normal University, Shanghai 200241, China\\ }

\smallskip

{\it  $^3$ Center for High Energy Physics, Peking University,\\
Beijing 100871, China}

\vspace*{0.8cm}
\end{center}
\vspace*{.5cm}

\noindent
In this note we present a detailed computation of the BPS states from M2-branes wrapping 2-cycles in local Calabi-Yau manifolds, at the level of brane quantization. For the cases of local Calabi-Yau threefolds, the supersymmetric quantum mechanics is the 1d $\mathcal{N}=4$ non-linear sigma model. We achieve systematic, physical interpretations of refined Gopakumar--Vafa (GV) invariants, first for the curve classes with fixed geometric genus, and then for curves whose geometric genus jumps. The BPS states of 5d $\mathcal{N}=1$ superconformal field theories are interpreted from the local geometries, which are presented in a number of case studies. In particular, we compute the refined GV invariants for generalized conifolds, local CY3 giving 5d $\mathcal{N}=1$ gauge theories and a singularity whose crepant resolution contains a singular divisor. For the cases of local Calabi-Yau fourfolds, we point out the supersymmetric quantum mechanics is of type $\mc{N}=2B$. We set up the framework for studying M2-branes wrapping 2-cycles, and discuss a few cases of M2-brane wrapping genus-0 curves with this new approach.
\newpage

\tableofcontents

\newpage

\section{Introduction}
\label{sec:intro}

In superstring/M/F-theory compactifications, an extremely important class of dynamical particles in the lower-dimensional physical theory arises from D-brane, membranes and other extended objects wrapping compact cycles in the extra-dimensional space. These states are crucial in the physical discussion of conifold transitions~\cite{Green:1988bp,Green:1988wa,Candelas:1989ug,Candelas:1989js,Strominger:1995cz,Greene:1995hu,Intriligator:2012ue,Gao:2025geq}, as well as in the M/F-duality to produce gauge fields, matter fields and other states in F-theory~\cite{Vafa:1996xn,Morrison:1996na,Morrison:1996pp,Witten:1996qb,Bershadsky:1996nh,Katz:1996xe,Intriligator:1997pq,Grimm:2010ks,Grimm:2011fx,Hayashi:2014kca,Weigand:2018rez,Chen:2024cvc,Sangiovanni:2024nfz}. In mathematics, such brane wrapping modes were used to define enumerative invariants for Calabi-Yau manifolds~\cite{Kontsevich:1994qz,Thomas:1998uj,Hosono:1999qc,Maulik:2003rzb,Pandharipande:2007kc,Pandharipande:2011jz}. The note would be focusing on the Gopakumar-Vafa (GV) invariants from M2-brane wrapping modes, initially defined for local Calabi-Yau threefolds in their unrefined forms~\cite{Gopakumar:1998ii,Gopakumar:1998jq,Katz:1999xq}. The later developments including the computation of refined GV invariants for local CY3~\cite{Nekrasov:2002qd,Hollowood:2003cv,Iqbal:2007ii,Huang:2010kf,Choi:2012jz,Huang:2013yta,Nekrasov:2014nea,Huang:2017mis,Kim:2020hhh}, the extension to compact CY3~\cite{Candelas:1990rm,Bershadsky:1993cx,Huang:2006hq,Huang:2020dbh,Huang:2025xkc}, as well as the more recent developments for local CY4~\cite{Klemm:2007in,Cao:2018wmd,Cao:2018ddu,Cao:2019fqq,Cao:2019tvv,Cao:2020vce,Cao:2020hoy,Bousseau:2020fus,Cao:2026ukb}. Such invariants shall be computed from topological string~\cite{Witten:1988xj,Witten:1991zz,Bershadsky:1993cx,Antoniadis:1993ze,Hori:2003ic,Aganagic:2003db,Marino:2004uf,Neitzke:2004ni,Huang:2006si,Iqbal:2007ii} and quiver supersymmetric quantum mechanics techniques~\cite{Denef:2002ru,Manschot:2010qz,Cordova:2014oxa,Hori:2014tda,Nekrasov:2017cih,Closset:2019juk,Duan:2020qjy} in a relatively systematic way.
 
However, after the developments of GV invariants on the more mathematical side, the initial physical motivations of the M2-brane wrapping modes have become more obscure. Witten's initial paper~\cite{Witten:1996qb} discussed a few examples of M2-branes wrapping curves $\Sigma$ on a CY3, and a detailed derivation of BPS states follow more recently, but only for the case of a rigid $\Sigma\cong\mb{P}^1$~\cite{Dedushenko:2014nya}. The interesting structure come from the geometric information such as the moduli space and obstruction space for the curve $\Sigma$~\cite{Katz:1999xq,Hosono:1999qc,Bryan:2000ve,Katz:2006gn,Choi:2012jz,Maulik:2016rip}, but the physical derivation for the M2-brane wrapping modes is still lacking in the literature. The physical interpretation for the GV invariants of CY4~\cite{Klemm:2007in,Cao:2018wmd,Cao:2019fqq}, in terms of the 3d $\mc{N}=2$ theories from M-theory on CY4~\cite{Gukov:1999ya,Haack:2001jz,Intriligator:2012ue,Najjar:2023hee,Najjar:2025rgt}, is even less clear.
 
In this note, we would first consider M-theory on $\mb{R}^{1,4}\times\CY$, with $\CY$ a local
(non-compact) Calabi--Yau threefold, which engineers a five-dimensional
$\mc{N}=1$ supersymmetric theory $\FT$ whose massive BPS particles arise from
M2-branes wrapping holomorphic two-cycles $\Sigma\subset\CY$~\cite{Seiberg:1996bd,Morrison:1996xf,Douglas:1996xp,Intriligator:1997pq,Katz:1996fh,Xie:2017pfl,Jefferson:2017ahm,Jefferson:2018irk,Closset:2018bjz,Bhardwaj:2018yhy,Bhardwaj:2018vuu,Apruzzi:2019vpe,Apruzzi:2019opn,Apruzzi:2019enx,Closset:2020scj,Eckhard:2020jyr,Closset:2020afy,Closset:2021lwy}, whose mass is given by 
\be
\label{eq:BPSmass}
m=T_{M2}\cdot A(\Sigma)\,,
\ee
which is fixed by the K\"ahler moduli of $\CY$. These states are classified by
representations of the 5d little group $SO(4)\cong SU(2)_L\times SU(2)_R$ (in the Lie algebra sense), and
their degeneracies are captured by the refined Gopakumar--Vafa invariants
$N^{Q}_{j_L,j_R}$~\cite{Gopakumar:1998ii,Gopakumar:1998jq,Katz:1999xq,Iqbal:2007ii,Choi:2012jz,Huang:2013yta} in the
homology class $Q=[\Sigma]$: the wrapped M2-brane produces particles
\be
\label{eq:intro-GV}
\left[\left(0,\tfrac12\right)\oplus 2(0,0)\right]\otimes (j_L,j_R)\,,
\ee
where we follow the convention of~\cite{Witten:1996qb} in which the broken
supersymmetries are $2(0,\tfrac12)$, so that the $j_L\leftrightarrow j_R$ labels
are exchanged relative to the mathematics literature.
 
Following the semiclassical quantization of solitons,
the BPS spectrum is read off from the geometric moduli space $\mc{M}_\Sigma$ of
the wrapped curve inside $\CY$~\cite{Katz:1996ht}, a K\"ahler manifold with Hodge numbers
$h^{p,q}$. When $\Sigma$ is a smooth genus-$g$ curve that stays non-degenerate
over all of $\mc{M}_\Sigma$, the wrapping modes assemble into
\be
\label{eq:intro-spectrum}
\left[\left(0,\tfrac12\right)\oplus 2(0,0)\right]^{g+1}
\otimes\big(H^*(\mc{M}_\Sigma),0\big)\,,
\ee
where the half-hypermultiplet factor counts the $g+1$ zero modes of the scalar
and holomorphic one-forms on $\Sigma$, and $H^*(\mc{M}_\Sigma)$ denotes the
Lefschetz decomposition of the cohomology of $\mc{M}_\Sigma$, with $SU(2)_L$
identified as the Lefschetz $SU(2)$~\cite{Witten:1996qb,Gopakumar:1998jq}, while the cases of degenerate $\mc{M}_\Sigma$ lead to the essential complications. On a further circle, M-theory reduces to
type IIA and the M2-brane becomes a D2-brane wrapping the same curve; the
little group is now $SO(3)\cong SU(2)$, and the relevant moduli space is the
\emph{refined} moduli space $\h{\mc{M}}\to\mc{M}$, a fibration by the moduli of
flat line bundles (generically $\op{Jac}(\Sigma)\cong T^{2g}$)~\cite{Gopakumar:1998jq,Katz:1999xq,Hosono:1999qc}.
 
The refined GV invariants are sensitive to the detailed geometry, while the
combinations
\be
N^{Q}_{j_R}=\sum_{j_L}(-1)^{2j_L}(2j_L+1)N^{Q}_{j_L,j_R}\,,\qquad
\sum_{j_R}N^{Q}_{j_R}[j_R]=\sum_{r\geq 0}N^{Q}_r\,I_r\,,
\ee
in the non-standard $SU(2)_R$ basis $I_r=I_1^{\otimes r}$,
$I_1=[(\tfrac12)\oplus 2(0)]$, $I_0=(0)$, define the unrefined GV invariants
$N^{Q}_r=n^r_\beta\in\mb{Z}$, which are invariant under smooth
deformations.
 
Despite of the physics interpretation of refined GV invariants reviewed above, the actual physical explanation of the M2-brane wrapping modes is really lacking in the literature, apart from the case of rigid $\Sigma$ with a zero-dimensional moduli space~\cite{Dedushenko:2014nya}. The aim of this note is to revisit the quantization of the M2-brane wrapping
modes directly from the eleven-dimensional supermembrane action~\cite{Bergshoeff:1987cm,Bergshoeff:1987qx,deWit:1988wri,deWit:1998tk}, rather than
only from the effective supersymmetric quantum mechanics. We first carry out the quantization for curves in a CY3, where the results can be compared with the known refined GV invariants, and then generalize it to curves in a CY4.

\paragraph{Worldline sigma models.}
In section~\ref{sec:NLSM}, we review the two types of 1d non-linear sigma models (NLSMs) that govern the deformation zero modes of a wrapped M2-brane. For a curve in a CY3, the M2-brane preserves four supercharges, and the zero modes are described by a 1d $\mc{N}=4$ NLSM whose target is the moduli space $\mc{M}$ of the curve, a K\"ahler manifold. Its supersymmetric ground states are given by the Dolbeault cohomology of $\mc{M}$, and the Lefschetz $SU(2)$ acts on pairs of fermions that are partners of the same bosonic field. For a curve in a CY4, only two supercharges are preserved, and we point out that the relevant model is the 1d $\mc{N}=2B$ NLSM, i.e.\ the dimensional reduction of a 2d $\mc{N}=(0,2)$ NLSM.

\paragraph{Quantization from the supermembrane action.}
In section~\ref{sec:M2quant}, we quantize an M2-brane wrapping a holomorphic curve $\Sigma$ of genus $g$ in a CY3, starting from the supermembrane action. The bosonic equation of motion is a covariant harmonic-map equation. Its linearization around $\Sigma$ is the Laplace operator on the normal bundle $N_{\Sigma|X}$, whose zero modes are the first-order deformations $H^0(\Sigma,N_{\Sigma|X})$ (section~\ref{subsec:membrane}). The fermion $\theta$ is a section of the pull-back of the eleven-dimensional spinor bundle. Since there is no covariantly constant spinor on a generic $\Sigma$, the worldvolume theory is topologically twisted, and the Calabi-Yau condition reduces the twisted spinor bundle to $(\mc{O}\oplus K_\Sigma^{-1})\otimes\bigoplus_{i}\Lambda^i N_{\Sigma|\mb{R}^4\times X}$. The four preserved supercharges realize the A-twist of a 3d $\mc{N}=4$ theory on $\mb{R}\times\Sigma$~\cite{Bullimore:2021auw,Ferrari:2019cni}, with bosonic and fermionic zero modes listed in table~\ref{tab:boson_fermion_zero}. In particular, the little group $SO(4)_J=SU(2)_1\times SU(2)_2$ of $\mb{R}^{1,4}$ acts on the fermionic zero modes in such a way that $SU(2)_1$ is exactly the Lefschetz $SU(2)$ of $\mc{M}$, which confirms the identification of \cite{Witten:1996qb}, while $SU(2)_2$ acts as the Lefschetz $SU(2)$ along the Jacobian of $\Sigma$, as proposed in ~\cite{Gopakumar:1998jq} (section~\ref{subsec:zeromodes}).

\paragraph{BPS spectrum.}
The zero modes in $H^{0,0}(\Sigma)$ give the universal half-hypermultiplet, those in $H^0(\Sigma,N_{\Sigma|X})$ and $H^1(\Sigma,N_{\Sigma|X})$ form an $\mc{N}=4$ NLSM on $\mc{M}$, and those in $H^{0,1}(\Sigma)$ generate the cohomology of the Jacobian of $\Sigma$. Since the genus of $\Sigma$ can jump over $\mc{M}$, the latter is described by the higher direct images $R^s\pi_*\mb{C}$ of the Jacobian fibration $\pi:\h{\mc{M}}\to\mc{M}$. We obtain the BPS spectrum
\be
\label{eq:intro-hatM}
\left[\left(0,\tfrac12\right)\oplus 2(0,0)\right]\otimes\bigoplus_{r,s}H^r(\mc{M},R^s\pi_*\mb{C})
=\left[\left(0,\tfrac12\right)\oplus 2(0,0)\right]\otimes H^*(\h{\mc{M}})\,,
\ee
where the equality follows from the Leray spectral sequence when $\mc{M}$ is smooth and $\h{\mc{M}}$ is reduced with sufficiently mild singularities~\cite{Pandharipande:2011jz}. It reproduces the result of  ~\cite{Gopakumar:1998jq}, which was obtained from D2-branes, and it reduces to \eqref{eq:intro-spectrum} when the genus of $\Sigma$ is constant over $\mc{M}$.

\paragraph{Obstructions.}
In section~\ref{subsec:obstruction}, we explain that the higher-order terms of the Nambu-Goto action play the role of obstructions. When $h^1(\Sigma,N_{\Sigma|X})\neq 0$, only the deformations in the kernel of the Kuranishi map $H^0(\Sigma,N_{\Sigma|X})\to H^1(\Sigma,N_{\Sigma|X})$ are genuine bosonic zero modes, and the spectrum has to be computed from the actual moduli space. We illustrate this with resolved threefold singularities. For the generalized conifold $x^2+y^2+z^2+w^{2n}=0$, the exceptional curve has the normal bundle $\mc{O}(-2)\oplus\mc{O}$, but its deformation is obstructed at order $n$, and the moduli space is the fat point $\op{Spec}\,\mb{C}[t]/t^n$. Consequently, an M2-brane wrapping this curve gives $n$ hypermultiplets, in agreement with the unrefined GV invariants~\cite{Collinucci:2021wty,Collinucci:2021ofd}. 

For the new examples, we consider canonical singularities whose crepant resolution is smooth but contains a singular compact exceptional divisor, such as $x^3+y^3+z^3+w^4=0$. Compared with the smooth $dP_6$, the $27$ isolated lines collapse to $9$ lines through the singular point of the divisor, which are in turn absorbed into a continuous family of lines parametrized by a cubic curve $T^2\subset\mb{P}^2$. However from the careful analysis of obstruction and hypercohomology of the singular moduli, we found that there are still $27$ hypermultiplets, as same as the smooth local $dP_6$ geometry. It leads to strong hints that M-theory on  $x^3+y^3+z^3+w^4=0$ still describes the rank-1 Seiberg $E_6$ theory in some particular phase, consistent with the speculations in \cite{Closset:2020afy}.

\paragraph{M2-branes versus D2-branes.}
In section~\ref{sec:M2D2}, we compare the M2-brane with the D2-brane on a further circle. The flat connections of the D2-brane worldvolume gauge field enlarge $\mc{M}$ to the refined moduli space $\h{\mc{M}}$, and the two extremal unrefined GV invariants are the Euler characteristics $N^Q_0=(-1)^{\dim_{\mb{C}}\h{\mc{M}}}\chi(\h{\mc{M}})$ and $N^Q_g=(-1)^{\dim_{\mb{C}}\mc{M}}\chi(\mc{M})$. We show that the 3d abelian duality between the D2-brane gauge field and the position of the M2-brane on the M-theory circle~\cite{Beasley:2014ila,Beasley:2014jta} does not map zero modes to zero modes: the Jacobian zero modes of the Maxwell theory are Pontryagin dual to the winding sectors of the compact scalar. Moreover, the worldvolume fluxes of D2--D0 bound states lift to the Kaluza-Klein momenta along the M-theory circle, so the full IIA/M-theory duality requires the sum over all D2--D0 bound states (section~\ref{subsec:abelian}). We also construct $\h{\mc{M}}$ for genus-one curves in a projective variety as a projective bundle (section~\ref{subsec:refined-projective}). For the cubic curves in $\mb{P}^2$, $\h{\mc{M}}$ is a $\mb{P}^8$-fibration over $\mb{P}^2$, which resolves the puzzle of the degree-three curves in local $\mb{P}^2$ described below. For higher genus, we conjecture a birational model of $\h{\mc{M}}$ via the Abel-Jacobi map.

\paragraph{Tests on 5d SCFT geometries.}
In section~\ref{sec:5dSCFT}, we test the framework against the refined GV invariants of local CY3 that engineer 5d SCFTs~\cite{Choi:2012jz,Huang:2013yta,Kim:2020hhh}. When the geometric genus of the curves jumps over the moduli space $\mc{M}$, the naive formula \eqref{eq:intro-spectrum} fails, and we compute the spectrum from $\h{\mc{M}}$ and the relative Hilbert schemes of points on the curves, with the refined version~\cite{Choi:2012jz} of the Katz-Klemm-Vafa (KKV) approach~\cite{Katz:1999xq}. This already happens for the cubic curves in local $\mb{P}^2$, where the cohomology of $\h{\mc{M}}$, the universal cubic curve, reproduces the known refined GV invariants~\cite{Choi:2012jz,Huang:2013yta} (section~\ref{subsec:localP2}). We then study the rank-2 theories $SU(3)_4$ and $SU(3)_5$ and the rank-3 theory $SU(4)_8$. For the genus-one curves, we construct $\h{\mc{M}}$ from restricted nested Hilbert schemes, which are blow-ups along the loci where the marked point collides with an attachment point, and we subtract the contributions of reducible curves. In this way, we exactly reproduce the refined GV invariants of ref.~\cite{Kim:2020hhh}, while the method can be generally applied to other local CY3 geometries. 

\paragraph{M2-branes in Calabi-Yau fourfolds.}
In section~\ref{sec:CY4}, we generalize the quantization to M2-branes wrapping holomorphic curves in a CY4 $X$, which give massive BPS particles of the 3d $\mc{N}=2$ theory from M-theory on $\mb{R}^{1,2}\times X$. Now the wrapped M2-brane preserves only two real supercharges, the massive little group is $SO(2)\cong U(1)$, and the zero modes form a twisted 3d $\mc{N}=2$ theory~\cite{Ferrari:2019cni} that reduces to a 1d $\mc{N}=2B$ SQM on the worldline. A crucial difference from the CY3 case is that $h^0(\Sigma,N_{\Sigma|X})\neq h^1(\Sigma,N_{\Sigma|X})$ in general, which means that the obstruction bundle is no longer dual to the tangent bundle of the moduli space. With the ground states of the $\mc{N}=2B$ NLSM, we find that the BPS spectrum is
\be
\label{eq:intro-CY4}
\left[(0)\oplus\left(-\tfrac12\right)\right]\otimes\bigoplus_i H^i\big(\mc{M},K_{\mc{M}}^{1/2}\big)
\ee
when $h^{0,1}(\Sigma)=h^1(\Sigma,N_{\Sigma|X})=0$, where $(q)$ denotes a state with $SO(2)$ charge $q$. In general, the zero modes in $H^{0,1}(\Sigma)$ and $H^1(\Sigma,N_{\Sigma|X})$ further twist the cohomology by the Jacobian and the obstruction bundles over $\mc{M}$, see eq.~\eqref{eq:BPS_cy4}. Compared with the CY3 case, the cohomology of $\mc{M}$ is now twisted by the square root of its canonical bundle, and the obstruction bundle enters the spectrum directly. 

We test this proposal on rational curves $\Sigma\cong\mb{P}^1$ with normal bundle $\mc{O}(m_1)\oplus\mc{O}(m_2)\oplus\mc{O}(m_3)$, $m_1+m_2+m_3=-2$, assuming that $\dim\mc{M}=h^0(\Sigma,N_{\Sigma|X})$. For $N_{\Sigma|X}=\mc{O}\oplus\mc{O}(-1)\oplus\mc{O}(-1)$, the moduli space $\mc{M}$ is a curve without an obstruction bundle, and \eqref{eq:intro-CY4} gives $h^0(\mc{M},K_{\mc{M}}^{1/2})$ chiral multiplets and $h^1(\mc{M},K_{\mc{M}}^{1/2})$ anti-chiral multiplets. For $N_{\Sigma|X}=\mc{O}\oplus\mc{O}\oplus\mc{O}(-2)$, $\mc{M}$ is a complex surface with an obstruction line bundle, which we argue to be $K_{\mc{M}}$ under several assumptions, similar to the result of ref.~\cite{Cao:2018ddu}. The spectrum then consists of $h^{1,0}(\mc{M})+h^{2,0}(\mc{M})$ vector-like pairs and $h^{0,0}(\mc{M})$ vector multiplets. In both cases, the spectrum agrees with that of the topologically twisted field theories in refs.~\cite{Beasley:2008dc,Intriligator:2012ue,Najjar:2023hee}.

\section{Non-linear sigma model}
\label{sec:NLSM}

M2-branes wrapped on a holomorphic curve $\Sigma$ in a CY3-fold $X$ give rise to massive BPS states in the 5d $\mathcal{N}=1$ supersymmetric field theory.
We are working in a limit where the mass of a wrapped M2-brane is much larger than the Planck mass~\cite{Dedushenko:2014nya}, and the resulting 5d state can be regarded as a particle described on its worldline by an $\mathcal{N}=4$ supersymmetric quantum mechanics (SQM).
This SQM can be obtained by dimensional reduction of the 3d $\mathcal{N}=2$ world-volume theory on the wrapped M2-brane \cite{Dedushenko:2014nya}.
When the deformation moduli space $\mathcal{M}$ of the holomorphic curve $\Sigma\subset X$ is non-trivial, the deformation zero modes of the M2-brane constitute an important sector of this SQM.
These modes can be described by a 1d $\mathcal{N}=4$ non-linear sigma model (NLSM) with target space $\mathcal{M}$. In this section, we review the basic properties of this NLSM \cite{Hori:2003ic,Tong:sqm}.

\subsection{1d $\mathcal{N}=4$ non-linear sigma model}
\label{subsec:1dn4}

The 1d $\mathcal{N}=4$ non-linear sigma model is a supersymmetric quantum mechanics described by maps from a 1d worldline $\mathcal{T}$ to a Kähler target manifold $M$.
Its bosonic and fermionic fields are given by
\begin{equation}
\begin{array}{ll}
\text{bosonic}  & \phi:\mathcal{T}\to M \\
\text{fermionic} & \psi,\bar{\psi}\in\Gamma(\mathcal{T},\phi^*TM\otimes \mathbb{C})
\end{array}
\end{equation}
Choose local complex coordinates $z^i,i=1,\cdots,d=\dim_{\mathbb{C}}M$.
The only non-zero components of the metric are $g_{i\bar{j}}$, and the only non-zero components of the Christoffel symbol are ${\Gamma^i}_{jk}=g^{i\bar{l}}\partial_{j}g_{k\bar{l}}$ and ${\Gamma^{\bar{i}}}_{\bar{j}\bar{k}}$.
The NLSM contains $d$ complex scalar fields $\phi^i,\bar{\phi}^{\bar{i}}$ and $2d$ complex fermions $\psi^i,\psi^{\bar{i}},\bar{\psi}^i,\bar{\psi}^{\bar{i}}$, satisfying $\psi^{i\dagger}=\bar{\psi}^{\bar{i}}$ and ${\bar{\psi}}^{i\dagger}=\psi^{\bar{i}}$. The Lagrangian is
\begin{equation}
L=g_{i\bar{j}}\dot{\phi}^i \dot{\bar{\phi}}^{\bar{j}}+ig_{i\bar{j}}\bar{\psi}^{\bar{j}}D_{t}\psi^i+ig_{i\bar{j}}\bar{\psi}^iD_{t}\psi^{\bar{j}}+R_{i\bar{j}k\bar{l}}\bar{\psi}^i \psi^k \psi^{\bar{j}}\bar{\psi}^{\bar{l}} \, .
\end{equation}
Here, $D_t\psi^i=\partial_t\psi^i+\Gamma^i_{jk}\dot{\phi}^j\psi^k$ and $D_t\psi^{\bar{j}}=\partial_t\psi^{\bar{j}}+\Gamma^{\bar{j}}_{\bar{k}\bar{l}}\dot{\bar{\phi}}^{\bar{k}}\psi^{\bar{l}}$. This theory has the following supersymmetry:
\be
\label{eq:1dn4-susy}
\begin{array}{ll}
\delta \phi^i=\epsilon_{+}\bar{\psi}^i-\epsilon_{-}\psi^i\, , & \delta \bar{\phi}^{\bar{i}}=-\bar{\epsilon}_{+}\psi^{\bar{i}}+\bar{\epsilon}_{-}\bar{\psi}^{\bar{i}}\, , \\
\delta \psi^i=i\bar{\epsilon}_{-}\dot{\phi}^i-\epsilon_{+}\Gamma^i_{jk}\bar{\psi}^j\psi^k\, , & \delta \bar{\psi}^{\bar{i}}=-i\epsilon_{-}\dot{\bar{\phi}}^{\bar{i}}-\bar{\epsilon}_{+}\Gamma^{\bar{i}}_{\bar{j}\bar{k}}\bar{\psi}^{\bar{j}}\psi^{\bar{k}}\, , \\
\delta \bar{\psi}^i=-i\bar{\epsilon}_{+}\dot{\phi}^i-\epsilon_{-}\Gamma^i_{jk}\bar{\psi}^j\psi^k\, , & \delta \psi^{\bar{i}}=i\epsilon_{+}\dot{\bar{\phi}}^{\bar{i}}-\bar{\epsilon}_{-}\Gamma^{\bar{i}}_{\bar{j}\bar{k}}\bar{\psi}^{\bar{j}}\psi^{\bar{k}}\, .
\end{array}
\ee
There are four conserved supercharges:
\begin{equation}
Q_{+}=\psi^i p_{i}\, , \, Q_{-}=\bar{\psi}^ip_{i} \, , \, \overline{Q}_{+}=\bar{\psi}^{\bar{j}}p_{\bar{j}} \, , \, \overline{Q}_{-}=\psi^{\bar{j}}p_{\bar{j}}\, ,
\end{equation}
with $p_{i}=\partial L /\partial \dot{\phi}^i$. The canonical commutation relations are
\begin{equation}
\begin{aligned}
&[\phi^i,p_{j}]=i\delta^i_{j} \, , \, [\bar{\phi}^{\bar{i}},p_{\bar{j}}]=i\delta^{\bar{i}}_{\bar{j}}\, ,\\
&\{ \psi^i,\bar{\psi}^{\bar{j}} \}=g^{i\bar{j}} \, , \, \{ \psi^{\bar{i}},\bar{\psi}^j \}=g^{\bar{i}j}\, .
\end{aligned}
\end{equation}
A standard quantization of this system proceeds by identifying the Hilbert space with the space of differential forms:
\begin{equation}
\mathcal{H}=\Omega(M)\otimes \mathbb{C}=\bigoplus_{p,q=0}^d \Omega^{p,q}(M) \, .
\end{equation}
The operators acting on the Hilbert space are
\begin{equation}
\begin{aligned}
&\bar{\psi}^i=dz^i\wedge \, , \, \bar{\psi}^{\bar{i}}=d\bar{z}^{\bar{i}}\wedge \, ,\\
&\psi^i=g^{\bar{j}i}i_{\partial /\partial \bar{z}^{\bar{j}}} \, , \, \psi^{\bar{i}}=g^{\bar{i}j}i_{\partial /\partial z^j}\, ,\\
&\phi^i=z^i\times \, , \, p_{i}=-i \partial_{i}\, , \, \bar{\phi}^{\bar{i}}=\bar{z}^{\bar{i}}\times \, , \, p_{\bar{i}}=-i\bar{\partial}_{\bar{i}} \, .
\end{aligned}
\end{equation}
where $i_{V}$ is the contraction of a differential form with the vector field $V$. Choose the vacuum to be the state annihilated by all $\psi^i$ and $\psi^{\bar{i}}$. This vacuum corresponds to a constant $(0,0)$-form, and a state created by $p$ operators $\bar{\psi}^i$ and $q$ operators $\bar{\psi}^{\bar{i}}$ corresponds to a $(p,q)$-form.

The four supercharges then become
\begin{equation}
Q_{-}=-i\partial \, , \, \overline{Q}_{+}=-i\bar{\partial} \, , \, \overline{Q}_{-}=Q_{-}^{\dagger}=i\partial ^{\dagger} \, , \, Q_{+}=\overline{Q}_{+}^{\dagger}=i\bar{\partial}^{\dagger} \, .
\end{equation}
The Hamiltonian is
\begin{equation}
H=\{ \overline{Q}_{+},\overline{Q}_{+}^{\dagger} \}=\{ Q_{-},Q_{-}^{\dagger} \}=\frac{1}{2}\{ Q,Q^{\dagger} \}  \, , \, Q\equiv i(Q_{-}+\overline{Q}_{+}) \, ,
\end{equation}
and is represented on a Kähler manifold by $\Delta_{\bar{\partial}}=\Delta_{\partial}=\frac{1}{2}\Delta$. In such a supersymmetric theory, every zero energy ground state is a supersymmetric state, and vice versa. The supersymmetric ground states of this NLSM are therefore represented by harmonic $(p,q)$-forms, whose space is isomorphic to the Dolbeault cohomology:
\begin{equation}
\mathcal{H}_0\cong \bigoplus_{p,q=0}^dH_{\bar{\partial}}^{p,q}(M) \, .
\end{equation}

A Lefschetz $SU(2)$ action exists on the differential forms of a Kähler manifold \cite{Griffiths:1994prl}. The raising operator $J_{+}$ is given by wedging with the Kähler form $\omega=ig_{i\bar{j}}dz^i\wedge d\bar{z}^{\bar{j}}$:
\begin{equation}
J_{+}=\omega \wedge=ig_{i\bar{j}}\bar{\psi}^i\bar{\psi}^{\bar{j}} \, .
\end{equation}
The lowering operator is its Hermitian conjugate:
\begin{equation}
J_{-}=J_{+}^{\dagger}=-i g^{i\bar{j}}i_{\partial / \partial z^i}i_{\partial /\partial \bar{z}^{\bar{j}}}=-ig_{i\bar{j}}\psi^i\psi^{\bar{j}}\, .
\end{equation}
Their commutator is the degree counting operator, which is related to the axial fermion number:
\begin{equation}
2J_{3}=[J_{+},J_{-}]=g_{i\bar{j}}(\bar{\psi}^{\bar{j}}\psi^i+\bar{\psi}^i\psi^{\bar{j}})-d \, , \, J_{3}\eta_{(p,q)}=\frac{1}{2}(p+q-d)\eta_{(p,q)} \, \text{for}\, \eta\in\Omega^{p,q}(M) \, .
\end{equation}
The four fermions are pairwise related by complex conjugation, whereas the two pairs are related by the Lefschetz $SU(2)$ action:
\begin{equation}
\begin{aligned}
&[J_{+},\psi^i]=i\bar{\psi}^i\, , \, [J_{+},\psi^{\bar{i}}]=-i\bar{\psi}^{\bar{i}} \, , \, [J_{+},\bar{\psi}^i]=0=[J_{+},\bar{\psi}^{\bar{j}}] \, ,\\
&[J_{-},\bar{\psi}^i]=-i\psi^i\, , \, [J_{-},\bar{\psi}^{\bar{i}}]=i\psi^{\bar{i}} \, , \, [J_{-},\psi^i]=0=[J_{-},\psi^{\bar{i}}]\, ,\\
&[J_{3},\psi^i]=-\frac{1}{2}\psi^i \, , \, [J_{3},\bar{\psi}^{\bar{i}}]=\frac{1}{2}\bar{\psi}^{\bar{i}} \, , \, [J_{3},\bar{\psi}^i]=\frac{1}{2}\bar{\psi}^{i}\, , \, [J_{3},\psi^{\bar{i}}]=-\frac{1}{2}\psi^{\bar{i}} \, .
\end{aligned}
\end{equation}
These relations can be summarized as
\begin{equation}
    \begin{array}{c@{\quad}c@{\quad}c}
\bar\psi^{i}\, (J_3=+\tfrac12)
&\overset{\dagger}{\longleftrightarrow}&
\psi^{\bar{i}}\, (J_3=-\tfrac12)
\\[-0.1em]
\substack{\downarrow\, J_-:\,-i\\[-0.15em]
           \uparrow\, J_+:\,+i}
&&
\substack{\downarrow\, J_+:\,-i\\[-0.15em]
           \uparrow\, J_-:\,+i}
\\[-0.1em]
\psi^{i}\, (J_3=-\tfrac12)
&\overset{\dagger}{\longleftrightarrow}&
\bar\psi^{\bar{i}}\, (J_3=+\tfrac12) \, .
\end{array}
\end{equation}
It is useful to recall the SUSY pairing relation:
\be
\delta \phi^i=\epsilon_{+}\bar{\psi}^i-\epsilon_{-}\psi^i\, ,  \delta \bar{\phi}^{\bar{i}}=-\bar{\epsilon}_{+}\psi^{\bar{i}}+\bar{\epsilon}_{-}\bar{\psi}^{\bar{i}} \, .
\ee
That is, the Lefschetz $SU(2)$ acts on a fermion doublet paired with the same bosonic field.


The cohomology can be organized into representations of the Lefschetz $SU(2)$ \cite{Hori:2003ic}.
A spin-$\mathbf{j}$ representation contains $2j+1$ states with $J_3$ eigenvalues $-j,-j+1,\ldots,j-1,j$.
For example, the cohomology of a torus $T^2$ forms the representation $\left(\mathbf{\frac{1}{2}}\right)+2(\mathbf{0})\equiv I_1$, and that of $T^{2g}$ forms $\left[\left(\mathbf{\frac{1}{2}}\right)+2(\mathbf{0})\right]^g=I_1^g$. The cohomology of $\mathbb{CP}^n$ forms $\left(\frac{\mathbf{n}}{\mathbf{2}}\right)$.

\subsection{1d $\mathcal{N}=2B$ non-linear sigma model}

The worldvolume theory of an M2-brane wrapped on a holomorphic curve $\Sigma$ in a Calabi–Yau fourfold preserves two real supercharges. The resulting massive BPS states in the 3d effective field theory can be described as particles whose dynamics are governed by an $\mathcal{N}=2$ SQM. This $\mathcal{N}=2$ SQM can be obtained by reducing on $\Sigma$ the topologically twisted 3d $\mathcal{N}=2$ worldvolume action of the M2-brane and typically it has the form of a non-linear sigma model.

Reference \cite{Hori:2003ic} describes two types of 1d NLSMs: $\mathcal{N}=2$ models on Riemannian manifolds and $\mathcal{N}=4$ models on Kähler manifolds. Both are described by a map from the worldline $\mathcal{T}$ to the target space $M$,
\begin{equation}
\phi:\mathcal{T}\to M
\end{equation}
together with fermionic fields $\psi,\bar{\psi}\in\Gamma(\mathcal{T},\phi^*TM\otimes \mathbb{C})$ that are the supersymmetric partners of the bosonic scalars. The field content of the $\mathcal{N}=2$ and $\mathcal{N}=4$ models is essentially the same. But the Kähler condition, that is the existence of complex structure and Hermitian metric, enhances the supersymmetry. Indeed, these two 1d NLSMs can be described as dimensional reduction of 2d $\mathcal{N}=(1,1)$ and $(2,2)$ supersymmetric field theory respectively. However, the resulting 1d $(1,1)$ NLSM is not the $\mathcal{N}=2$ SQM obtained from the M2-brane reduction. From the 2d perspective, another $\mathcal{N}=2$ SQM can be obtained by dimensional reduction of a 2d $\mathcal{N}=(0,2)$ supersymmetric field theory. These two different types of theories correspond to two types of representations of the $\mathcal{N}=2$ SQM \cite{Lawrie:2016axq}: the $\mathcal{N}=2A$ theory which is the dimensional reduction of 2d $\mathcal{N}=(1,1)$, and the $\mathcal{N}=2B$ theory which is the dimensional reduction of 2d $\mathcal{N}=(0,2)$.The 1d $\mathcal{N}=2B$ NLSM is the model relevant to the M2-brane wrapping modes.

In 2d $\mathcal{N}=(0,2)$ supersymmetry, there are two kinds of chiral superfields \cite{Hori:2003ic}: a chiral multiplet contains a complex scalar and a complex fermion, whereas a Fermi multiplet contains a complex fermion and a complex auxiliary scalar field. 1d supermultiplets are denoted by $(n, N, N-n)$ where $n$ is the number of real bosonic degrees of freedom,
$N$ the number of supersymmetries and $N-n$ the number of auxiliary fields. After reduction to 1d, a 2d $\mathcal{N}=(0,2)$ chiral multiplet becomes a $(2,2,0)$ multiplet, whereas a Fermi multiplet becomes a $(0,2,2)$ multiplet. An $\mathcal{N}=2B$ NLSM \cite{Hori:2014tda} consists of maps $\phi:\mathcal{T}\to M$ into a Kähler target space $M$, their fermion partners $\psi\in\Gamma(\mathcal{T},\phi^*TM)$, and fermions $\eta\in\Gamma(\mathcal{T},\phi^*E),\bar{\eta}\in\Gamma(\mathcal{T},\phi^*E^*)$ where $E$ is a holomorphic bundle on $M$ and $E^*$ is its dual. In terms of complex variables $\phi^i,\bar{\phi}^{\bar{i}} ,\psi^i,\bar{\psi}^{\bar{i}}$ with $i,\bar{i}=1,\cdots,d=\dim_{\mathbb{C}}M$ and $\eta^A,\bar{\eta}_{A}$ with $A=1,\cdots,r=\text{rank}(E)$, the Lagrangian is \cite{Melnikov:2011ez, Ivanov:2010ki}
\begin{equation}
L=g_{i\bar{j}}\dot{\phi}^i \dot{\bar{\phi}}^{\bar{j}}+ig_{i\bar{j}}\bar{\psi}^{\bar{j}}D_{t}\psi^i+i\bar{\eta}_{A}D_{t}\eta^A+(\mathcal{F}_{i\bar{j}})^A_{B}\bar{\eta}_{A}\eta^B\psi^i\bar{\psi}^{\bar{j}}\, ,
\end{equation}
where $(\mathcal{F}_{i\bar{j}})^A_{B}$ is the (1,1) component of the curvature of the Hermitian connection $(\mathcal{A}_{j})^A_{B}$ on $E$, and
\begin{equation}
D_{t}\psi^i=\partial_{t}\psi^i+\Gamma^i_{jk}\dot{\phi}^j\psi^k \, , \, D_{t}\eta^A=\partial_{t}\eta^A+(\mathcal{A}_{j})^A_{B}\dot{\phi}^j\eta^B \, .
\end{equation}
The supercharges are
\begin{equation}
Q_1=ig_{i\bar{j}}p^i\bar{\psi}^{\bar{j}} \, , \, Q_2=ig_{i\bar{j}}\bar{p}^{\bar{j}}\psi^i \, .
\end{equation}

When there are no Fermi multiplets, this $\mathcal{N}=2B$ NLSM can be regarded as an $\mathcal{N}=1$ NLSM whose supersymmetry is enhanced by the Kähler condition. The $\mathcal{N}=1$ NLSM has equal numbers of real bosonic scalars and real fermions. Its target space is a Riemannian manifold and its Lagrangian is \cite{Tong:sqm}
\begin{equation}
L=\frac{1}{2}g_{\mu \nu}\dot{\phi}^\mu \dot{\phi}^\nu+\frac{i}{2}g_{\mu \nu}\psi^\mu D_{t}\psi^\nu \, .
\end{equation}
 The single real supercharge is $Q=ig_{\mu \nu}\dot{\phi}^\mu \psi^\nu$. Canonical quantization requires $\{ \psi^\mu,\psi^\nu \}=g^{\mu \nu}$, implying that these fermions can be identified with gamma matrices. Under the identification $\phi^\mu=x^\mu \times,\dot{\phi}_{\mu}=-i\nabla_{\mu},\psi^\mu=\frac{1}{\sqrt{2}}\gamma^\mu$, the supercharge $Q=\frac{1}{\sqrt{2}}\gamma^\mu\nabla_{\mu}$ is the Dirac operator. The Hilbert space is the space of spinors on $M$, and the supersymmetric ground states are given by Dirac zero modes.

When the target space $M$ is Kähler, the Kähler condition again enhances the supersymmetry to $\mathcal{N}=2B$. We further assume that $M$ is spin. Canonical quantization gives
\begin{equation}
    [\phi^i, p_{j}]=i\delta^i_{j} \, , \, [\bar{\phi}^{\bar{i}},\bar{p}_{\bar{j}}]=i\delta^{\bar{i}}_{\bar{j}}  \, , \, \{ \psi^i,\bar{\psi}^{\bar{j}} \}=g^{i\bar{j}} \, , \, \{ \eta^A,\bar{\eta}_{B}\}=\delta^A_{B} \, .
\end{equation}
with all other commutators vanishing. First, we omit the Fermi multiplet sector. As in the preceding real case, the Hilbert space of this system can be identified with the space of spinors on the Kähler manifold $M$, namely, $S\simeq K_M^{1/2}\otimes\Lambda^{0,\bullet}T^*M$. The bosonic operators are identified as in the preceding case, and the fermionic operators are
\begin{equation}
    \bar{\psi}^{\bar{i}}=\frac{1}{\sqrt{ 2 }}\bar{\gamma}^{\bar{i}}=d\bar{z}^{\bar{i}}\wedge \, , \, \psi^i=\frac{1}{\sqrt{ 2 }}\gamma^i=g^{i\bar{j}} i_{\partial /\partial \bar{z}^{\bar{j}}}\, .
\end{equation}
The supercharge $Q=\frac{1}{\sqrt{ 2 }}\gamma^\mu D_{\mu}$ becomes $Q=i\bar{p}_{\bar{j}}\bar{\psi}^{\bar{j}}+ip_{i}\psi^i=\bar{\partial}+\bar{\partial}^{\dagger}$; that is, it becomes the Dirac operator acting on the spinor bundle
\begin{equation}
    S\simeq K_{M}^{1/2}\otimes (\bigoplus_{p=0}^d \Lambda^{0,p}T^*M)\,.
\end{equation}
The two components  $i\bar{p}_{\bar{j}}\bar{\psi}^{\bar{j}}=Q_{1}$ and $ip_{i}\psi^i=Q_{2}$ are conserved separately, thereby yielding two supercharges. The supersymmetric ground states of this NLSM are $H^{0,\bullet}(M,K_{M}^{1/2})$.

We now include the Fermi multiplets. They satisfy the anticommutation relation $\{ \eta^A,\bar{\eta}_{B} \}=\delta^A_{B}$. If the vacuum is chosen to be annihilated by every $\eta^A$, i.e., $\eta^A|0\rangle_E=0$ for all $A$, then the $\bar{\eta}_B$ are creation operators valued in the dual bundle $E^*$. The operators $\eta^A$ and $\bar{\eta}_A$ can again be identified with the gamma matrices of $SO(2r)\supset U(r)$: $\eta^A=\frac{1}{\sqrt{ 2 }}\gamma^A,\bar{\eta}_{A}=\frac{1}{\sqrt{ 2 }}\bar{\gamma}_{A}$. The generators of the Cartan subalgebra of $\mathfrak{so}(2r)$ are
\begin{equation}
    H_{A}=\frac{1}{4}[\bar{\gamma}_{A},\gamma^A]=\frac{1}{2}(\bar{\eta}_{A}\eta^A-\eta^A\bar{\eta}_{A})=\bar{\eta}_{A}\eta^A-\frac{1}{2} \, ,
\end{equation}
where no sum over $A$ is implied. Their eigenvalues on the vacuum form the weight $\left(-\frac{1}{2},\ldots,-\frac{1}{2}\right)$. The full Hilbert space of the Fermi multiplets is generated by applying the $\bar{\eta}_A$ to the vacuum. Each application of $\bar{\eta}_A$ raises the weight of $H_A$ by $+1$. Thus, the complete weight system is $\left(\pm\frac{1}{2},\ldots,\pm\frac{1}{2}\right)$, corresponding to the spinor representation of $SO(2r)$. Upon restriction to $U(r)\subset SO(2r)$, the spinor representation becomes $\bigoplus_{l=0}^r \Lambda^l\mathbb{C}^r \otimes(\det \mathbb{C}^r)^{-1/2}$. This result can be understood as follows. The preceding Cartan generators become the diagonal rotation of $U(r)$. Under a diagonal $U(r)$ rotation $U$ for which the frame of $E$ transforms as $e_{A}\to e^{i\alpha_{A}}e_{A}$, we have $R(U)=\exp\left( -i\sum_{A}\alpha_{A} H_{A} \right)$, and
\begin{equation}
R(U)|0\rangle_{E}=\exp\left( \frac{i}{2}\sum_{A}\alpha_{A} \right)|0\rangle_{E}=(\det U)^{1/2}|0\rangle_{E} \, .
\end{equation}
Thus, the vacuum $|0\rangle_{E}$ belongs to the bundle $(\det E)^{1/2}=(\det E^*)^{-1/2}$. Acting on the vacuum with the creation operators gives rise to Fermi multiplet states valued in the bundle
\begin{equation}
    (\det E)^{1/2}\otimes\bigoplus_{s=0}^r \Lambda^{s}E^*\,.
\end{equation}

Combining these two sectors, the full Hilbert space and the space of supersymmetric ground states can be identified as \cite{Hori:2014tda, Ferrari:2019cni}
\begin{equation}
\begin{aligned}
    \mathcal{H}&=\bigoplus_{p=0}^d \bigoplus_{s=0}^r \Omega^{0,p}(M,(K_{M}\otimes \det E)^{1/2}\otimes \Lambda^sE^*) \, , \\
    \mathcal{H}_0&=\bigoplus_{p=0}^d \bigoplus_{s=0}^r H^{p}(M,(K_{M}\otimes \det E)^{1/2}\otimes \Lambda^s E^*) \, .
\end{aligned}
\label{eq:N=2B_spec}
\end{equation}
The vacuum line bundle $(K_{M}\otimes \det E)^{1/2}$ can also be determined using the method in \cite{Anderson:2013sia}. We take the creation operators to be $\bar{\psi}^{\bar{i}}\in T^{*0,1}M$ and $\bar{\eta}_A\in E^*$, and the vacuum $|0\rangle\in\mathcal{L}$ is annihilated by all annihilation operators. The opposite, or dual, Fock vacuum $|0\rangle^{\mathrm{op}}\in\mathcal{L}^{\mathrm{op}}$ is defined by applying all creation operators. These two vacua are related by symmetry according to
\begin{equation}
\mathcal{L}^*=\mathcal{L}^{\text{op}}=(\det T^{*0,1}M)\otimes \det E^* \otimes  \mathcal{L} \, .
\end{equation}
Therefore, the vacuum line bundle should satisfy
\begin{equation}
\mathcal{L}^2=K_{M}\otimes  \det E \, .
\end{equation}
Cases may arise in which neither $K_M^{1/2}$ nor $(\det E)^{1/2}$ exists, even though $(K_M\otimes\det E)^{1/2}$ is well defined. The Hilbert space requires only the existence of $(K_{M}\otimes \det E)^{1/2}$.

\section{Semi-classical quantization in the CY3-fold}
\label{sec:M2quant}

We quantize a single M2-brane whose worldvolume is $T\times\Sigma$, with
$T=\mb{R}$ the time direction and $\Sigma\subset X$ a holomorphic curve of
genus $g$. There are two versions of the supermembrane action, the Polyakov-type
action of~\cite{deWit:1998tk} and the Nambu--Goto-type action
of~\cite{Bergshoeff:1987cm}; the former is used below. Throughout, $\alpha,\beta$
are worldvolume indices, $\mu,\nu$ target-space indices, and we split the target
coordinates into those tangent and normal to $\Sigma$.

\subsection{Supermembrane action}
\label{subsec:membrane}

\paragraph{Target superspace.}
The action maps the three-dimensional worldvolume, with coordinates $\zeta^i$
($i=0,1,2$), into the eleven-dimensional target superspace with coordinates
$Z^M=(X^\mu(\zeta),\theta^\alpha(\zeta))$. The dynamical background data are the
supervielbein $E_M^A$ and the antisymmetric three-form superfield $B_{MNP}$,
\be
\label{eq:membrane-action}
S[Z(\zeta)]=\int\dd^3\zeta\left[-\sqrt{-g(Z(\zeta))}
-\frac{1}{6}\varepsilon^{ijk}\,\Pi_i^A\Pi_j^B\Pi_k^C\,B_{CBA}(Z(\zeta))\right]\,,
\ee
where the pulled-back frame one-form and the induced metric are
\be
\Pi^A_i=\frac{\d Z^M}{\d\zeta^i}\,E_M^A\,,\qquad
g_{ij}=\Pi^r_i\Pi^s_j\,\eta_{rs}\,.
\ee
The membrane couples to the eleven-dimensional supergravity background
$(e_\mu^r,\psi_\mu^a,C_{\mu\nu\rho})$; in a convenient gauge the supervielbein
reads~\cite{deWit:1998tk}
\be
\label{eq:supervielbein}
\ba
E_\mu^r&=e_\mu^r+2\bar\theta\Gamma^r\psi_\mu
+\bar\theta\Gamma^r\Big[-\tfrac14\hat\omega_\mu^{st}\Gamma_{st}
+T_\mu{}^{\nu\rho\sigma\lambda}\hat F_{\nu\rho\sigma\lambda}\Big]\theta
+\mc{O}(\theta^3)\,,\\
E_\mu^a&=\psi_\mu^a-\tfrac14\hat\omega_\mu^{rs}(\Gamma_{rs}\theta)^a
+(T_\mu{}^{\nu\rho\sigma\lambda}\theta)^a\hat F_{\nu\rho\sigma\lambda}
+\mc{O}(\theta^3)\,,\\
E_\alpha^r&=-(\bar\theta\Gamma^r)_\alpha+\mc{O}(\theta^3)\,,\qquad
E_\alpha^a=\delta_\alpha^a+M_\alpha^a+\mc{O}(\theta^3)\,,
\ea
\ee
where $M_\alpha^a$ collects the $F\theta^2$ pieces left implicit
in~\cite{deWit:1998tk}, and the three-form superfield is
\be
\label{eq:Bfield}
\ba
B_{\mu\nu\rho}&=C_{\mu\nu\rho}-6\bar\theta\Gamma_{[\mu\nu}\psi_{\rho]}
-3\bar\theta\Gamma_{[\mu\nu}\Big[-\tfrac14\hat\omega_{\rho]}^{rs}\Gamma_{rs}
+T_{\rho]}{}^{\sigma\lambda\kappa\tau}\hat F_{\sigma\lambda\kappa\tau}\Big]\theta
-12\bar\theta\Gamma_{\sigma[\mu}\psi_\nu\,\bar\theta\Gamma^\sigma\psi_{\rho]}
+\mc{O}(\theta^3)\,,\\
B_{\mu\nu\alpha}&=(\bar\theta\Gamma_{\mu\nu})_\alpha
-\tfrac83\bar\theta\Gamma^\rho\psi_{[\mu}(\bar\theta\Gamma_{\nu]\rho})_\alpha
+\tfrac43(\bar\theta\Gamma^\rho)_\alpha\bar\theta\Gamma_{\rho[\mu}\psi_{\nu]}
+\mc{O}(\theta^3)\,,\\
B_{\mu\alpha\beta}&=(\bar\theta\Gamma_{\mu\nu})_{(\alpha}(\bar\theta\Gamma^\nu)_{\beta)}
+\mc{O}(\theta^3)\,,\qquad
B_{\alpha\beta\gamma}=(\bar\theta\Gamma_{\mu\nu})_{(\alpha}(\bar\theta\Gamma^\mu)_\beta
(\bar\theta\Gamma^\nu)_{\gamma)}+\mc{O}(\theta^3)\,.
\ea
\ee
The action is invariant under superspace diffeomorphisms with parameter
$\epsilon$ obeying $\delta Z^M E_M^r=2\bar\theta\Gamma^r\epsilon$,
$\delta Z^M E_M^a=\epsilon^a$, and under a fermionic $\kappa$-symmetry,
\be
\delta Z^M E_M^r=0\,,\qquad \delta Z^M E_M^a=(1-\Gamma)^a{}_b\kappa^b\,,\qquad
\ba
\delta X^\mu&=\bar\kappa_-\Gamma^\mu\theta-\bar\kappa_-\Gamma^\nu\theta\,\bar\theta\Gamma^\mu\psi_\nu+\cdots\,,\\
\delta\theta&=\kappa_-+\psi_\mu\bar\theta\Gamma^\mu\kappa_-+\cdots\,,
\ea
\ee
whose projector removes half of the $32$ components of $\theta$,
\be
\label{eq:kappa-projector}
\Gamma=\frac{\varepsilon^{ijk}}{6\sqrt{-g}}\,\Pi_i^r\Pi_j^s\Pi_k^t\,\Gamma_{rst}\,,
\qquad g=\op{det}g_{ij}\,,\qquad \Gamma^2=\1\,.
\ee
Setting $\psi=C=0$ and dropping $\mc{O}(\theta^3)$ terms, the pulled-back frame
one-form becomes
\be
\label{eq:Nambu-momentum}
\Pi_i^r=e_\mu^r\big(\d_iX^\mu-\bar\theta\Gamma^\mu\nabla_i\theta\big)+\mc{O}(\theta^3)\,,
\ee
with the induced Dirac operator equal to the pull-back of the eleven-dimensional
spin connection,
\be
\label{eq:worldvol-Dirac}
\nabla_i\theta=\Big(\d_i+\frac{1}{4}(\d_iX^\mu)\,\omega_\mu^{st}\Gamma_{st}\Big)\theta\,,
\qquad \nabla_i=(\d_iX^\mu)\nabla_\mu\,.
\ee
Thus $\theta$ is a section of the pull-back of the $11d$ spinor bundle over the
membrane worldvolume, and~\eqref{eq:worldvol-Dirac} is its equation of motion.

\paragraph{Equation of motion of $X$.}
Ignoring  $\theta$, the bosonic part of the Lagrangian reduces to the Nambu--Goto form
\be
\mc{L}_{\op{bos}}=\sqrt{-\op{det}(g_{\alpha\beta})}
=\sqrt{-\op{det}\big(G_{\mu\nu}(X)\,\d_\alpha X^\mu\d_\beta X^\nu\big)}\,.
\ee
Denote the Levi--Civita connections of the induced metric $g$ and of the target
metric $G$ by
\be
\Gamma^{\gamma}_{g,\alpha\beta}=\tfrac12 g^{\gamma\delta}(\d_\alpha g_{\beta\delta}+\d_\beta g_{\alpha\delta}-\d_\delta g_{\alpha\beta})\,,\qquad
\Gamma^{\rho}_{G,\mu\nu}=\tfrac12 G^{\rho\sigma}(\d_\mu G_{\nu\sigma}+\d_\nu G_{\mu\sigma}-\d_\sigma G_{\mu\nu})\,,
\ee
and pull the latter back to the worldvolume,
$\Gamma^{\rho}_{G,\alpha\nu}=(\d_\alpha X^\mu)\Gamma^{\rho}_{G,\mu\nu}$, which has
the form of a Hermitian connection. Using
$\delta\mc{L}_{\op{bos}}=\tfrac12\sqrt{-g}\,g^{\alpha\beta}\delta g_{\alpha\beta}$,
the equation of motion is
\be
\d_\alpha\!\left[\sqrt{-g}\,g^{\alpha\beta}G_{\mu\nu}\d_\beta X^\nu\right]
-\tfrac12\sqrt{-g}\,g^{\alpha\beta}(\d_\mu G_{\rho\nu})\d_\alpha X^\rho\d_\beta X^\nu=0\,.
\ee
Expanding the first term with the connections above,
\be
\ba
\d_\alpha\!\left[\sqrt{-g}\,g^{\alpha\beta}G_{\mu\nu}\d_\beta X^\nu\right]
=\sqrt{-g}\Big[&G_{\mu\nu}(\d_\alpha+\Gamma^{\rho}_{g,\alpha\rho})(g^{\alpha\beta}\d_\beta X^\nu)
+(\d_\rho G_{\mu\nu})g^{\alpha\beta}\d_\alpha X^\rho\d_\beta X^\nu\Big]\,,
\ea
\ee
contracting with $G^{\sigma\mu}/\sqrt{-g}$, and using the symmetry of
$g^{\alpha\beta}\d_\alpha X^\rho\d_\beta X^\nu$ in $(\rho,\nu)$, the equation
collapses to the covariant harmonic-map equation
\be
\label{eq:harmonic-map}
(\d_\alpha+\Gamma^{\rho}_{g,\alpha\rho})(g^{\alpha\beta}\d_\beta X^\sigma)
+\Gamma^{\sigma}_{G,\alpha\nu}(g^{\alpha\beta}\d_\beta X^\nu)=0\,.
\ee
The first term is the worldvolume-covariant divergence of
$g^{\alpha\beta}\d_\beta X^\sigma$, and $\Gamma^{\sigma}_{G,\alpha\nu}$ plays the
role of a connection on the target bundle. Because $G$ depends on $X$ itself,
this is not literally a vector bundle over the worldvolume: a metric on a bundle
cannot depend on the fiber coordinate.

From the full Nambu-Goto action, from the \cite{Becker:2006dvp}, together with the SUSY constraint, we can obtain that the solution is the moduli space of the holomorphic or special lagrangian cycle in the Calabi-Yau manifold.

\paragraph{Approximation around the wrapped curve.}
To compare with the K\"ahler-potential form of supersymmetry in
Section~\ref{subsec:Nambu_NLSM}, expand around the embedded curve. Split the
target coordinates as those tangent to $\Sigma$, $X^i=X^i(\sigma^\alpha)$, fixed
by diffeomorphisms, and those normal to it, $X^m=X_0^m+\delta X^m(\sigma^\alpha)$,
\be
g_{0,\alpha\beta}=\d_\alpha X^i\d_\beta X^j\,G_{ij}\,,\qquad
\delta g_{\alpha\beta}=\d_\alpha X^m\d_\beta X^n\,G_{mn}\,.
\ee
Using $\op{det}(1+M)=1+\op{tr}M+\tfrac12(\op{tr}M)^2-\tfrac12\op{tr}(M^2)+\cdots$,
the Nambu-Goto Lagrangian expands as
\be
\label{eq:Nambu-expansion}
\ba
\mc{L}=\sqrt{-\op{det}(g_0+\delta g)}
&=\sqrt{-g_0}\,\sqrt{\op{det}\!\big(1+g_0^{-1}\delta g\big)}\\
&=\sqrt{-g_0}\Big[1+\tfrac12 g_0^{\alpha\beta}\d_\alpha X^m\d_\beta X^n G_{mn}
+\mc{O}(\delta g^3)\Big]\,.
\ea
\ee
The quadratic piece $\tfrac12 G_{mn}\,g_0^{\alpha\beta}\d_\alpha X^m\d_\beta X^n$
is precisely the bosonic kinetic term of a non-linear sigma model into the
normal directions. Repeating the variation with the fixed background metric
$g_0$ yields
\be
\label{eq:Kahler-eom}
(\d_\alpha+\Gamma^{\rho}_{g_0,\alpha\rho})(g_0^{\alpha\beta}\d_\beta X^\sigma)
+\Gamma^{\sigma}_{G,\alpha\nu}(g_0^{\alpha\beta}\d_\beta X^\nu)=0\,,
\ee
which has the same structure as~\eqref{eq:harmonic-map}. Freezing the dependence
of $\Gamma_G$ on $\delta X$, i.e. $\Gamma_G\approx\Gamma_G|_{X=X_0}$,
equation~\eqref{eq:Kahler-eom} becomes the Laplace/harmonic operator on the
normal bundle $N_{\Sigma|X}$ with fiber coordinates $X^m$ and Hermitian metric
$G_{mn}$; its zero modes are the holomorphic sections of $N_{\Sigma|X}$, i.e. the
bosonic collective coordinates of the wrapped curve. 

With this approximation, in the following discussion, we will show that the solution is related to the $h^0(\Sigma,N_{\Sigma|X})$, which can be different from the moduli space of the holomorphic or special lagrangian cycle. This difference comes from the higher order term $\mc{O}(\delta g^3)$, which play the role of the obstruction. We will return to this topic at the end of this section.

\subsection{From Nambu-Goto action to NLSM}
\label{subsec:Nambu_NLSM}

The harmonic operator~\eqref{eq:Kahler-eom} strongly suggests that the reduced
theory is a supersymmetric non-linear sigma model whose target is the moduli
space of $\Sigma$. We now show how the Nambu-Goto action is reorganized into a series of fields in the NLSM, and argue that its equation of motion is
equivalent to the approximation~\eqref{eq:Kahler-eom}.

The $\mathcal{N}=4$ NLSM comes from the D-term of the K\"ahler potential
\be
\label{eq:Dterm}
\mc{L}_D=\int\dd^4\theta\,\mc{K}
=g_{i\bar j}\,\d\phi^i\!\cdot\!\d\overline{\phi^j}
+g_{i\bar j}\,\overline{\psi^j}\,\sigma(\nabla_g\psi)^i+\cdots\,,
\ee
with $g_{i\bar j}=\d_i\bar\d_j\mc{K}$ the K\"ahler metric and $\nabla_g$ the
pull-back of its Levi--Civita connection. This form is quite different from the Nambu-Goto action, where the fermion field is an 11d spinor on the target space instead of some spinors on the worldvolume at the first glance. But in the following discussion we will show how to transform the target space spinor into the worldvolume spinors. 

For simplicity, we will only focus on the locally flat frame and ignore the connection, which is essential to determine the number of the zero modes. But as we shown above, all fields can be understood as sections of bundles over the worldvolume. So we can just use the theorem of the decomposition of the spinor bundle to determine the bundle where the fields belong to.

For the M2-brane on the complex curve, in order to have covariant constant spinors, which generate the SUSY charge, we need to do topological twist. Such topological twist will make the SUSY variation different from the standard NLSM, and will allow fermions to have no bosonic pairs. Such net fermions will generate the coherent sheaves over the moduli space and change the spectrum of the ground state.

\paragraph{Spinor bundle}

The fields in the 0+1d NLSM come from the semi-classical quantization\cite{Shifman:2012zz,Gao:2025geq} of the zero modes of the time-independent equation of motion. For the 11d spinor field, its equation of motion is the covariant translation of the pull-back bundle from the 11d spinor bundle to the 3d word-volume. So we need to analysis this pull-back bundle at first.

On a K\"ahler manifold $X$ the spinor bundle factorizes as
\be
\label{eq:spin-bundle}
\op{Spin}(X)=K_X^{1/2}\otimes\bigoplus_{i=0}^n\Lambda^{0,i}T^*X\,.
\ee
Writing the pre-twist 11d spinor as the product of a 3d spinor and a 8d spinor $\psi=\psi_3\otimes\psi_8$. Here
$\psi_3\in K_{\Sigma}^{1/2}\otimes (\mc{O}\oplus\Omega^{0,1}(\Sigma))$ and $\psi_8\in K_{N_{\Sigma|X}}^{1/2}\otimes \bigoplus_{i=0}^4\Lambda^{i}N_{\Sigma|\mb{R}^4\times X}$,
the twist absorbs the normal factor into the worldvolume spinor,
$\tilde\psi_3\in K_{\Sigma}^{1/2}\otimes K_{N_{\Sigma|\mb{R}^4\times X}}^{1/2}\otimes(\mc{O}\oplus\Omega^{0,1}(\Sigma))$, $\tilde{\psi}_8\in \bigoplus_{i=0}^4\Lambda^{i}N_{\Sigma|\mb{R}^4\times X}$. The Calabi--Yau condition $K_{\Sigma}^{1/2}\otimes K_{N_{\Sigma|X}}^{1/2}=\left( K_X|_{\Sigma}\right)^{\frac{1}{2}}=\mc{O}$
then yields the expected trivially-twisted pull-back spinor bundle
\begin{equation}
    (\mc{O}\oplus K^{-1}_{\Sigma})\otimes \bigoplus_{i=0}^4\Lambda^{i}N_{\Sigma|\mb{R}^4\times X}\,.
\end{equation}

\paragraph{Gamma matrix.}

Before we proceed, we need to review the eight- and eleven-dimensional Clifford algebra used
above~\cite{Park:2022pjv,Becker:2006dvp}. In a suitable basis the $8d$ gamma
matrices are off-diagonal with real $8\times8$ blocks $r^i$, and $\gamma^*$ is
diagonal,
\be
\gamma^i=\begin{pmatrix}0 & r^i\\ (r^i)^T & 0\end{pmatrix}\,,\qquad
\gamma^*=\begin{pmatrix}\1 & 0\\ 0 & -\1\end{pmatrix}\,,
\ee
so $\gamma^i$ exchanges the two chiralities $8_s\leftrightarrow 8_c$: writing an
$8_s$ spinor as $(\eta^T,0)$ and an $8_c$ spinor as $(0,\eta^T)$, the projectors
\be
P_\pm=\tfrac12(1\pm\gamma^*)
\ee
pick out $8_s$ or $8_c$. One has $\gamma^i=(\gamma^i)^T=(\gamma^i)^\dagger$ and,
in complex coordinates $m_a,a=1,\cdots,4$, $\gamma^{\bar m_a}=(\gamma^{m_a})^\dagger$. On the K\"ahler
normal space the algebra takes the oscillator form~\cite{Witten:1988ze}
\be
\{\gamma^{m_a},\gamma^{\bar m_b}\}=2G^{m_a\bar m_b}\,,\qquad
\{\gamma^{m_a},\gamma^{m_b}\}=\{\gamma^{\bar m_a},\gamma^{\bar m_b}\}=0\,,
\ee
where we assume the metric is diagonal, that is $G^{m_{a}\bar{m}_{b}}=0$ for $a\neq b$. So the $\frac{1}{\sqrt{2}}\gamma^{\bar m_a}$ act as creation and the $\frac{1}{\sqrt{2}}\gamma^{m_b}$ as annihilation
operators. Choosing a Clifford vacuum $|0\rangle\in 8_c$ with $\gamma^{m_a}|0\rangle=0$,
the $8d$ Majorana spinor is spanned by the $16$ states
\be
\begin{aligned}
&|0\rangle\, ,\\
&|a\rangle=2^{-1/2}\gamma^{\bar m_{a}}|0\rangle\, ,\\ 
&|ab\rangle=2^{-1}\gamma^{\bar m_a}\gamma^{\bar m_b}|0\rangle \, , \, a<b \, ,\\
&|abc\rangle=2^{-3/2}\gamma^{\bar m_a}\gamma^{\bar m_b}\gamma^{\bar m_c}|0\rangle\, , \, a<b<c\, ,\\
&|1234\rangle=2^{-2}\gamma^{\bar m_{1}}\gamma^{\bar m_{2}}\gamma^{\bar m_{3}}\gamma^{\bar m_{4}}|0\rangle\,,
\end{aligned}
\ee
i.e. by the $(0,p)$-forms $\Omega^{\bar{m}_1\dots \bar{m}_p}_{N_{\Sigma|\mb{R}^4\times X}}$; the odd-form subset $\{|a\rangle,|abc\rangle\}$ spans $8_s$. Taking complex conjugate, the vacuum becomes
\be
\gamma^{m^{a}}|0\rangle=0\, \xrightarrow{\mathrm{c.c.}}\, \gamma^{\bar{m}^{a}}|0\rangle^*=0\implies |0\rangle^*\sim|1234\rangle \, .
\ee
The normalization of these two states are related by $\langle 1234|1234\rangle=\prod_{a=1}^4 G^{m_{a}\bar{m}_{a}}\langle 0|0\rangle$. After fixing an arbitrary phase, we find $|0\rangle^*=\frac{1}{\mathcal{G}}|1234\rangle$ where $\mathcal{G}=\sqrt{ \prod_{a=1}^4 G^{m_{a}\bar{m}_{a}} }$. Other complex conjugation relations can be obtained as for example
\be
\begin{aligned}
&|1\rangle^*=\frac{\gamma^{m_{1}}}{\sqrt{ 2 }}|0\rangle^*=\frac{G^{m_{1}\bar{m}_{1}}}{\mathcal{G}}|234\rangle \, ,\\
&|12\rangle^*=\frac{{\gamma^{m_{1}}\gamma^{m_{2}}}}{2}|0\rangle^*=-\frac{G^{m_{1}\bar{m}_{1}}G^{m_{2}\bar{m}_{2}}}{\mathcal{G}}|34\rangle \, .
\end{aligned}
\ee
We fix the normalization as $\langle 0|0\rangle=1$. The norm of each state can be computed, e.g.
\be
\langle a|a\rangle=\frac{1}{2}\langle 0| \gamma^{m_{a}}\gamma^{\bar{m}_{a}}|0\rangle=G^{m_{a}\bar{m}_{a}} \, .
\ee

The $32\times 32$
eleven-dimensional gamma matrices factorize as
\be
\Gamma^\alpha=\tilde\gamma^\alpha\otimes\gamma^*\,,\qquad \Gamma^i=\1\otimes\gamma^i\,,
\ee
with the $3d$ $SO(1,2)$ gammas realized by real Pauli matrices,
\be
\tilde\gamma^0=i\sigma_2\,,\qquad \tilde\gamma^1=\sigma_1\,,\qquad \tilde\gamma^2=\sigma_3\,.
\ee
We choose the following basis for a 3d Dirac spinor
\be
v_{\pm}=\frac{1}{\sqrt{ 2 }} \begin{pmatrix}
1 \\
\pm i
\end{pmatrix} \, .
\ee
It satisfies $v_{\pm}^*=v_{\mp}$, and the orthogonal relation
\be
v_{\pm}^{\dagger}\tilde{\gamma}^0v_{\pm}=\pm i \, , \, v_{\mp}^{\dagger}\tilde{\gamma}^0 v_{\pm}=0 \, .
\ee
Recall that the $3d$ part corresponds to $(\mc{O}\oplus \Omega^{0,1}(\Sigma))$ in the whole pull-back spinor bundle, we can identify $v_{+}$ with $\Gamma(\Sigma,\mathcal{O}_{\Sigma})$ and $v_{-}$ with $\Gamma(\Sigma,K_{\Sigma}^{-1})$.

So we know that for each spinor component $\psi_{\bar{m}_1\dots \bar{m_p},\pm}$ above, it should correspond to the bundle $(\mc{O}\oplus \Omega^{0,1}(\Sigma))\otimes\Omega^{\bar{m}_1\dots \bar{m}_p}_{N_{\Sigma|\mb{R}^4\times X}}$. Using the $8d$ metric $G^{m_a\bar m_a}$ one can transform a section of $\Omega^{\bar{m}_1\dots \bar{m}_p}_{N_{\Sigma|\mb{R}^4\times X}}$ into a section of $\Lambda^pN_{\Sigma|\mb{R}^4\times X}$. Thus the zero modes of the pull-back spinor bundle are determined by $H^0\left(\Sigma,\wedge^p N_{\Sigma|\mb{R}^4\times X}\right)\oplus H^1\left(\Sigma,\wedge^p N_{\Sigma|\mb{R}^4\times X}\right)$.

The $\kappa$-symmetry projector along the membrane worldvolume is then
$\Gamma_{\op{M2}}=\Gamma^0\Gamma^1\Gamma^2=\1\otimes\gamma^*$, so gauge-fixing
$P_-\Theta=0$ leaves the $8d$ spinor in $8_s$, the basis underlying
expansion~\eqref{eq:Theta-expansion}.

\paragraph{From the $11d$ spinor to $3d$ superfields.}
The leading kinetic terms of the Nambu-Goto action are
\be
\label{eq:LNambu-leading}
\mc{L}=\sqrt{-g_0}\left(\tfrac12 G_{ij}\,g_0^{\alpha\beta}\d_\alpha X^i\d_\beta X^j
+\bar\Theta\,\Gamma^\alpha D_\alpha\Theta+\cdots\right)\,,
\ee
After fixing the gauge, the remaining supersymmetry variation~\cite{Bagger:2006sk} is
\be
\label{eq:susy-Theta}
\delta\Theta=G_{ij}\d_\alpha X^i\,\Gamma^\alpha\Gamma^j\,\epsilon_\Theta\,,\qquad
\delta X^i=\bar\epsilon_\Theta\Gamma^i\Theta\,.
\ee
To form superfields, $\Theta$ must be reorganized into $3d$ spinors carrying the
same internal (normal-bundle) index as the bosons, i.e. into chiral multiplets
$\Phi^{m_a}=X^{m_a}+\theta\psi^{m_a}+\theta\theta F^{m_a}$ with $\delta X^i=\bar\epsilon\psi^i$,
$\delta\psi^i=\tilde\gamma^\alpha\d_\alpha X^i\epsilon$. Fixing the
$\kappa$-symmetry with the worldvolume projector $\Gamma_{\op{M2}}=\Gamma^0\Gamma^1\Gamma^2=\1\otimes\gamma^*$ leaves the $8d$ spinor part in $8_s$, i.e. $\Theta$
sits in the $(2,8_s)$ representation of $SO(1,2)\times SO(8)$. Using the K\"ahler
Clifford vacuum $|0\rangle$ with $\gamma^{m_a}|0\rangle=0$, the $8_s$ (odd-form)
components expand $\Theta$ as
\be
\label{eq:Theta-expansion}
\Theta=\sum_{s=\pm}v_{s}\otimes (\psi_{\bar m_a,s}|a\rangle
+\psi_{\bar m_a\bar m_b\bar m_c,s}|abc\rangle)\,,
\ee
The 11d Majorana condition is imposed by $\Theta^*=\Theta$, which is in components given by
\be
\begin{aligned}
&\psi_{\bar{m}_{1},\pm}^*=\frac{\mathcal{G}}{G^{m_{1}\bar{m}_{1}}}\psi_{\bar{m}_{2}\bar{m}_{3}\bar{m}_{4},\mp} \, , \, \psi_{\bar{m}_{2},\pm}^*=-\frac{\mathcal{G}}{G^{m_{2}\bar{m}_{2}}}\psi_{\bar{m}_{1}\bar{m}_{3}\bar{m}_{4},\mp} \, , \\
&\psi_{\bar{m}_{3},\pm}^*=\frac{\mathcal{G}}{G^{m_{3}\bar{m}_{3}}}\psi_{\bar{m}_{1}\bar{m}_{2}\bar{m}_{4},\mp} \, , \, \psi_{\bar{m}_{4},\pm}^*=-\frac{\mathcal{G}}{G^{m_{4}\bar{m}_{4}}}\psi_{\bar{m}_{1}\bar{m}_{2}\bar{m}_{3},\mp} \,. \\
\end{aligned}
\ee

\paragraph{Reduction of the kinetic term.}
With $\bar\Theta=\Theta^\dagger\Gamma^0$ and $\bar\psi_{m_a}=\psi_{\bar m_a}^T$,
$(\gamma^{m_a})^\dagger=\gamma^{\bar m_a}$, one has
\be
\bar\Theta=\sum_{s=\pm}v_{s}^\dagger\otimes (\psi_{\bar m_a,s}^*\langle a|
+\psi_{\bar m_a\bar m_b\bar m_c,s}^*\langle abc|)\,.
\ee
Since $\langle0|\gamma^{\bar m}=\gamma^m|0\rangle=0$ and unequal form-degrees have
vanishing overlap, the Dirac term becomes a sum of decoupled sigma-model kinetic
terms,
\be
\begin{aligned}
    \bar\Theta\,\Gamma^\alpha D_\alpha\Theta
=\sum_{s_1,s_2=\pm} (&G^{m_a\bar m_a}\psi_{m_a,s_1}^*v_{s_1}^\dagger \tilde\gamma^0\tilde\gamma^\alpha v_{s_2} D_\alpha\psi_{\bar m_a,s_2}\\
&+G^{m_1\bar m_1}G^{m_2\bar m_2}G^{m_3\bar m_3}\,
\psi_{m_1m_2m_3,s_1}^*v_{s_1}^\dagger \tilde\gamma^0\tilde\gamma^\alpha v_{s_2} D_\alpha\psi_{\bar m_1\bar m_2\bar m_3,s_2})\,,
\end{aligned}
\ee
after a suitable normalization of $\psi$. This reproduces the sigma-model
fermion kinetic term in~\eqref{eq:Dterm} and confirms that the equation of
motion of the superfield formalism and the  harmonic
operator~\eqref{eq:Kahler-eom} are indeed equivalent.

\paragraph{Preserved SUSY.}On the worldvolume $\mathcal{T}\times\Sigma$ of the M2-brane wrapped on $\Sigma\subset X$, there are four preserved supercharges. Since a covariant constant spinor cannot exist on general $\Sigma$, this 3d theory on $\mathcal{T}\times\Sigma$ with four supercharges should be a topological twisted theory. The topological twist is implemented by absorbing the normal factor into the worldvolume spinor as stated before, and such that an $\mathcal{N}=4$ SQM on $\mathcal{T}$ is preserved.

The SUSY variation \eqref{eq:susy-Theta} is parametrized by an 11d spinor $\epsilon_{\Theta}$ in the representation $(2,8_c)$, which can be expanded in the basis as
\be
\epsilon_{\Theta}=\sum_{s=\pm} v_{s}\otimes (\epsilon_{s}|0\rangle+\epsilon_{\bar{m}_{a}\bar{m}_{b},s}|ab\rangle+\epsilon_{\bar{m}_{1}\bar{m}_{2}\bar{m}_{3}\bar{m}_{4},s}|1234\rangle) \, .
\ee
subject to the Majorana condition:
\be
\begin{aligned}
\epsilon_{\pm}^*&=\mathcal{G}\epsilon_{\bar{m}_{1}\bar{m}_{2}\bar{m}_{3}\bar{m}_{4},\mp} \, ,\, \epsilon_{\bar{m}_{1}\bar{m}_{2},\pm}^*=-\frac{\mathcal{G}}{G^{m_{1}\bar{m}_{1}}G^{m_{2}\bar{m}_{2}}}\epsilon_{\bar{m}_{3}\bar{m}_{4},\mp} \, ,\\
\epsilon_{\bar{m}_{1}\bar{m}_{3},\pm}^*&=\frac{\mathcal{G}}{G^{m_{1}\bar{m}_{1}}G^{m_{3}\bar{m}_{3}}}\epsilon_{\bar{m}_{2}\bar{m}_{4},\mp} \, ,\, \epsilon_{\bar{m}_{1}\bar{m}_{4},\pm}^*=-\frac{\mathcal{G}}{G^{m_{1}\bar{m}_{1}}G^{m_{4}\bar{m}_{4}}}\epsilon_{\bar{m}_{2}\bar{m}_{3},\mp} \, ,\\
\end{aligned}
\ee
On the M2-brane wrapped on a holomorphic curve in the Calabi-Yau threefold, only $\frac{1}{4}$ of these parameters are preserved. The preserved SUSY parameters are covariant constant spinors on $\Sigma$ which should belongs to sections of the trivial line bundle of $\Sigma$:
\be
\begin{aligned}
    \epsilon_+,\epsilon_{\bar z_{1}\bar z_{2},+}&\in \Gamma(\Sigma,\mc{O}_\Sigma) \, , \\
    \epsilon_{\bar w_1\bar w_2,-}, \epsilon_{\bar z_{1}\bar z_{2}\bar w_1\bar w_2,-}&\in\Gamma(\Sigma,K_{\Sigma}^{-1}\otimes \Lambda^2N_{\Sigma |X})=\Gamma(\Sigma,\mathcal{O}_\Sigma) \, .
\end{aligned}
\ee
where we used the Calabi-Yau condition $\Lambda^2N_{\Sigma |X}\cong K_{\Sigma}$. There are two complex SUSY parameters $(\epsilon_+,\epsilon_{\bar z_{1}\bar z_2,+})$ whose complex conjugates are $(\epsilon_{\bar w_1\bar w_2,-}, \epsilon_{\bar z_{1}\bar z_{2}\bar w_1\bar w_2,-})$. So we obtain
\be
\epsilon_{\text{pres}}=\epsilon_{+}v_{+}\otimes |0\rangle+\epsilon_{\bar{z}_{1}\bar{z}_{2},+}v_{+}\otimes |12\rangle-\frac{G^{z_{1}\bar{z}_{1}}G^{z_{2}\bar{z}_{2}}}{\mathcal{G}}\epsilon_{\bar{z}_{1}\bar{z}_{2},+}^*v_{-}\otimes |34\rangle+\frac{1}{\mathcal{G}}\epsilon_{+}^*v_{-}\otimes |1234\rangle \, .
\ee
This kind of preserved SUSY on $\mathcal{T}\times\Sigma$ is the A-twist of 3d $\mathcal{N}=4$ theory in \cite{Bullimore:2021auw,Ferrari:2019cni}. The 1d $\mathcal{N}=4$ SQM on the worldline has supermultiplets obtained by dimensional reduction of 2d $\mathcal{N}=(2,2)$ chiral multiplets.

Then the preserved SUSY variations of the bosonic fields are
\be
\delta X^m=\bar{\epsilon}_{\text{pres}}\Gamma^m\Theta=-\epsilon_{\text{pres}}^{\dagger}(\tilde{\gamma}^0\otimes \gamma^m)\Theta \, , \, \delta X^{\bar{m}}=-\epsilon_{\text{pres}}^{\dagger}(\tilde{\gamma}^0\otimes \gamma^{\bar{m}})\Theta \, .
\ee
Expand the spinor fields in the basis and we find the SUSY variations of the bosonic fields are
\be
\label{eq:susy-variations}
\begin{aligned}
\delta X^{z_1}&=-i\sqrt2\left(\epsilon_+^{  *}\psi_{\bar z_1,+}^{ }-\epsilon_{\bar z_1\bar z_2,+}^{ }\psi_{\bar z_2,+}^{  *}\right),\\
\delta X^{z_2}&=-i\sqrt2\left(\epsilon_+^{  *}\psi_{\bar z_2,+}^{ }+\epsilon_{\bar z_1\bar z_2,+}^{ }\psi_{\bar z_1,+}^{  *}\right),\\
\delta X^{\bar z_1}&=+i\sqrt2\left(\epsilon_+^{ }\psi_{\bar z_1,+}^{  *}
-\epsilon_{\bar z_1\bar z_2,+}^{  *}\psi_{\bar z_2,+}^{ }\right),\\
\delta X^{\bar z_2}&=+i\sqrt2\left(
\epsilon_+^{ }\psi_{\bar z_2,+}^{  *}
+\epsilon_{\bar z_1\bar z_2,+}^{  *}\psi_{\bar z_1,+}^{ }
\right),\\
\delta X^{w_1}
&=-i\sqrt2\,
\left(
G^{w_1\bar w_1}\epsilon_+^{  *}\psi_{\bar w_1,+}^{ }
- \sqrt{ G^{w_{1}\bar{w}_{1}}G^{w_{2}\bar{w}_{2}} }\,
\epsilon_{\bar z_1\bar z_2,+}^{  *}\psi_{\bar w_2,-}^{  *}
\right),\\
\delta X^{w_2}
&=-i\sqrt2\,
\left(
G^{w_2\bar w_2}\epsilon_+^{  *}\psi_{\bar w_2,+}^{ }
+ \sqrt{ G^{w_{1}\bar{w}_{1}}G^{w_{2}\bar{w}_{2}} }\,
\epsilon_{\bar z_1\bar z_2,+}^{  *}\psi_{\bar w_1,-}^{  *}
\right),\\
\delta X^{\bar w_1}
&=+i\sqrt2\,
\left(
G^{w_1\bar w_1}\epsilon_+^{ }\psi_{\bar w_1,+}^{  *}
- \sqrt{ G^{w_{1}\bar{w}_{1}}G^{w_{2}\bar{w}_{2}} }\,
\epsilon_{\bar z_1\bar z_2,+}^{ }\psi_{\bar w_2,-}^{ }
\right),\\
\delta X^{\bar w_2}
&=+i\sqrt2\,
\left(
G^{w_2\bar w_2}\epsilon_+^{ }\psi_{\bar w_2,+}^{  *}
+ \sqrt{ G^{w_{1}\bar{w}_{1}}G^{w_{2}\bar{w}_{2}} }\,
\epsilon_{\bar z_1\bar z_2,+}^{ }\psi_{\bar w_1,-}^{ }
\right)\,,
\end{aligned}
\ee
where we have substituted $\psi_{\bar{z}_{1}\bar{w}_{1}\bar{w}_{2},-}\sim\psi_{\bar{z}_{2},+}^*,\psi_{\bar{z}_{1}\bar{z}_{2}\bar{w}_{1},+}\sim\psi_{\bar{w}_{2},-}^*$ and so on, and we set the flat $\mathbb{R}^4$ metric to be unit. Thus we obtain the SUSY pairing relations:
\be
\label{eq:susy-pairing}
\begin{aligned}
&X^{w_{1}} \longleftrightarrow \psi_{\bar{w}_{1},+}, \psi_{\bar{w}_{2},-}^*\sim \psi_{\bar{z}_{1}\bar{z}_{2}\bar{w}_{1},+}\, ,\\
&X^{\bar{w}_{1}} \longleftrightarrow  \psi_{\bar{w}_{1},+}^*\sim \psi_{\bar{z}_{1}\bar{z}_{2}\bar{w}_{2},-},\psi_{\bar{w}_{2},-}\, ,\\
&X^{w_{2}} \longleftrightarrow \psi_{\bar{w}_{2},+}, \psi_{\bar{w}_{1},-}^*\sim \psi_{\bar{z}_{1}\bar{z}_{2}\bar{w}_{2},+}\, ,\\
&X^{\bar{w}_{2}} \longleftrightarrow  \psi_{\bar{w}_{2},+}^*\sim \psi_{\bar{z}_{1}\bar{z}_{2}\bar{w}_{1},-},\psi_{\bar{w}_{1},-}\, .\\
\end{aligned}
\ee

\paragraph{Little group}
Decompose the $R$-symmetry as $SO(8)\supset SO(4)_J\times SO(4)_N$, with $z_i$ ($i=1,2$)
the complex coordinates of the $\mb{R}^{1,4}$ little group $SO(4)_J$ and $w_i$
those of the normal directions. The preceding $8d$ complex coordinates are $(m_{1},m_{2},m_{3},m_{4})=(z_{1},z_{2},w_{1},w_{2})$. Writing $z_1=x^1+ix^3$, $z_2=x^2+ix^4$, the two
$SU(2)$ factors of $SO(4)_J$ are generated by
\be
\ba
&J_{1,+}=T^{z_1z_2}\,,\quad J_{1,-}=-T^{\bar z_1\bar z_2}\,,\quad
J_{1,0}=\tfrac12(T^{z_1\bar z_1}+T^{z_2\bar z_2})\,,\\
&J_{2,+}=T^{z_2\bar z_1}\,,\quad J_{2,-}=T^{z_1\bar z_2}\,,\quad
J_{2,0}=\tfrac12(-T^{z_1\bar z_1}+T^{z_{2}\bar z_{2}})\,,
\ea
\ee
with $T^{ij}=\tfrac14[\gamma^i,\gamma^j]$, whose action on the $(0,p)$-form
basis is
\be
[T^{z_iz_j},\gamma^{\bar z_j}]=\gamma^{z_i}\,,\quad
[T^{z_i\bar z_j},\gamma^{z_j}]=\gamma^{z_i}\,,\quad
[T^{z_i\bar z_j},\gamma^{\bar z_i}]=-\gamma^{\bar z_j}\,,\quad
[T^{\bar z_i\bar z_j},\gamma^{z_j}]=\gamma^{\bar z_i}\,.
\ee
From this we can obtain the charge under the little group of all the relevant fields as shown in Table \ref{tab:boson_fermion_zero}. The first $SU(2)_1$ has an action on the fermionic fields:
\be
\begin{aligned}
    J_{1,+}&: \psi_{\bar{w}_i,\pm}\to\psi_{\bar{z}_{1}\bar{z}_{2}\bar{w}_{i},\pm}\, ,\\
    J_{1,-}&:\psi_{\bar{z}_{1}\bar{z}_{2}\bar{w}_{i},\pm}\to \psi_{\bar{w}_i,\pm} \, .
\end{aligned}
\ee

\subsection{Zero modes}
\label{subsec:zeromodes}
The quantization of the Nambu-Goto action is quite difficult. But if we focus on the ground state, we can only quantize with semi-classical approximation \cite{Witten:1996qb,Shifman:2012zz,Gao:2025geq} that $x\rightarrow x(t)$, $\psi\rightarrow \psi(t)$. Then we can obtain a generalized NLSM, whose ground states are given by the cohomology group.

\paragraph{Bosonic versus fermionic zero modes.}

The bosonic zero modes are the geometric deformations of $\Sigma$ inside $\CY$,
the holomorphic sections of the normal bundle,
\be
n_b=h^0(\Sigma,N_{\Sigma|X})\,,
\ee
which coordinatize (a patch of) the moduli space $\mc{M}$ and the two complex flat direction $z_1,z_2$ in the Minkowski space $\mb{R}^4$. From the twist above, we can convert the spinor
components of~\eqref{eq:Theta-expansion} into worldvolume $(p,q)$-forms with the
covariantly-constant $\Omega$ on the CY manifold and $G_{\sigma\bar\sigma}$ gives the fermion zero
modes of Table~\ref{tab:boson_fermion_zero}. 

$\psi_{\bar{z}\bar{w}_1\bar{w}_2,\pm}$ corresponds to the $8d$ 2-form bundle $A^{\bar{w}_1\bar{w}_2}$. Here we ignore the index $\bar{z}$ because it is on the flat space $\mb{R}^4$ and does not change the cohomology group. By contract with $\Omega$ and $G$, we can obtain the 1-form bundle $\Omega^{1,0}(\Sigma)$ by $s^{\sigma}=A^{\bar{w}_1\bar{w}_2}\Omega^{\sigma w_1 w_2}G_{w_1\bar{w}_1}G_{w_2\bar{w}_2}$. So the related zero modes from $(\mc{O}\oplus \Omega^{0,1}(\Sigma))\otimes \Omega^{1,0}(\Sigma)$ gives the zero modes from $H^{1,0}(\Sigma)$ and $H^{1,1}(\Sigma)\simeq H^{0,0}(\Sigma)$. $\psi_{\bar{z},\pm}$ corresponds to the $8d$ trivial bundle $\mc{O}_{\Sigma}$ . So the final zero modes are $H^{0,0}(\Sigma)$ and $H^{0,1}(\Sigma)\simeq H^{1,0}(\Sigma)$. The $H^{0,0}(\Sigma)$ zero modes paired with $X^z$ and $X^{\bar{z}}$ constitute $\mathcal{N}=4$ $(4,4,0)$ chiral multiplets, whereas the $H^{0,1}(\Sigma)$ zero modes are $\mathcal{N}=4$ $(0,4,4)$ Fermi multiplets. 
They do not generate new moduli space, but only generate the basis of the bundle over the moduli space.

$\psi_{\bar{w},\pm}$ and $\psi_{\bar{z}_1\bar{z}_2\bar{w},\pm}$ both correspond to the conjugate of the $8d$ normal bundle $\bar{N}_{\Sigma|X}$. In the following, for convenience, we just write normal bundle $N_{\Sigma|X}$, whose cohomology is conjugate to $\bar{N}_{\Sigma|X}$. The related zero modes are $H^{0}(\Sigma,N_{\Sigma|X})$ and $H^1(\Sigma,N_{\Sigma|X})$. In CY3-fold, for the complex curve $\Sigma$, we have $h^0(\Sigma,N_{\Sigma|X})= h^1(\Sigma,N_{\Sigma|X})$. They have the same number of zero modes with $X^{w}$ and $X^{\bar{w}}$. So they are $(2,4,2)$ chiral multiplets and generate $H^{0}(\Sigma,N_{\Sigma|X})$ $\mathcal{N}=4$ NLSM. 

The fermion zero modes can be summarized as
\be
\begin{aligned}
&\psi_{\bar{z}_{i},+}\in H^0(\Sigma) \, , \, \psi_{\bar{z}_{i},-}\in H^1(\Sigma)\, ,\\
&\psi_{\bar{w}_{i},+}\in H^0(\Sigma,N_{\Sigma |X}) \, , \, \psi_{\bar{w}_{i},-}\in H^1(\Sigma,N_{\Sigma |X})\, ,\\
&\psi_{\bar{z}_{i}\bar{w}_{1}\bar{w}_{2},+}\in H^0(\Sigma,K_{\Sigma}) \, , \, \psi_{\bar{z}_{i}\bar{w}_{1}\bar{w}_{2},-}\in H^1(\Sigma,K_{\Sigma}) \, ,\\
&\psi_{\bar{z}_{1}\bar{z}_{2}\bar{w}_{i},+}\in H^0(\Sigma,N_{\Sigma |X}) \, , \, \psi_{\bar{z}_{1}\bar{z}_{2}\bar{w}_{i},-}\in H^1(\Sigma,N_{\Sigma |X}) \, .
\end{aligned}
\ee

\begin{table}[h]
\centering
\begin{tabular}{c|c|c|c|c}
 Boson & Fermion & $\op{SU}(2)_1\times \op{SU}(2)_2$ of fermion&Bosonic zero modes&Fermionic zero modes\\\hline
   $X^{z_i}$, $X^{\bar z_i}$ & $\psi_{\bar{z}_i,+},\psi_{\bar{z}_i\bar{w}_1\bar{w}_2,-}$ & $\left(0,\pm\tfrac12\right)$, $\left(0,\pm\tfrac12\right)$&$H^{0,0}(\Sigma)^{\oplus 4}$&$H^{0,0}(\Sigma)^{\oplus 4}$\\
  & $\psi_{\bar z_i,-},\psi_{\bar z_i\bar w_1 \bar w_2,+}$ & $\left(0,\pm\tfrac12\right)$, $\left(0,\pm\tfrac12\right)$&&$H^{0,1}(\Sigma)^{\oplus 4}$\\
$X^{w_i}$  & $\psi_{\bar{w_i},+},\psi_{\bar z_1\bar z_2 \bar w_i,+}$ & $\left(-\tfrac12,0\right)$, $\left(\tfrac12,0\right)$&$H^{0}(\Sigma,N_{\Sigma|X})$& $H^{0}(\Sigma,N_{\Sigma|X})^{\oplus 2}$\\
$X^{\bar{w}_i}$   & $\psi_{\bar w_j,-}, \psi_{\bar{z}_1\bar{z}_2\bar{w}_j,-}$ &$\left(-\tfrac12,0\right)$, $\left(\tfrac12,0\right)$ & $H^{0}(\Sigma,N_{\Sigma|X})$ & $H^{1}(\Sigma,N_{\Sigma|X})^{\oplus 2}$
\end{tabular}
\caption{The bosons and fermions related by the SUSY variation in the $3d$
worldvolume theory for the M2-branes on curves in CY3-fold. Here $z_i$ are the complex coordinates of $SO(4)_J$ and
$w_i$ are those of $SO(4)_N$, with $i\neq j$. For a curve in a CY3-fold, $h^{0}(\Sigma,N_{\Sigma|X})=h^{1}(\Sigma,N_{\Sigma|X})$}
\label{tab:boson_fermion_zero}
\end{table}

\paragraph{The Lefschetz group.}
Following~\cite{Witten:1996qb}, $SU(2)_1$ is identified with the Lefschetz
$SU(2)$ of $\mc{M}$. The unbroken structure group of the flat $\mb{R}^{1,4}$
directions is $SO(4)_J=SU(2)_1\times SU(2)_2$, under which the preserved supercharges sit
in $2(\frac{1}{2},0)$; since every fermionic mode is obtained by acting a broken
supercharge on a bosonic mode, the fermionic zero modes -- and with them the
$(p,q)$-cohomology of $\mc{M}$ -- carry a non-trivial $SU(2)_1$ action. The
three generators act on forms as
\be
\label{eq:Lefschetz-gen}
J_3:\ \Omega^{p,q}\mapsto\frac{p+q-\op{dim}\mc{M}}{2}\,\Omega^{p,q}\,,\qquad
J_+:\ \Omega^{p,q}\mapsto J\wedge\Omega^{p,q}\,,\qquad J_-=J_+^\dagger\,.
\ee
Using the dictionary $\bar\psi^i\leftrightarrow\dd z^i$,
$\psi^i\leftrightarrow g^{\bar j i}\d_{\bar j}$~\cite{Hori:2003ic}, the
$(p,q)$-form cohomology takes the form $(\bar\psi^i)^p(\bar\psi^{\bar i})^q|0\rangle$,
on which the axial $U(1)_A$ charge equals $J_3=(p+q-\op{dim}\mc{M})/2$; here the
fermions transform under the vector/axial symmetries as
\be
\bar\psi^i\mapsto e^{i(-\alpha+\beta)}\bar\psi^i\,,\quad
\bar\psi^{\bar i}\mapsto e^{i(\alpha+\beta)}\bar\psi^{\bar i}\,,\quad
\psi^i\mapsto e^{i(\alpha-\beta)}\psi^i\,,\quad
\psi^{\bar i}\mapsto e^{i(-\alpha-\beta)}\psi^{\bar i}\,.
\ee
A raising operator that exchanges $\psi^i\to\bar\psi^i$ is
$J_+\psi^i=g_{l\bar i}\bar\psi^l\bar\psi^{\bar i}\psi^i$, which in form language
gives
\be
g_{l\bar i}\,\dd z^l\dd\bar z^{\bar i}\,(g^{\bar j i}\d_{\bar j})
=g_{l\bar i}g^{\bar i i}\,\dd z^l=\delta^i_l\,\dd z^l=\dd z^i\,,
\ee
whose coefficient $g_{l\bar i}\dd z^l\dd\bar z^{\bar i}$ is precisely the K\"ahler
form. Thus $J_+$ is the Lefschetz raising operator. From
Table~\ref{tab:boson_fermion_zero}, $SU(2)_1$ acts non-trivially only on the
$\Phi^{w},\Phi^{\bar w}$ fermions,
\be
\ba
J_{1,+}:\ &\gamma^{\bar z_1}\gamma^{\bar z_2}\gamma^{\bar w_i}|0\rangle\to\gamma^{\bar w_i}|0\rangle\,,\quad
\psi_{\bar w_i,\pm}\to\psi_{\bar z_1\bar z_2\bar w_i,\pm}\,,\\
J_{1,-}:\ &\gamma^{\bar w_i}|0\rangle\to\gamma^{\bar z_1}\gamma^{\bar z_2}\gamma^{\bar w_i}|0\rangle\,,\quad
\psi_{\bar z_1\bar z_2\bar w_i,\pm}\to\psi_{\bar w_i,\pm}\,,
\ea
\ee
which is exactly the $J_\pm$ action that raises and lowers the Lefschetz degree.
The worldvolume construction therefore realizes the Lefschetz $SU(2)$ as a
subgroup $\op{SU}(2)_1$ of the $R$-symmetry $SO(4)_J$, confirming Witten's identification.

Similarly, we can find that the $H^{0,0}(\Sigma)$ and $H^{0,1}(\Sigma)$ parts are in the spin-$\frac{1}{2}$ representation under $\op{SU}(2)_2$. This confirm the statement in that the $\op{SU}(2)_2$ is the Lefschetz group of the Jacobian fibration direction. \cite{Gopakumar:1998jq}

\subsection{BPS spectrum}
From Table~\ref{tab:boson_fermion_zero}, the spectrum related $H^0(\Sigma,N_{\Sigma|X})$ sector is the spectrum of a NLSM with the target space as the moduli space $\mc{M}$. And the spin is determined by
the Lefschetz decomposition $\sum_j a_j(j,0)$ of the graded algebra
$H^{*,*}(\mc{M})$. 

For The $H^{0,0}(\Sigma)$ part, although it is not a standard NLSM, it is on the flat space, so the bosonic field will not take place in the ground state. Then we can start from a vacuum $|0\rangle$ with spin $-\frac{1}{2}$. Acting the $\psi_{\bar{z}_1,+}$,$\psi_{\bar{z}_2,+}$ on it we can obtain the state  $|0\rangle$, $\psi_{\bar{z}_1,+}|0\rangle$, $\psi_{\bar{z}_1,+}|0\rangle$ and $\psi_{\bar{z}_2,+}\psi_{\bar{z}_1,+}|0\rangle$. The representation is a half-hypermultiplet $\left[\left(0,\tfrac12\right)+2(0,0)\right]$. 

The $H^{0,1}(\Sigma)$ part is special. Locally it looks like $2g$ free fermions for a genus $g$ curve $\Sigma$. So the related ground state is $\psi_{i_1,+}\psi_{i_2,+}\dots \psi_{i_p,+}|0\rangle$. This generate the basis of the p-form of the Jacobian $\op{Jac}(\Sigma_g)$. Over the moduli space, the arithmetic genus $p_a$ of the curve $\Sigma$ is constant, whereas the geometric genus $g$ may change as the curve may degenerate. In the simple cases of curves with fixed geometric genus, we can obtain the BPS spectrum \cite{Witten:1996qb,Gopakumar:1998jq}
\be
\label{eq:final-spectrum}
\left[\left(0,\tfrac12\right)\oplus 2(0,0)\right]^{g+1}
\otimes\big(\bigoplus_{n}H^n(\mc{M}),0\big)\,.
\ee

In general cases, the Jacobian is not simply a bundle, we need to use the sheaf language to describe this.
$R^q \pi_*\mb{F}$ is sheafification of the presheaf $U\rightarrow H^q(\pi^{-1}(U),\mb{F})$. If we take $\mb{F}=\mb{C}$, $\pi:\h{\mc{M}}\rightarrow \mc{M}$. We can choose arbitrary open set $U$ where $\pi^{-1}(U)=U\times F$. This means that at each point $p\in U$, the Jacobian of the curve $\Sigma_p$ at $p$ is of the same dimension, i.e. the genus $g$ is constant over $U$. Then we can obtain $R^q \pi_*\mb{F}(U)=H^q(U\times F,\mb{F})$
\begin{equation}
    R^q \pi_*\mb{C}(U)=H^q(U\times \op{Jac}(\Sigma),\mb{C})=H^q(\op{Jac}(\Sigma),\mb{C})\simeq \wedge^q H^1(\Sigma)\,.
    \label{eq:spec_sheaf_coho}
\end{equation}
So we can see that at least for the region with constant genus, this sheaf is compatible with our intuition. 

With all above discussion, we can obtain the ground state spectrum 
\begin{equation}
    \left[\left(0,\tfrac12\right)+2(0,0)\right]\otimes \bigoplus_{r,s}H^r(\mc{M}, R^s \pi_*\mb{C})\,,
    \label{eq:rs_coho}
\end{equation}
A state in $H^r(\mc{M}, R^s \pi_*\mb{C})$ has $SU(2)_L\times SU(2)_R$ weight $(\frac{r-d}{2},\frac{s-p_a}{2})$, where $d=\dim_{\mathbb{C}}\mathcal{M}$ and $p_a$ is the arithmetic genus. After arranging these states with corresponding weights into $(j_L,j_R)$ representations, we can find the 5d BPS spectrum.
From the Leray spectral sequence, when $\mc{M}$ is non-singular and $\h{\cal M}$ is reduced with sufficiently mild singularities \cite{Pandharipande:2011jz}, we can obtain that
\begin{equation}
    H^n(\h{\mathcal{M}},\mb{C})=\bigoplus_{r+s=n}H^r(\mc{M}, R^s \pi_*\mb{C})\,.
\end{equation}
We can see that after taking diagonal representation of $SU(2)_{\rm diag}\subset SU(2)_L\times SU(2)_R$, the above spectrum (\ref{eq:rs_coho}) is equivalent to the result of diagonal spectrum in \cite{Gopakumar:1998jq}
\begin{equation}
    \left[\left(\tfrac12\right)+2(0)\right]\otimes \bigoplus_{n}H^{n}(\h{\mc{M}})\,.
    \label{eq:spec_hatM}
\end{equation}
When the moduli space is simple enough that the geometric genus of $\Sigma$ is constant over the whole $\mc{M}$, we can obtain the previous simpler result \eqref{eq:final-spectrum}.

The refined Gopakumar-Vafa invariants are defined as the coefficients in the BPS spectrum of a M2-brane wrapped on curve $\Sigma$ in the class $\beta\in H_2(X,\mathbb{Z})$
\be
\left[(0,\tfrac12)\oplus2(0,0)\right]\otimes \bigoplus_{j_{L},j_{R}}N_{j_{L},j_{R}}^\beta [(j_{L},j_{R})]\,  .
\ee
In section \ref{sec:5dSCFT} we consider some examples of local Calabi-Yau threefolds that engineer 5d SCFTs and compute some refined GV invariants of those geometries.

\subsection{Obstruction}
\label{subsec:obstruction}

In the above discussion, we focused on the linearized equation (\ref{eq:Kahler-eom}). But the original equation (\ref{eq:harmonic-map}) has some additional higher order corrections to the linearized equation (\ref{eq:Kahler-eom}). These corrections will play the role of obstructions. Other than the trivial zero modes from $H^{0,0}(\Sigma)$, the normal deformation $H^0(\Sigma,N_{\Sigma|X})$ and the Jacobian $H^{0,1}(\Sigma)$ are  likely to be obstructed.
\begin{Theorem}\cite[Theorem~1.1]{Hartshorne:2010DT}
\label{thm:hilb}
Let $Y$ be a non-singular closed subvariety of a non-singular projective variety
$X$ over a field $k$. Then:
\begin{enumerate}
\item There is a Hilbert scheme $H$ parametrizing closed subschemes of $X$ with
the same Hilbert polynomial $P$ as $Y$, with a universal family
$W\subset X\times H$ flat over $H$: for any scheme $T$ and any
$W'\subset X\times T$ flat over $T$ with fibers of Hilbert polynomial $P$, there
is a unique $\varphi:T\to H$ with $W'=W\times_H T$.
\item The Zariski tangent space of $H$ at the point $[Y]$ is $H^0(Y,N_Y)$.
\item If $H^1(Y,N_Y)=0$ then $H$ is smooth at $[Y]$, of dimension $h^0(Y,N_Y)$.
\item In general, $\dim_{[Y]}H\geq h^0(Y,N_Y)-h^1(Y,N_Y)$.
\end{enumerate}
\end{Theorem}

Whenever $h^1(\Sigma,N_{\Sigma|X})\neq 0$ the first-order deformations may fail to
integrate to finite ones. The obstruction is measured by the Kuranishi
map~\cite{kodaira2005complex}
\be
\kappa:\ H^0(\Sigma,N_{\Sigma|X})\longrightarrow H^1(\Sigma,N_{\Sigma|X})\,,
\ee
and only those first-order deformations $\alpha$ with $\kappa(\alpha)=0$ extend
to actual deformations. Consequently $h^1$ only bounds the dimension of the
moduli space,
\be
\label{eq:dim-bound}
h^0(\Sigma,N_{\Sigma|X})-h^1(\Sigma,N_{\Sigma|X})\leq\op{dim}\mc{M}\leq h^0(\Sigma,N_{\Sigma|X})\,.
\ee
When the inequality is strict, the true number of bosonic zero modes is smaller
than the naive $h^0$, and the spectrum~\eqref{eq:final-spectrum} must be computed
with the genuine, obstructed moduli space $\mc{M}=\ker\kappa$ rather than the
linearized one. Although Theorem~\ref{thm:hilb} is stated for projective
varieties, every non-compact Calabi--Yau geometry arises as a non-compact limit
of a compact one, so we expect the same conclusion to hold.

For the Jacobian $H^{0,1}(\Sigma)\simeq H^1(\Sigma,\mc{O})$, the obstruction is $H^2(\Sigma,\mc{O})$, which is empty from the dimension of $\Sigma$. So we do not need to worry about the obstruction of the Jacobian.

\paragraph{Supersymmetric quantum mechanics and hypercohomology.} 
The obstructions to extending a
first-order deformation order by order are collected by the Kuranishi map of
section~\ref{subsec:obstruction},
\be
\label{eq:cone-kappa-def}
\kappa:\ H^0(\Sigma,N_{\Sigma|X})\longrightarrow H^1(\Sigma,N_{\Sigma|X})\,,
\ee
which starts at second order, and near $\Sigma$ the moduli space is
$\kappa^{-1}(0)$~\cite{kodaira2005complex}. Through~\eqref{eq:cone-Serre}, $\kappa$
is a holomorphic one-form on $A$. This self-duality of the deformation-obstruction
complex makes the derived moduli space $(-1)$-shifted symplectic, as for the vortex
moduli spaces of ref.~\cite{Bullimore:2021auw}, and the
one-form is locally exact~\cite{Katz:2000ab}.

In the following cases, $\mc{M}=\kappa^{-1}(0)$ are singular. We can add the NLSM of section~\ref{subsec:1dn4} with target $A$ and
a holomorphic superpotential $W$, determined below. The moduli space of
the resulting 1d $\mc{N}=4$ SQM~\cite{Safronov:2021tsf} is the critical
locus~\cite{Bullimore:2021auw}
\be
\label{eq:cone-critical}
\mc{M}=\{dW=0\}\subset A\,.
\ee
If the superpotential satisfies 
\begin{equation}
    \kappa=dW\,,
\end{equation}
The new Landau-Ginzburg model with target space $A$ and superpotential $W$ is equivalent to the NLSM with the target space $\mc{M}$ when $\mc{M}$ is smooth. In fact, we can find that when assume this equivalence is still true for the singular $\mc{M}$, the computation with the new LG model is compatible with the mathematical computation.

The Hilbert space of the LG model is still given by the differential forms on $A$, and the supercharges
of section~\ref{subsec:1dn4} are deformed
to~\cite{Safronov:2021tsf}
\be
\label{eq:cone-supercharges}
\ba
&\overline{Q}_+=-i\big(\bar\partial+\partial W\wedge\big)\,,\qquad
Q_-=-i\big(\partial+\bar\partial\,\overline W\wedge\big)\,,\\
&Q=i\big(Q_-+\overline{Q}_+\big)=d+d\big(W+\overline W\big)\wedge\,.
\ea
\ee
The supersymmetric ground states are the cohomology of
$Q$, the de Rham differential twisted by the
Morse function $2\,\mathrm{Re}\,W$~\cite{Witten:1982im}. This cohomology is computed
by the perverse sheaf $\phi_W$ of vanishing cycles of $W$, which is supported on
$\mc{M}$, so
that~\cite{Bullimore:2021auw}
\be
\label{eq:cone-hypercohomology}
\mc{H}_0\cong\mb{H}^\bullet(\mc{M},\phi_W)\,.
\ee
For $W=0$, \eqref{eq:cone-hypercohomology} reduces to the cohomology of
section~\ref{subsec:1dn4}, with a $(p,q)$-form in degree
$F=p+q-\op{dim}A=2J_3$, so the degree
$F$ of $\mb{H}^F(\mc{M},\phi_W)$ measures the Lefschetz spin.

Below we illustrate the effect of the obstruction on two families of resolved
threefold singularities where the naive counting fails.

\subsubsection{Generalized conifold}
\label{subsubsec:singular-divisor}

The generalized conifold is the hypersurface
\be
x^2+y^2+z^2+w^{2n}=0\,,
\ee
which after $u=ix-y$, $v=ix+y$ becomes
\be
X:\ uv=(z+iw^n)(z-iw^n)\subset\mb{C}^4_{(u,v,z,w)}\,.
\ee
Its small resolution is the blow-up
\be
\widetilde X:\ \left\{\ba
(z-iw^n)\lambda_0&=u\lambda_1\\
(z+iw^n)\lambda_1&=v\lambda_0
\ea\right.\ \subset\mb{C}^4_{(u,v,z,w)}\times\mb{P}^1_{[\lambda_0:\lambda_1]}\,,
\ee
covered by two charts
\be
U_0=\{\lambda_0\neq0\}:\big(\gamma=\tfrac{\lambda_1}{\lambda_0},u,w\big)\,,\qquad
U_\infty=\{\lambda_1\neq0\}:\big(\beta=\tfrac{\lambda_0}{\lambda_1},v,w\big)\,,
\ee
glued on the overlap by
\be
\beta=\gamma^{-1}\,,\qquad v=\gamma(z+iw^n)=\gamma^2 u+2i\gamma w^n\,,\qquad w=w\,.
\ee
The exceptional curve is $C=\{u=w=0\}\subset U_0=\{v=w=0\}\subset U_\infty$.
Linearizing the transition along $C$, the nonlinear term $2i\gamma w^n$ drops out
and the normal bundle is
\be
N_{C/X}\cong\mc{O}(-2)\oplus\mc{O}\,,\qquad h^0(C,N_{C/X})=h^1(C,N_{C/X})=1\,.
\ee
In the $U_0$ patch the single first-order deformation is a shift in $w$, moving
$C$ to $\{u=0,\,w=t\}$; on the overlap this gives
\be
v=2i\gamma t^n=2i\beta^{-1}t^n\,,\qquad w=t\,,
\ee
which is not holomorphic unless $t^n=0$. The deformation is therefore killed by
the obstruction: the Kuranishi map is
\be
\kappa(t\cdot e)=c\,t^n\cdot\eta\,,
\ee
with $e,\eta$ generators of $H^0,H^1$, and the moduli space is the fat point
\be
\mc{M}=\op{Spec}\,\mb{C}[t]/t^n\,.
\ee
Equivalently, the exceptional locus $\{u=w=0\}$ is really $\{u=w^n=0\}$, i.e. $n$
rigid $\mb{P}^1$'s coinciding, so an M2-brane wrapping $C$ yields $n$
hypermultiplets. This is consistent with the unrefined GV
invariants~\cite{Collinucci:2021wty,Collinucci:2021ofd} and with the Higgs
branch (magnetic quiver) of the 4d $\mc{N}=2$ SCFT obtained from IIB on the
generalized conifold~\cite{Closset:2020scj,Wang:2026jad}.

\subsubsection{Case with a singular exceptional divisor: $x^3+y^3+z^3+w^4=0$}
\label{sec:3334}

Consider canonical threefold singularities $X$ admitting a smooth crepant
resolution $\widetilde X$ but a \emph{singular} compact exceptional divisor. For
$x^3+y^3+z^3+w^4=0$ the resolution $(x,y,z,w;\delta)$ gives a smooth
$\widetilde X$,
\be
x^3+y^3+z^3+w^4\delta=0\,,
\ee
with the singular compact divisor
\be
\label{eq:singS}
S:\ \delta=x^3+y^3+z^3=0\,,
\ee
singular at $x=y=z=0$. Compared to the smooth del Pezzo $dP_6$,
$x^3+y^3+z^3+w^3=0$ in $\mb{P}^3_{[x:y:z:w]}$, the Picard number of the singular
$S$ drops to $1$~\cite{Closset:2020afy,Closset:2021lwy}. The $27$ lines
($\mb{P}^1$'s with self-intersection $-1$) of $dP_6$ are the isolated linear
solutions to
\be
\ba
(x+y)(x+\omega y)(x+\omega^2 y)&=(z+w)(z+\omega w)(z+\omega^2 w)=0\,,\\
(x+z)(x+\omega z)(x+\omega^2 z)&=(y+w)(y+\omega w)(y+\omega^2 w)=0\,,\\
(x+w)(x+\omega w)(x+\omega^2 w)&=(y+z)(y+\omega z)(y+\omega^2 z)=0\,,
\ea
\ee
($\omega^3=1$); in the degenerate limit~\eqref{eq:singS} they collapse to the
$9$ lines
\be
\ba
(x+y)(x+\omega y)(x+\omega^2 y)&=z=0\,,\\
(x+z)(x+\omega z)(x+\omega^2 z)&=y=0\,,\\
(y+z)(y+\omega z)(y+\omega^2 z)&=x=0\,,\label{eq:9lines}
\ea
\ee
all passing through the singular point $x=y=z=0$. However, these $9$ lines are
not isolated on $S$. The line $C_q$ joining $(0,0,0)$ to a point
$q=(x_0,y_0,z_0)\neq(0,0,0)$ with $x_0^3+y_0^3+z_0^3=0$, up to
$(x_0,y_0,z_0)\sim(\lambda x_0,\lambda y_0,\lambda z_0)$, lies on $S$, so these
lines form a continuous family parametrized by the compact cubic
$E=\{x_0^3+y_0^3+z_0^3=0\}\cong T^2\subset\mb{P}^2$. All $C_q$ have the same
class, the generator of $H_2(\widetilde X,\mb{Z})\cong\mb{Z}$, which we call
charge $1$. The $9$ lines~\eqref{eq:9lines} are the $C_q$ with $x_0y_0z_0=0$,
i.e.\ with $q$ one of the $9$ flexes of $E$. If the moduli space were the smooth
$\mc{M}=T^2$, the NLSM of section~\ref{subsec:1dn4} on it would give
$\mc{H}^{\mathrm{BPS}}_1=[(\tfrac12,0)]\oplus2[(0,0)]$, i.e.\ one vector multiplet
and two hypermultiplets with $n^0_1=0$. We now show that the deformations of
$C_q$ transverse to $T^2$ are obstructed, so that the 1d $\mc{N}=4$ SQM of the
M2-brane acquires a superpotential, and that its supersymmetric ground states
are $27$ hypermultiplets.
 
\paragraph{Normal bundle.} Without loss of generality we take $q=(1,b,c)$ with
$1+b^3+c^3=0$; the other points of $E$ follow by permuting $x,y,z$. We cover $C_q$
by the two patches
\be
\label{eq:cone-patches}
U_w=\{w=1\}:\ (x,y,z)\,,\qquad
U_x=\{x=1\}:\ (y',z',w',\delta')=\Big(\frac{y}{x},\frac{z}{x},\frac{1}{x},x\,\delta\Big)\,,
\ee
on which $\widetilde X$ is $\delta=-(x^3+y^3+z^3)$ and $1+y'^3+z'^3+w'^4\delta'=0$,
respectively, and $C_q=\{(x,y,z)=(s,bs,cs)\}$ and
$C_q=\{(y',z',w',\delta')=(b,c,s^{-1},0)\}$. On $U_w$ we use the normal coordinates
$\zeta_\perp=b^2(y-bx)+c^2(z-cx)$, normal to $S$, and
$\zeta_\parallel=\mu(y-bx)+\nu(z-cx)$, along $S$, with constants $\mu c^2-\nu b^2=1$.
On $U_x$, $b^2(y'-b)+c^2(z'-c)=-w'^4\delta'/3$ to linear order, so we use $\delta'$
and $\zeta'_\parallel=\mu(y'-b)+\nu(z'-c)$. On the overlap $U_w\cap U_x$,
\be
\label{eq:cone-transition}
\delta'=-x\,(x^3+y^3+z^3)=-3s^3\,\zeta_\perp+O(\zeta^2)\,,\qquad
\zeta'_\parallel=s^{-1}\,\zeta_\parallel\,,
\ee
so the linearized transition functions give, for every $q\in E$,
\be
\label{eq:cone-normal}
N_{C_q/\widetilde X}\cong\mc{O}(1)\oplus\mc{O}(-3)\,,\qquad
h^0(C_q,N_{C_q/\widetilde X})=h^1(C_q,N_{C_q/\widetilde X})=2\,.
\ee
The two bosonic zero modes in $H^0(C_q,N_{C_q/\widetilde X})$ are
$\zeta_\parallel=v+\alpha s$ and $\zeta_\perp=0$, i.e.
\be
\label{eq:cone-deformation}
(x,y,z)=\big(s,\ bs+c^2(v+\alpha s),\ cs-b^2(v+\alpha s)\big)\,.
\ee
The lines~\eqref{eq:cone-deformation} with $\alpha=0$ and arbitrary $(q,v)$ form a
smooth complex surface $A\supset E$, whose tangent space at $(q,0)$ is
$H^0(C_q,N_{C_q/\widetilde X})$. The obstruction space
$H^1(C_q,N_{C_q/\widetilde X})=H^1(C_q,\mc{O}(-3))$ is spanned by the classes
$\eta_1$ and $\eta_2$ of $\delta'=w'^{-1}$ and $\delta'=w'^{-2}$ on the overlap. Since
$N^\vee\otimes K_{C_q}\cong N$ for a curve in a CY3, Serre duality gives
\be
\label{eq:cone-Serre}
H^1(C_q,N_{C_q/\widetilde X})\cong H^0(C_q,N_{C_q/\widetilde X})^\vee\,,
\ee
and since $\zeta'_\parallel=\alpha+v\,w'$, $\eta_1$ and
$\eta_2$ are dual to the modes $\alpha$ and $v$.

\paragraph{Kuranishi map and superpotential.} The mode $\alpha$ moves $C_q$ within the family and
is unobstructed, so we determine $\kappa$ and $W$ from the mode $v$ in three steps.
 
\emph{(i) Obstruction.} Inserting~\eqref{eq:cone-deformation} with $\alpha=0$ and
using $1+b^3+c^3=0$,
\be
\label{eq:cone-cubic}
x^3+y^3+z^3=-3bc\,s\,v^2+(c^6-b^6)\,v^3\,.
\ee
This only determines $\delta$ on $U_w$, and $y'=b+c^2v\,w'$ and $z'=c-b^2v\,w'$ are
holomorphic on $U_x$. However, with $s=w'^{-1}$,
\be
\label{eq:cone-obstruction}
\delta'=-s\,(x^3+y^3+z^3)=3bc\,v^2\,w'^{-2}-(c^6-b^6)\,v^3\,w'^{-1}
\ee
has poles at $w'=0$, so the translated line does not extend to $w'=0$ in
$\widetilde X$ unless $bc\,v^2=(c^6-b^6)\,v^3=0$.
 
\emph{(ii) Kuranishi map.} Since $\delta'$ is the coordinate of the summand
$\mc{O}(-3)$ on $U_x$, the poles in~\eqref{eq:cone-obstruction} are the obstruction
class
\be
\label{eq:cone-kappa}
\kappa(q,v)=3bc\,v^2\,\eta_2-(c^6-b^6)\,v^3\,\eta_1\,,
\ee
at second order in $v$ along $\eta_2$ and at third order along $\eta_1$. It cannot be
removed at higher order: any curve near $C_q$ is a graph $(s,Y(s),Z(s))$ in $U_w$,
holomorphy of $y'=Y/s$ and $z'=Z/s$ at $w'=0$ forces $Y,Z$ to be linear in $s$, and
holomorphy of $\delta'=-s\,(s^3+Y^3+Z^3)$ then forces $s^3+Y^3+Z^3=0$. Such a line
lies in the cone $S$, hence passes through its vertex and is one of the $C_{q'}$.
 
\begin{figure}[ht]
\centering
\begin{tikzpicture}[x={(1cm,0cm)},y={(0.75cm,0.5cm)},z={(0cm,1cm)},font=\small]
\def\h{0.07}
\fill[blue!8] (0,0,\h) -- (5,0,\h) -- (5,4,\h) -- (0,4,\h) -- cycle;
\fill[blue!25] (0,0,-\h) -- (5,0,-\h) -- (5,0,\h) -- (0,0,\h) -- cycle;
\fill[blue!18] (5,0,-\h) -- (5,4,-\h) -- (5,4,\h) -- (5,0,\h) -- cycle;
\draw[blue!60!black] (0,0,-\h) -- (5,0,-\h) -- (5,4,-\h) (0,0,-\h) -- (0,0,\h)
  (5,0,-\h) -- (5,0,\h) (5,4,-\h) -- (5,4,\h);
\draw[blue!60!black,postaction={decorate},decoration={markings,mark=at position 0.33 with {\arrow{>}}}] (0,0,\h) -- (5,0,\h);
\draw[blue!60!black,postaction={decorate},decoration={markings,mark=at position 0.33 with {\arrow{>}}}] (0,4,\h) -- (5,4,\h);
\draw[blue!60!black,postaction={decorate},decoration={markings,mark=at position 0.33 with {\arrow{>>}}}] (0,0,\h) -- (0,4,\h);
\draw[blue!60!black,postaction={decorate},decoration={markings,mark=at position 0.33 with {\arrow{>>}}}] (5,0,\h) -- (5,4,\h);
\begin{scope}[teal!70!black,thick]
\foreach \y in {0.667,2,3.333}{
  \draw (0.833,\y,\h) -- (4.167,\y,\h);
  \draw (4.167,\y,\h) -- (5,\y,\h) (0,\y,\h) -- (0.833,\y,\h);
}
\foreach \x in {0.833,2.5,4.167}{
  \draw (\x,0.667,\h) -- (\x,3.333,\h);
  \draw (\x,3.333,\h) -- (\x,4,\h) (\x,0,\h) -- (\x,0.667,\h);
}
\end{scope}
\foreach \x in {0.833,2.5,4.167}{
  \foreach \y in {0.667,2,3.333}{
    \draw[red!70!black,thick] (\x,\y,\h) circle (0.3);
    \draw[red!65!black,line width=1.8pt,line cap=round] (\x,\y,-0.18) -- (\x,\y,0.18);
  }
}
\draw[->] (0,4,\h) -- (0,4,1.0) node[left] {$v$};
\node at (2.5,-0.55,-\h) {$E\cong T^2$};
\draw[<-] (5.05,1,0) -- (6.9,1,0) node[right] {$\mc{M}\simeq E\times\op{Spec}\,\mb{C}[v]/v^2$};
\draw[teal!70!black,thick] (8.55cm,2.45cm) -- (9.05cm,2.45cm);
\node[right] at (9.15cm,2.45cm) {$H^1(E\setminus\{9\text{ flexes}\},\mb{L})\cong\mb{C}^{18}$};
\draw[red!70!black,thick] (8.8cm,1.85cm) circle (0.22);
\node[right] at (9.15cm,1.85cm) {$\mb{C}^9$ at the flexes};
\end{tikzpicture}
\caption{The moduli space $\mc{M}$ over $E\cong T^2$, a square with opposite sides
identified, and the $27$ ground states~\eqref{eq:cone-H0} as $1$-cycles. Away from
the $9$ flexes, $\mc{M}$ is $E$ doubled in the $v$-direction, and it becomes triple
at the flexes. The $18$ states of $H^1_c(E\setminus\{9\text{ flexes}\},\mb{L})$ are
spanned by the arcs between neighboring flexes, carrying the differences $v_k-v_l$
of the transverse vacua, and the $9$ states of $\mb{C}^9$ are the circles of the
annuli at the flexes.}
\label{fig:cone-moduli}
\end{figure}
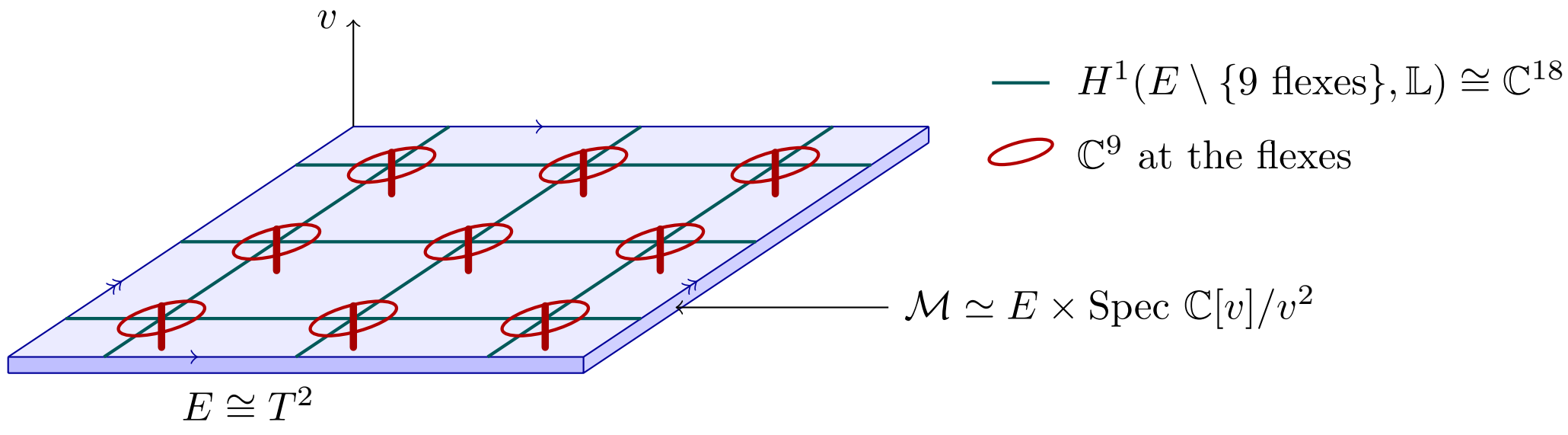

\emph{(iii) Superpotential.} Since $\eta_1$ and $\eta_2$ are dual to $\alpha$ and
$v$, the condition $\kappa=dW$ requires
\be
\label{eq:cone-dW}
\partial_vW=3bc\,v^2\,,\qquad \partial_\alpha W=-(c^6-b^6)\,v^3\,,
\ee
where $\partial_\alpha=c^2\partial_b$, with $b$ a local coordinate on $E$ where
$c\neq0$, because the mode $\alpha$ moves $q=(1,b,c)$ to
$(1,b+c^2\alpha,c-b^2\alpha)$. The first equation gives $W=bc\,v^3$ up to a function
on $E$, which the second sets to a constant, since $dc/db=-b^2/c^2$ and
$b^3+c^3=-1$ give
\be
\label{eq:cone-W}
W=bc\,v^3\,,\qquad
\partial_\alpha W=c^2\Big(c+b\,\frac{dc}{db}\Big)v^3=(c^3-b^3)\,v^3=-(c^6-b^6)\,v^3\,.
\ee
Hence $\mc{M}=\{bc\,v^2=(c^6-b^6)\,v^3=0\}$ is supported on $E$. The function $bc$
has simple zeros at the flexes $b=0$ and $c=0$, and the flexes with $x_0=0$ follow
by permuting $x,y,z$. Away from the $9$ flexes, $W\simeq v^3$ and $\mc{M}$ is
locally $E\times\op{Spec}\,\mb{C}[v]/v^2$, i.e.\ $E$ doubled in the $v$-direction
as for the generalized conifold with $n=2$, while near a flex, $W\simeq u\,v^3$, with $u$ a local coordinate
on $E$, and $\mc{M}$ is locally $\op{Spec}\,\mb{C}[u,v]/(uv^2,v^3)$, triple at $u=0$,
see figure~\ref{fig:cone-moduli}.
 
\paragraph{Ground states.} Since $\phi_W$ is supported on $\mc{M}$, we compute the
hypercohomology~\eqref{eq:cone-hypercohomology} from the stalks of $\phi_W$ at the
points of $E$ and from how they vary along $E$, see figure~\ref{fig:cone-moduli}. The
stalk at $P$ is the reduced cohomology, i.e.\ the cohomology modulo constants, of the
Milnor fiber $\{W=\epsilon\}$ near $P$ for small $\epsilon\neq0$~\cite{Milnor:1968},
and a class of degree $k$ contributes to $\mb{H}^F$ with
$F=k+1-\op{dim}A$~\cite{Dimca:2004}.
\begin{itemize}
\item Away from the flexes, the Milnor fiber of $W=bc\,v^3$ consists of the three
points $v_k=\omega^k(\epsilon/bc)^{1/3}$, so the stalk is $\mb{C}^2$ at $F=-1$, the
two vacua of the LG model~\cite{Hori:2003ic}. Around a flex, $bc$ winds once around
zero and the $v_k$ are permuted cyclically. Hence the restriction of $\phi_W$ to
$E\setminus\{9\text{ flexes}\}$ is a rank-two local system $\mb{L}$ at $F=-1$, with
monodromy eigenvalues $\omega,\omega^2$ around each flex.
\item At a flex, the Milnor fiber of $W\simeq u\,v^3$ is the annulus $u=\epsilon/v^3$,
so the stalk is its $H^1=\mb{C}$, at $F=0$.
\end{itemize}
Splitting $E$ into $E\setminus\{9\text{ flexes}\}$ and the $9$ flexes gives a long
exact sequence for $\mb{H}^\bullet(\mc{M},\phi_W)$~\cite{Dimca:2004}. As $\mb{L}$
has no invariant vector, its cohomology on $E\setminus\{9\text{ flexes}\}$, with or
without compact support, is concentrated in degree one, with dimension
$-2\,\chi(E\setminus\{9\text{ flexes}\})=18$~\cite[section~3.H]{Hatcher:AT}, and it
contributes to $F=-1+1=0$. All terms thus sit at $F=0$, and the sequence reduces to
\be\nonumber
0\to H^1\big(E\setminus\{9\text{ flexes}\},\mb{L}\big)\to\mb{H}^0(\mc{M},\phi_W)
\to\mb{C}^9\to0\,.
\ee
Hence $\mb{H}^{F}(\mc{M},\phi_W)=0$ for $F\neq0$, and
\be
\label{eq:cone-H0}
\mb{H}^{0}(\mc{M},\phi_W)\cong H^1\big(E\setminus\{9\text{ flexes}\},\mb{L}\big)
\oplus\mb{C}^9\cong\mb{C}^{18}\oplus\mb{C}^9\,.
\ee
All $27$ states have $J_3=0$, and the Jacobian of $C_q\cong\mb{P}^1$ is a point.
Including the universal half-hypermultiplet, the BPS spectrum of charge $1$ is
\be
\label{eq:cone-BPS}
\left[\left(0,\tfrac12\right)\oplus2(0,0)\right]\otimes27\,(0,0)\,,
\ee
i.e.\ $\mc{H}^{\mathrm{BPS}}_1=27\,[(0,0)]$ and $N^1_{0,0}=27$: $27$
hypermultiplets and no vector multiplet, with $n^0_1=27$ and
$n^{g\geq1}_1=0$.\footnote{As a check, the Witten index, i.e.\ the Euler
characteristic of~\eqref{eq:cone-hypercohomology}~\cite{Bullimore:2021auw},
is the Euler characteristic of $\mc{M}$ weighted by the Behrend function
$\nu=1-\chi(F_P)$~\cite{Behrend:2005wzg}, with $F_P$ the Milnor fiber at $P$, which
is the genus-zero GV invariant~\cite{Katz:2006gn}: $\nu=-2$ away from the flexes and
$\nu=1$ at the flexes give $n^0_1=(-2)\cdot\chi(E\setminus\{9\text{ flexes}\})+9=27$.}
The naive result above corresponds to the trivial local system on $T^2$, whose $H^0$
and $H^2$ form the vector multiplet; for $\mb{L}$ these classes are absent. Thus the
M2-brane on the singular divisor reproduces the $27$ hypermultiplets of the smooth
$dP_6$: $18$ of them come from the two transverse vacua, which are permuted around
the $9$ flexes, and one comes from each flex.

As the physical conclusion, the genus-0 refined GV invariants $n_1^0=27$ and $n_1^{g\geq 1}=0$ reproduce the results of a local dP$_6$ surface. This is a strong evidence for the 5d $\mc{N}=1$ SCFT from M-theory on $X:$ $x^3+y^3+z^3+w^4=0$ to be interpreted as the rank-1 Seiberg $E_6$ theory (coupled to a free hypermultiplet~\cite{Closset:2020afy} from the Higgs branch analysis), in a phase where the flavor mass parameters are frozen. 
 
The singularities $x^2+y^4+z^4+w^5=0$ and $x^2+y^3+z^6+w^7=0$, whose compact
divisors are cones over elliptic curves of degree two and one, can be analyzed in
the same way, leading to the similar physical results, i.e. the 5d $\mc{N}=1$ SCFT from M-theory on the singularity $x^2+y^4+z^4+w^5=0$ and $x^2+y^3+z^6+w^7=0$ are rank-1 Seiberg $E_7$ and $E_8$ theories respectively, coupled to a free hypermultiplet~\cite{Closset:2020afy}, in a phase where the flavor mass parameters are frozen.

\section{Relation between M2 and D2}
\label{sec:M2D2}

Compactifying M-theory on a further circle $S^1_M$ maps it to type IIA, under which the M2-brane wrapping $\Sigma$ becomes a D2-brane wrapping the same curve. The 5d little group $SO(4)$ reduces to the 4d little group $SO(3)\cong SU(2)_{\rm diag}\subset SU(2)_L\times SU(2)_R$, and the M-theory picture of Section~\ref{sec:M2quant} translates into a D2-brane picture with an extra worldvolume gauge field. In this section, we make this relation precise and identify the refined moduli space controlling the D2 spectrum.

\subsection{From the geometric to the refined moduli space}
\label{subsec:M2D2-refined}

The new ingredient on the D2 side is the worldvolume $U(1)$ gauge field, whose flat connections enlarge the geometric moduli space $\mc{M}$ to a fibration
\be
    \h{\mc{M}}\longrightarrow\mc{M}\,,
\ee
with fiber over a smooth genus-$g$ curve $\Sigma_g$ the moduli space of flat line bundles, generically the Jacobian $\op{Jac}(\Sigma_g)\cong T^{2g}$. When singular
curves are included, one must use the compactified Jacobian instead, as discussed in \cite{Yau:1995mv,Bershadsky:1995qy,Maulik:2016rip}.
Under the reduction, $SU(2)_L$ still remains the Lefschetz group of the base $\mc{M}$, whereas $SU(2)_R$ becomes the Lefschetz group of the fiber, and the 4d little group is the diagonal $SU(2)=\op{diag}(SU(2)_L\times SU(2)_R)$. A state $(j_L,j_R)$ of $SO(4)$ decomposes as $j_L\otimes j_R$ under the diagonal $SU(2)$, and the D2 spectrum is read from the cohomology of $\h{\mc{M}}$~\cite{Gopakumar:1998jq,Katz:1999xq},
\be
    \label{eq:HMhat}
    H^*(\h{\mc{M}})=\sum N^{Q}_{j_b,j_f}\,[j_b,j_f]\,,
\ee
where $j_b=j_L$ and $j_f=j_R$ are the base and fiber quantum numbers. For a non-degenerate genus-$g$ curve the type IIA spectrum takes the form
\be
    \left[\left(\tfrac12\right)+2(0)\right]\otimes H^*(\h{\mc{M}})\,,
\ee
where, relative to~\eqref{eq:final-spectrum}, the $g$ extra half-hypermultiplet factors are absorbed into the cohomology of the fiber $\op{Jac}(\Sigma_g)$. The two extremal unrefined invariants are Euler characteristics,
\be
    \label{eq:euler}
    N_0^Q=(-1)^{\dim_{\mb{C}}\h{\mc{M}}}\chi(\h{\mc{M}})\,,\qquad N_g^Q=(-1)^{\dim_{\mb{C}}\mc{M}}\chi(\mc{M})\,.
\ee
The second follows because the top $SU(2)_R$ spin $g/2$ pairs the fiber classes $(1,k_f,\dots,k_f^g)$ with the cohomology of $\mc{M}$, so that $(H^*(\mc{M}),g/2)$ appears and $N_g^Q=\sum_{j_L}(-1)^{2j_L}(2j_L+1)N^Q_{j_L,g/2}$. After writing the number of spin-$j_L$ states through the Betti numbers $b_i$ of $\mc{M}$ ($i=2j_L+\dim_{\mb{C}}\mc{M}$), the alternating sum telescopes to $(-1)^{\dim_{\mb{C}}\mc{M}}\chi(\mc{M})$. The computation method for all genus unrefined and refined GV invariants via deformation moduli space and its Jacobian fibration was developed by Katz-Klemm-Vafa in \cite{Katz:1999xq}, and later was refined in \cite{Choi:2012jz}. We review this approach and compute some examples of local Calabi-Yau 3-folds in section \ref{sec:5dSCFT}.

\subsection{Abelian duality and the lift of D2 to M2}
\label{subsec:abelian}

The precise map between the D2 gauge field and the M2 position on $S^1_M$ is given by the 3d abelian duality~\cite{Beasley:2014ila,Beasley:2014jta}. Classically, on the $3d$ worldvolume $M=\mb{R}\times\Sigma$ we have
\be
    \label{eq:abelian-duality}
    \star F=e^2\,\dd\phi\,,
\ee
which relates the field strength $F=\dd A$ of the worldvolume $U(1)$ to a scalar $\phi:M\to S^1_M$ encoding the M2 position. On the Maxwell side the classical configurations satisfy $\dd^\dagger F=\dd F=0$, solved by $F=\sum_{\omega_i\in H^2(M,\mb{Z})}f_i\omega_i$. Fixing a class $F_0$, the residual freedom $F_0\to F_0+\dd\eta$ for closed $\eta$ has moduli space
\be
    \label{eq:JSigma}
    \mathscr{J}_\Sigma=H^1(\Sigma,\mb{R})/H^1(\Sigma,\mb{Z})\cong U(1)^{b_1(\Sigma)}\cong T^{2g}
\ee
which is the fiber of $\h{\mc{M}}\to\mc{M}$. Viewing $\eta$ (with $\dd\eta=0$) as a flat connection, $\mathscr{J}_\Sigma$ is the space of flat bundles on $\Sigma$. On the scalar side, $\dd\phi=\star F=\sum_{\eta_i\in H^1(M,\mb{Z})}f_i\eta_i$ defines a harmonic map~\cite{eells1964harmonic}, with $\dd\phi=2\pi\eta$ for each $\eta\in H^1(\Sigma,\mb{Z})$, and the only modulus of $\phi$ is a constant shift in $S^1_M$, so the space of classical solutions is
\be
    \mathcal{S}_\phi=\bigsqcup_{\omega\in H^1(\Sigma,\mb{Z})}S^1_M\cong H^1(\Sigma,\mb{Z})\times S^1\,.
\ee
Since $\mathscr{J}_\Sigma=\Hom(\pi_1(\Sigma),U(1))/U(1)=\Hom(H_1(\Sigma,\mb{Z}),U(1))$, the two field spaces are Pontryagin dual,
\be
    \label{eq:pontryagin}
    \widetilde{\mathcal{S}}_\phi=H^1(\Sigma,\mb{Z})\xLongleftrightarrow{\ \text{Pontryagin}\ }\mathscr{J}_\Sigma\,,
\ee
so in $3d$ the duality exchanges the zero modes of the Maxwell theory with the winding sectors of the scalar theory rather than zero modes with zero modes and the difference is manifest in the partition functions of~\cite{Broda:2002wg,Beasley:2014ila}, summarized in Table~\ref{tab:partition}.

\begin{table}[h]
\centering
\begin{tabular}{c|c|c|c|c}
 & Saddle points & Zero modes & Massive modes & Ghosts\\\hline
Maxwell theory & $H^2(\Sigma,\mb{Z})$ & $\mathscr{J}_\Sigma$ & $\Delta_1$ & $\Delta_0$\\\hline
Scalar theory  & $H^1(\Sigma,\mb{Z})$ & $S^1_M$              & $\Delta_0$ & --
\end{tabular}
\caption{Structure of the partition functions of $3d$ Maxwell theory and its
dual $3d$ scalar theory on $M=\mb{R}\times\Sigma$~\cite{Beasley:2014ila}.}
\label{tab:partition}
\end{table}

Sweeping out all Maxwell saddle points requires including all D2--D0 bound states, i.e. all worldvolume fluxes
\be
    \frac{1}{2\pi}\int_\Sigma F=k\ \Longrightarrow\ c_1(L)\to c_1(L)+k[\omega_\Sigma]\,,
\ee
which lift, on the M-theory side, to all KK momenta along $S^1_M$, since a D0-brane is a longitudinal KK mode~\cite{Townsend:1995kk,Obers:1998fb}. Indeed $k=\tfrac{1}{2\pi}\int_\Sigma F=\tfrac{e^2}{2\pi}\int_\Sigma\star\dd\phi$ is the Noether charge of the $S^1_M$ shift, i.e. the quantized momentum. Thus the sum over D2--D0 bound states lifts to the sum over KK modes, and, after quantization, the classical and quantum labels are exchanged between the Maxwell and scalar descriptions. A single D2-brane without D0-branes therefore probes only the zero-momentum sector, and the naive lift lacks a manifest $\h{\mc{M}}\to\mc{M}$ structure. Hence the full IIA/M-theory duality requires the sum over all D2--D0 bound states. On the topological Hilbert space $\mathscr{H}_\Sigma^{\op{top}}=\sum_{(m,\omega)}\mb{C}\ket{m,\omega}$ the Lefschetz map $\mathscr{L}:k\mapsto j_f=(k-g)/2$ ($k=p+q$ the form degree on the Jacobian) yields the isomorphism $\mathscr{H}^{(\frac{k-g}{2},0)}\cong\bigwedge^k\mathscr{H}_\Sigma^{\op{top},m=0}$. For an isolated smooth genus-$g$ curve, where $\mc{M}$ is a point and $\h{\mc{M}}\cong\mathscr{J}_\Sigma$, the Poincar\'e polynomial is~\cite{zhao2023gopakumar}
\be
    P(\h{\mc{M}};t)=(1+t)^{2g}\,,
\ee
whose binomial coefficients $\binom{2g}{k}$ reproduce the expected spin degeneracies.

\subsection{Refined moduli space of a curve in a projective variety}
\label{subsec:refined-projective}

The refined moduli space $\h{\mc{M}}$ can be constructed geometrically. For a genus-$1$ (elliptic) curve $C[\tau]\subset X$ with a choice of flat bundle, the data of a flat connection on the torus $T^2\cong\mb{C}/(\mb{Z}+\tau\mb{Z})$ are a point $z\in C[\tau]$, so the pair is $(C[\tau],z)$. Writing the section as $C[\tau]\in\Gamma(X,\mc{O}(D))$ with $\Gamma(X,\mc{O}(D))=\langle s_1,\dots,s_n\rangle$ and using the projective rescaling $(a_1,\dots,a_n)\sim(ka_1,\dots,ka_n)$, the pair becomes a point of
\be
    \label{eq:Cp-pair}
    \{(a_1:\cdots:a_n;p_0)\in\mb{P}^{n-1}\times X\ |\ \textstyle\sum_i a_i s_i(p_0)=0\}\,,
\ee
where the marked point $p_0$ carries the position on $C[\tau]$. The linear
constraint carves a hyperplane in each $\mb{P}^{n-1}$ fiber, so
\be
\label{eq:E-fibration}
\mb{P}^{|\Gamma(X,\mc{O}(D))|-2}\hookrightarrow E\to X\,,
\ee
a projectivization of a rank-$(n-1)$ bundle over $X$. By the projective-bundle
(Leray--Hirsch) formula~\cite{nunez_projective_bundle},
\be
\label{eq:proj-bundle}
H^*(\h{\mc{M}})=\frac{H^*(X)[\alpha]}{\langle\alpha^{|\Gamma(X,\mc{O}(D))|-1}\rangle}\,.
\ee
For the anticanonical $\mc{O}(3)$ curves of $\mb{P}^2$, where the defining cubic
$f(X,Y,Z)=a_1X^3+a_2X^2Y+\cdots$ has $|\Gamma(\mb{P}^2,\mc{O}(3))|=\binom{5}{2}=10$
monomials, requiring $f(X_0,Y_0,Z_0)=0$ imposes one linear constraint that cuts
$\mb{P}^9$ down to a $\mb{P}^8$, so $\h{\mc{M}}$ is a $\mb{P}^8$-fibration over
$\mb{P}^2$ and
\be
\label{eq:Mhat-P2}
H^*(\h{\mc{M}})=\frac{H^*(\mb{P}^2)[\alpha]}{\langle\alpha^9\rangle}
=H^*(\mb{P}^2)\otimes H^*(\mb{P}^8)\,,
\ee
which resolves the degree-3 paradox of Section~\ref{sec:5dSCFT}. For higher
genus, the Abel--Jacobi map $\op{Sym}^g(C)\to\op{Pic}^0(C)\cong J(C)$,
$\{p_1,\dots,p_g\}\mapsto\mc{O}(p_1+\cdots+p_g-g\cdot O)$, lets one trade the
flat bundle for $g$ unordered marked points on $C$; since $\op{Sym}^g(C)$ is only
birationally isomorphic to $J(C)$, one is led to conjecture a birational model
\be
\mb{P}^{|\Gamma(X,\mc{O}(D))|-1-g}\hookrightarrow E\to\op{Sym}^g(X)\,,\qquad
\alpha^g:E\ \dashrightarrow\ \h{\mc{M}}\,.
\ee

\section{Examples in 5d SCFTs}
\label{sec:5dSCFT}

We test the framework against the $g\leq 1$ refined GV invariants of a series of local
Calabi--Yau geometries engineering 5d SCFTs, comparing with~\cite{Huang:2013yta}.

The refined GV invariants are defined as the coefficients in the BPS spectrum of a M2-brane wrapped on curve $\Sigma$ in the class $\beta\in H_2(X,\mathbb{Z})$
\be
\left[(0,\tfrac12)\oplus2(0,0)\right]\otimes \bigoplus_{j_{L},j_{R}}N_{j_{L},j_{R}}^\beta [(j_{L},j_{R})]\,  ,
\ee
Recall that we follow the convention of~\cite{Witten:1996qb} in which the broken
supersymmetries are $2(0,\tfrac12)$, so that the $j_L\leftrightarrow j_R$ labels
are exchanged relative to the mathematics literature.
Aside from the universal $\left[  \left(  0,\frac{1}{2}  \right)+2(0,0) \right]$ part, we use the following representations to denote the BPS spectrum:
\be
\mathcal{H}_{\beta}^{\mathrm{BPS}}\equiv \bigoplus_{j_{L},j_{R}}N_{j_{L},j_{R}}^\beta [(j_{L},j_{R})]
=\bigoplus_{j_R}[(R_{j_R},j_R)] \, .
\ee
The unrefined GV invariants are defined as the signed trace over the $SU(2)_L$:
\be
\bigoplus_{j_R}\left[\sum_{j_L}(-1)^{2j_L}(2j_L+1)
N^\beta_{j_L,j_R}\right][j_R]=\bigoplus_{g\geq 0}n_{\beta}^g I_g \, , \quad I_g\equiv \left[(\tfrac12)\oplus2(0)\right]^{\otimes g}\, .
\ee
Curves in the class $\beta$ have fixed arithmetic genus $p_{a}$. It is the geometric genus of the smoothing of the curve. Suppose $\Sigma$ is a curve with $\delta$ nodes such that the normalization $\tilde\Sigma$ has $n$ components of genera $g_1,\cdots,g_n$, then
\be
p_a=\sum_{i=1}^n (g_i-1)+\delta+1\, .
\ee
If throughout the moduli space $\mathcal{M}$, the geometric genus $g$ of the curve always equal to its arithmetic genus, then using the M-theory method we can obtain the BPS spectrum from the cohomology of $\mathcal{M}$:
\be
\left[(0,\tfrac12)\oplus2(0,0)\right]^{ g+1}\otimes [(H^*(\mathcal{M}),0)] \, .
\ee
However for most $p_a\geq 1$ cases this formula cannot work as the geometric genus of the curve often degenerates somewhere on the moduli space. For example the curve can develop a nodal singularity or it can decompose into two irreducible components.
The Type IIA interpretation argues that instead we can consider the cohomology of $\h{\mathcal{M}}$ which is a Jacobian fibration $\mathrm{Jac}(\Sigma)\to \h{\mathcal{M}}\to \mathcal{M}$. The BPS spectrum of the 4d $\mathcal{N}=2$ theory are labeled by their representations under 4d massive little group $SO(3)\cong SU(2)_{\mathrm{diag}}\subset SU(2)_{L}\times SU(2)_{R}$:
\be
\left[ \left( \frac{1}{2} \right)\oplus 2(0) \right]\otimes H^*(\h{\mathcal{M}}) \, .
\ee
The Lefschetz decomposition $H^*(\h{\mathcal{M}})$ only contain the information of the $SU(2)_{\mathrm{diag}}$ representation
\be
H^*(\h{\mathcal{M}})=(\mathcal{H}^{\mathrm{BPS}}_{\beta})_{\mathrm{diag}} \, .
\ee
Over a generic point of $\h{\mathcal{M}}$, it is a $T^{2g}$ fiber over $\mathcal{M}$, corresponding to the highest spin representation $(H^*(\mathcal M),p_a/2)$. To recover all the $[R_{j_{R}}]$ for $0\leq j_{R}\leq \frac{p_{a}}{2}-1$, we can use the KKV approach. This method is originally imposed by Katz, Klemm and Vafa in \cite{Katz:1999xq}, to compute the unrefined GV invariants for some local CY 3-folds. Later in \cite{Choi:2012jz}, the authors proposed that this method readily extends to compute the refined invariants. The key point is use the $\theta$ class of the Jacobian $\mathrm{Jac}(\Sigma)$, which acts only on the $SU(2)_{R}$ part as the $SU(2)_{R}$ raising operator, i.e. $\theta([(j_{L},j_{R})])=[j_{L}]\otimes \theta([j_{R}])$. If we know the $(\theta^{p_{a}-k}\mathcal{H}^{\mathrm{BPS}}_{\beta})_{\mathrm{diag}}$ for all $0\leq k\leq p_{a}$, then we can recover the $SU(2)_{L}\times SU(2)_{R}$ spectrum from these diagonal representations. In the following paragraphs we summarize the KKV approach to the refined GV invariants. 

\paragraph{The KKV approach.}
The Jacobian $\mathrm{Jac}(\Sigma)$ of a smooth genus $g$ curve $\Sigma$ has the cohomology of $T^{2g}$
\be
H^*(\mathrm{Jac}(\Sigma))=I_{g}=\bigoplus_{i=0}^g \left\{ \binom{2g}{g-i}- \binom{2g}{g-i-2} \right\} \left[ \frac{i}{2} \right]\, .
\ee
Consider the Hilbert scheme of $k$ points on $\Sigma$ that is $\Sigma$ together with the choice of $k$ points on it, denoted by $\mathrm{Hilb}^k(\Sigma)$. It has the cohomology
\be
H^*(\mathrm{Hilb}^k(\Sigma))=\bigoplus_{i} \binom{2g}{k-i}\left[ \frac{i}{2} \right]\, .
\ee
Thus the $SU(2)$ representations satisfy
\be
H^*(\mathrm{Hilb}^g(\Sigma))=H^*(\mathrm{Jac}(\Sigma))\oplus H^*(\mathrm{Hilb}^{g-2}(\Sigma))\, .
\ee
The cohomology class $\theta$ of the Riemann theta-divisor of $\mathrm{Jac}(\Sigma)$ is a Kähler class \cite{Griffiths:1994prl}. It acts as the Lefschetz $SU(2)$ raising operator on the cohomology ring of the Jacobian:
\be
\theta\left( \left[ \frac{k}{2} \right] \right)=\left\{\begin{array}{cl}
\left[ \frac{{k-1}}{2} \right] & k>0 \\
0 & k=0
\end{array}  \right. \quad .
\ee
Using the action of $\theta$ we can find more relations
\be
H^*(\mathrm{Hilb}^k(\Sigma))=\theta^{g-k}H^*(\mathrm{Jac}(\Sigma))\oplus H^*(\mathrm{Hilb}^{k-2}(\Sigma)) \, , \quad 0\leq k\leq g\, ,
\ee
with the convention that $H^*(\mathrm{Hilb}^{k-2}(\Sigma))=0$ for $k<2$ and $\mathrm{Hilb}^0(\Sigma)$ being a point.

The Abel-Jacobi map $f_{k}:\mathrm{Hilb}^k(\Sigma)\to \mathrm{Jac}(\Sigma)$ gives a map between the cohomology of the Hilbert scheme and the Jacobian. Now promote this map to maps between fibration over $\mathcal{M}$ as $\hat{f}_{k}: \widetilde{\Sigma}^{[k]}\to \h{\mathcal{M}}$, where $\widetilde{\Sigma}^{[k]}$ denotes the moduli space of holomorphic curves of fixed class $\beta$ with arithmetic genus $g$ together with the choice of $p$ points on it. We suppose that $\widetilde{\Sigma}^{[k]}$ is smooth, then the assertion of \cite{Choi:2012jz,Katz:1999xq} was that
\be
\label{eq:kkv}
H^*(\widetilde{\Sigma}^{[k]})=(\theta^{p_{a}-k}\mathcal{H}_{\beta}^{\mathrm{BPS}})_{\mathrm{diag}}\oplus H^*(\widetilde{\Sigma}^{[k-2]})+\text{correction terms}\, ,
\ee
where the correction terms come from reducible curves. These correction terms can be fixed by the theory of stable pairs \cite{Choi:2012jz}. For some simple cases, we can also use geometric arguments to find the correction terms. Suppose that the curve $\Sigma$ can degenerate to two or more curve components. The reducible curve has the same arithmetic genus as $\Sigma$. Besides it has $k$ nodes which can be either nodes on curve components or the intersection point between different components. Then the correction term given by this reducible configuration begins at $k$ nodes level (corresponding to refined GV invariants with $j_R\leq (p_a-k)/2$) and equals to the tensor product of the cohomology of each curve component.
Using the above equalities, the KKV approach is to compute all $H^*(\widetilde{\Sigma}^{[k]})$ and possible correction terms so as to obtain all $(\theta^{p_{a}-k}\mathcal{H}^{\mathrm{BPS}}_{\beta})_{\mathrm{diag}}$, then one can read off the refined $SU(2)_{L}\times SU(2)_{R}$ from these diagonal representations. When the arithmetic genus is relatively small and the deformation moduli space $\mathcal{M}$ can be computed, the KKV approach is convenient. For large arithmetic genus cases, the relative Hilbert scheme can be singular and we need to consider the correction terms arise from the singular locus \cite{Katz:1999xq}. To compute higher genus refined GV invariants, other approaches are more convenient, such as the refined topological vertex \cite{Iqbal:2007ii} and the blow-up equation \cite{Kim:2020hhh}.

\paragraph{Local surfaces.}
We limit ourselves to local Calabi-Yau threefolds. They are the total space $\mathrm{Tot}(K_{S})\to S$ of a complex surface $S$ with ample anti-canonical bundle. In these cases with the additional condition \eqref{eq:negative-degree} that we will argue below, the moduli space $\mathcal{M}$ is easy to compute. 

For a holomorphic curve $\Sigma \subset S$, the adjunction formula gives
\be
N_{\Sigma /S}=\mathcal{O}_{\Sigma}(\Sigma)\, , \, K_{\Sigma}=(K_{S}+\Sigma)|_{\Sigma}\, .
\ee
Consequently, Serre duality on $\Sigma$ implies
\be
H^1(\Sigma,N_{\Sigma /S})^* \cong H^0(\Sigma,K_{\Sigma}\otimes N_{\Sigma /S}^*)=H^0(\Sigma,K_{S}|_{\Sigma})\, .
\ee
The normal bundle of $\Sigma$ inside the whole Calabi-Yau manifold $X$ splits:
\be
N_{\Sigma /X}=N_{\Sigma /S}\oplus K_{S}|_{\Sigma}\, .
\ee
Combining these identities,
\be
h^0(\Sigma,N_{\Sigma /X})=h^0(\Sigma,N_{\Sigma /S})+ h^1(\Sigma,N_{\Sigma /S})\, .
\ee
And Serre duality in $X$ gives $h^0(\Sigma,N_{\Sigma /X})=h^1(\Sigma,N_{\Sigma /X})$. We focus on the cases when $h^0(\Sigma,N_{\Sigma /X})=h^0(\Sigma,N_{\Sigma /S})$, that is all the deformations of the curve in $X$ are deformations in $S$ and the curve cannot have deformation modes leaving the surface. In such cases the computation can be done in the local surface. As $h^1(\Sigma,N_{\Sigma /S})=h^0(\Sigma,N_{\Sigma /X})-h^0(\Sigma,N_{\Sigma /S})=0$, the deformations inside of $S$ are unobstructed and the deformation moduli space is the linear system $\mathbb{P}^{h^0(\Sigma,N_{\Sigma /S})}$. We can use the following fact to compute its dimension
\be
h^0(\Sigma,N_{\Sigma /S})=h^0(S,\mathcal{O}_{S}(\Sigma))-1 \, .
\ee
This relation comes from the sequence $0\rightarrow \mathcal{O}_{S}\rightarrow \mathcal{O}_{S}(\Sigma)\rightarrow \mathcal{O}_{\Sigma}(\Sigma)\rightarrow 0$, together with $\mathcal{O}_{\Sigma}(\Sigma)\simeq N_{\Sigma /S}$ and $h^1(S,\mathcal{O}_{S})=0$.

Now we ask when the relation $h^0(\Sigma,N_{\Sigma /X})=h^0(\Sigma,N_{\Sigma /S})$ holds. From the normal bundle sequence
\be
0 \rightarrow N_{\Sigma/ S} \rightarrow N_{\Sigma/ X} \rightarrow K_{S}|_{\Sigma} \rightarrow 0 \,,
\ee
We have the long exact sequence
\be
0 \rightarrow H^0\left(N_{{\Sigma} / S}\right) \rightarrow H^0\left(N_{\Sigma / X}\right) \rightarrow H^0\left(\left.K_S\right|_{\Sigma}\right) \xrightarrow{\delta} H^1\left(N_{\Sigma / S}\right) \rightarrow \cdots \, .
\ee
So the condition for $h^0(\Sigma,N_{\Sigma /X})=h^0(\Sigma,N_{\Sigma /S})$ is equivalent to
\be
 h^0(\Sigma,K_S|_{\Sigma})=0\,.
\ee
As a line bundle on a general curve has no global section if its degree is negative, we obtain the sufficient condition
\be
\label{eq:negative-degree}
\mathrm{deg}(K_{S}|_{\Sigma})=K_{S}\cdot\Sigma<0 \, .
\ee
On rational curves it is also necessary, whereas on other general curves the necessary condition need to be checked case by case. We only consider the curves satisfying \eqref{eq:negative-degree} so our method can apply.

The local Calabi-Yau threefold geometries which engineer 5d $\mathcal{N}=1$ SCFTs are studied and classified in \cite{Jefferson:2018irk}. They are locally given by the so-called shrinkable surfaces that are conjectured to have the form $S=S_{1}\cup S_{2}\cup \cdots\cup S_{r}$, where the irreducible components $S_{i}$ are either equal to $\mathbb{P}^2$ or a blowup $\mathrm{Bl}_{p}\mathbb{F}_{n}$ of a Hirzebruch surface at $p$ points. The rank-1 geometries are the del Pezzo surfaces $\mathrm{dP}_{n},0\leq n\leq 8$ and Hirzebruch $\mathbb{F}_{0}$. All their curves satisfy \eqref{eq:negative-degree}. The GV invariants of these surfaces are originally studied in \cite{Katz:1999xq}. The refined GV invariants of local $\mathbb{P}^2$ and local $\mathbb{F}_0$ are computed in \cite{Choi:2012jz} using the refined KKV approach. The rank-2 geometries are classified as two surface components glued along a rational curve. We choose three examples: the $SU(3)_{4}$ theory given by $\mathbb{F}_{5}\cup\mathbb{F}_{1}$, the $SU(3)_{5}$ theory given by $\mathbb{F}_{6}\cup\mathbb{F}_{0}$, and the $SU(4)_8$ theory given by $\mathbb{F}_{10}\cup\mathbb{F}_{8}\cup\mathbb{F}_{0}$. Their refined GV invariants can be computed via blow-up equation, the $g\leq 2$ results are in \cite{Kim:2020hhh}. Our computation matches the results therein.

In the following cases, the relative Hilbert scheme of a curve will generally be a projective space fiber over the Hilbert scheme of the smooth projective surface $S$. The Betti numbers of $\mathrm{Hilb}^n(S)$ can be computed by the Göttsche’s formula \cite{Göttsche:1994}. If we use $P_n(t)$ to denote the Poincaré polynomial of $\mathrm{Hilb}^n(S)$,
\be
P_n(t):= P(\mathrm{Hilb}^n(S);t)=\sum_{r=0}^{2n}b_{2r}(\mathrm{Hilb}^n(S))t^r \, ,
\ee
and denote $b=b_2(S)$, then the Göttsche’s formula is
\be
\sum_{n=0}^{\infty}P_n(t)q^n
=\prod_{m=1}^{\infty}
\frac{1}
{(1-t^{m-1}q^m)(1-t^mq^m)^b(1-t^{m+1}q^m)}.
\ee

\subsection{Local $\mb{P}^2$}
\label{subsec:localP2}

We review the known line, conic, and cubic spectra of Choi–Katz–Klemm \cite{Choi:2012jz} to exhibit the preceding physics spectrum discussion.
For $X=\mc{O}(-3)\to\mb{P}^2$ the curve classes are $dH$, a degree-$d$ curve
being the zero locus of $\sum_{i+j+k=d}a_{ijk}x_0^ix_1^jx_2^k=0$ with
$(d^2+3d+2)/2$ coefficients, so the linear system and genus are
\be
\mc{M}=|dH|=\mb{P}^{(d^2+3d)/2}\,,\qquad p_a(dH)=\frac{(d-1)(d-2)}{2}\,.
\ee
The latter formula computes the arithmetic genus of
every divisor in the linear system, including singular members. 

\paragraph{Curve degree $d=1,2$.}
For a smooth $\Sigma\subset S\subset X=\operatorname{Tot}(K_S)$ the zero section splits the normal bundle as
$N_{\Sigma/X}=N_{\Sigma/S}\oplus K_S|_\Sigma$. For a line these two summands are $\mathcal O_{\mathbb{P}^1}(1)$ and $\mathcal O_{\mathbb{P}^1}(-3)$; for a smooth conic they
are $\mathcal O_{\mathbb{P}^1}(4)$ and $\mathcal O_{\mathbb{P}^1}(-6)$.
The negative summand has no section, while the positive summand has,
respectively, two and five sections and no $H^1$. These are exactly the
tangent dimensions of $\mathbb{P}^2$ and $\mathbb{P}^5$. Since degree $d=1,2$ curves in $\mathbb{P}^2$ all have arithmetic genus $p_a=0=g$, the M-theory approach gives the BPS spectrum and the refined GV:
\be
\begin{array}{ll}
 \mathcal{H}^{\mathrm{BPS}}_1=[(1,0)]\, ,  & \quad N^1_{1,0}=1,\\
 \mathcal{H}^{\mathrm{BPS}}_{2}=[(\tfrac52,0)]\, ,  &\quad N^{2}_{5/2,0}=1.
\end{array}
\ee
Although a degree-$2$ curve can degenerate to a double line or two intersecting lines, its geometric genus stays zero, so the M-theory argument holds. In both cases $p_a=g=0$ implies $\h{\mc{M}}=\mc{M}$.

\paragraph{Curve degree $d=3$.}
A degree-$3$ curve has $p_a=1$, and applying~\eqref{eq:final-spectrum} naively with
$\mc{M}=\mb{P}^9$ gives
\be
\mathcal{H}^{\mathrm{BPS}}_{3}=\left[(0,\tfrac12)\oplus2(0,0)\right]\otimes(\tfrac92,0)
=(\tfrac92,\tfrac12)\oplus2(\tfrac92,0),
\ee
i.e. $N^3_{9/2,1/2}=1$, $N^3_{9/2,0}=2$, which contradicts the correct values
$N^3_{9/2,1/2}=1$, $N^3_{3,0}=1$ of~\cite{Huang:2013yta}. The failure is that thegenus jumps over $\mc{M}$, so~\eqref{eq:final-spectrum} breaks down.
The resolution uses the type IIA moduli space $\h{\mathcal{M}}$. A flat line bundle on a smooth torus is equivalent to the choice of a point $p$ on the torus via the Abel-Jacobi map \cite{Griffiths:1994prl}. So the Jacobian of a smooth cubic is the cubic itself. When the cubic curve develops singularities, we should use the compactified Jacobian to build the moduli space \cite{Yau:1995mv}. The result is that for $p_a=1$ curves, the type IIA moduli space $\h{\mathcal{M}}$ is equivalent to the universal curve
\be
\h{\mathcal{M}}\simeq \widetilde{\Sigma}^{[1]}=\{(\Sigma,p):p\in\Sigma \}\, .
\ee
The universal curve can be constructed as follows: pick any point $p\in\mathbb{P}^2$, through this point there is a $\mathbb{P}^8$ family of cubic curves. Thus $\h{\mc{M}}$ is
a $\mb{P}^8$-fibration over $\mb{P}^2$, whose diagonal-$SU(2)$ content
$[1]\otimes[4]=[3]\oplus[4]\oplus[5]$ gives
\be
H^*(\h{\mc{M}})=(\tfrac92,\tfrac12)\oplus(3,0)\,,
\ee
reproducing $N^3_{9/2,1/2}=1$, $N^3_{3,0}=1$. Once the genus is non-constant, the D2/refined-moduli-space description is the reliable one. This construction of the universal curve can be viewed as a special case of \eqref{eq:kkv}.

\paragraph{Curve degree d=4: reducible correction.} The computation of degree-4 curve is straightforward, except the reducible correction to $R_0$. A degree-4 curve has $p_a=3$. Using \eqref{eq:kkv}, we can obtain the spectrum $R_{3/2},R_1,R_{1/2},R_0$ from the Lefschetz cohomology $H^*(\mathcal{M})$ and $H^*(\widetilde{\Sigma}^{[k]})$ with $k=1,2,3$. At the three nodes level, a possible reduction is $4\to3+1$: the quartics become unions of lines and cubic curves which have three intersection points as nodes. The moduli space of such reducible configuration is $\mathbb{P}^2\times\mathbb{P}^9$ with cohomology $[1]\otimes[\frac{9}{2}]$. The correction in \eqref{eq:kkv} is given by
\be
H^*(\widetilde{\Sigma}^{[3]})=(\mathcal{H}_{\beta}^{\mathrm{BPS}})_{\mathrm{diag}}\oplus H^*(\widetilde{\Sigma}^{[1]})+H^*(\mathbb{P}^2\times\mathbb{P}^9)\, .
\ee

\subsection{Rank-2 example: $SU(3)_4$ theory}
\label{subsec:SU3}

For the 5d $SU(3)_4$ theory the resolved geometry is $S=S_1\cup S_2=\mathbb{F}_5\cup\mathbb{F}_1$. The gluing curves are
\be
G_1=e_1,\qquad G_2=e_2+2f_2,\qquad G_1^2=-5,\quad G_2^2=3.
\ee
Their normal degrees add to $-2$, as required for a rational gluing curve in a Calabi-Yau threefold. The primitive curve classes are $e_{2},f_{1},f_{2}$. We use $\mathbf{d}=(d_{1},d_{2},d_{3})$ to denote the curve class $d_{1}e_{2}+d_{2}f_{1}+d_{3}f_{2}$. 

For a curve class $ae+bf$ in the Hirzebruch surface $\mathbb{F}_{n}$, the following formulas are used in the computation:
\be
\label{ex:hirzebruch}
\begin{aligned}
h^0(\mathbb{F}_{n},\mathcal{O}_{\mathbb{F}_{n}}(ae+bf))&=\sum_{i=0}^{a}\max(b-ni+1,0),\\
h^1(\mathbb{F}_{n},\mathcal{O}_{\mathbb{F}_{n}}(ae+bf))&=\sum_{i=0}^{a}\max(ni-b-1,0),\\
p_a(ae+bf)&=1+\frac{-na^2+2ab+(n-2)a-2b}{2}.
\end{aligned}
\ee
The adjunction formula gives
\be
\mathcal O_X(S)|_{S_1}=K_{S_1}+G_1=-e_1-7f_1,\quad \mathcal O_X(S)|_{S_2}=K_{S_2}+G_2=-e_{2}-f_{2}.
\ee
The first line bundle has negative degree on the negative section $e_{1}$ and on fiber $f_{1}$, and hence on every effective irreducible curve of $\mathbb{F}_{5}$. The second has negative degree on every irreducible curve of $\mathbb{F}_{1}$ other than the rigid $e_2$. Hence for curves with non-trivial deformations, the deformation moduli space is the same as the linear system of the curve in local surface. We have $h^0(\mathbb{F}_{n},\mathcal{O}_{\mathbb{F}_{n}}(e))=1$, implying that the linear system of the curve $e_{1}$ is zero-dimensional i.e. $e_{1}$ is rigid in $\mathbb{F}_{5}$; and $h^0(\mathbb{F}_{5},\mathcal{O}_{\mathbb{F}_{5}}(f_{1}))=2$, leading to the deformation moduli space $\mathbb{P}^1$. Inside of $S_{2}=\mathbb{F}_{1}$, the curve class $e_{2}+2f_{2}$ has linear system $\mathbb{P}^4$. So the gluing curve $G$ is the rigid $e_{1}\subset S_{1}$, and a specific curve $G_{2}$ in the class $e_{2}+2f_{2}$ of $S_{2}$.

A distribution of the $(d_{1},d_{2},d_{3})$ cycle between the surfaces is
\be
C_1=xe_1+d_{2}f_1,\quad C_2=(d_1-x)e_2+(d_3-2x)f_2.
\ee
For the range $0\leq d_{2}\leq3$ considered below, $x>0$ gives $C_1\cdot e_1=d_{2}-5x<0$. An effective divisor with negative intersection against an irreducible curve must contain that curve. In this range, all $x$ copies of the rigid $e_{1}$ are fixed. The generic configurations can therefore be described by a core $D=d_1e_2+d_3f_2\subset S_2$ and $k\equiv d_{2}$ fibers in $S_1$ attached to it. For example consider $\mathbf{d}=(1,1,2)$, it can be described as $C_{1}=f_{1},C_{2}=e_{2}+2f_{2}$ or $C_{1}'=e_{1}+f_{1}$ with no $C_{2}'$, whereas the latter description is a specific case in the family of the first configuration.

Each fiber of $S_1$ meets the gluing curve once, so an unordered set of $k$ tails is specified by a length-$k$ divisor $Z\subset G\simeq \mathbb{P}^1$. The parameter space is $B_{k}=\mathrm{Sym}^{k}G=\mathbb{P}^{k}$. The core must pass through $Z$, namely
\be
\mathcal{M}_{D,k}=\{(Z,[s]):Z\in B_{k},\ 0\ne s\in H^0(S_2,\mathcal O_{S_{2}}(D)),s|_{Z}=0\}\, .
\ee
Denote $h^0(S_{2},\mathcal{O}_{S_{2}}(D))=N+1$, then the linear system of the core curve $D$ in $S_2$ is $\mathbb{P}^{N}$. Requiring the core curve is connected to the $k$ fibers imposes $k$ linear conditions if $k\leq D\cdot G$, thus the moduli space $\mathcal{M}_{D,k}$ is a $\mathbb{P}^{N-k}$ fiber over the base $\mathbb{P}^{k}$ with
\be
H^*(\mathcal{M}_{D,k})=\left[ \frac{k}{2} \right]\otimes \left[ \frac{{N-k}}{2} \right]\, .
\ee

\paragraph{Genus-zero curves.}
First take $D=e_2+mf_2$. For $m\geq1$,
\be
H^0 (S_{2},\mathcal{O}_{S_{2}}(D))=2m+1,\quad p_a (D)=0,\quad D\cdot G=m+1,\quad D-G=(m-2) f_2.
\ee
A rational core with tails attached at one point each still has genus zero. For the displayed range $1\leq m\leq3$ and $0\leq k\leq\min(3,m+1)$, 
\be
\mathcal{H}^{\mathrm{BPS}}_{(1, k, m)}=\bigoplus_{j=|m-k|}^{m}[(j,0)].
\ee
For $m=0$ the core $e_2$ is rigid and meets $G$ once. Both $k=0$ and $k=1$ have a point as their moduli space and give $[(0,0)]$. The other rational core needed is $D=2H=2e_2+2f_2$:
\be
h^0 (2H)=3+2+1=6,\quad p_a (2H)=0,\quad 2H\cdot G=4,\quad 2H-G=e_2,\quad h^1 (e_2)=0.
\ee
Consequently $\mathcal{H}^{\mathrm{BPS}}_{(2,k,2)}=\left([k/2]\otimes[(N-k)/2],0\right)$ for $0\leq k\leq3$. The genus-zero results are listed in Table~\ref{tab:SU3-g0}. They agree with
\cite[Table~5]{Kim:2020hhh}

\begin{table}[htbp]
\centering\small
\begin{tabular}{ccc l}
\toprule
$(d_1,k,d_3)$ & base & fiber & $\mathcal{H}^{\mathrm{BPS}}$\\
\midrule
$(1,0,0),(1,1,0)$ & point & point & $[(0,0)]$\\
$(1,0,1)$ & point & $\mathbb{P}^2$ & $[(1,0)]$\\
$(1,1,1)$ & $\mathbb{P}^1$ & $\mathbb{P}^1$ & $[(0,0)]\oplus[(1,0)]$\\
$(1,2,1)$ & $\mathbb{P}^2$ & point & $[(1,0)]$\\
$(1,0,2)$ & point & $\mathbb{P}^4$ & $[(2,0)]$\\
$(1,1,2)$ & $\mathbb{P}^1$ & $\mathbb{P}^3$ & $[(1,0)]\oplus[(2,0)]$\\
$(1,2,2)$ & $\mathbb{P}^2$ & $\mathbb{P}^2$ & $[(0,0)]\oplus[(1,0)]\oplus[(2,0)]$\\
$(1,3,2)$ & $\mathbb{P}^3$ & $\mathbb{P}^1$ & $[(1,0)]\oplus[(2,0)]$\\
$(1,0,3)$ & point & $\mathbb{P}^6$ & $[(3,0)]$\\
$(1,1,3)$ & $\mathbb{P}^1$ & $\mathbb{P}^5$ & $[(2,0)]\oplus[(3,0)]$\\
$(1,2,3)$ & $\mathbb{P}^2$ & $\mathbb{P}^4$ & $[(1,0)]\oplus[(2,0)]\oplus[(3,0)]$\\
$(1,3,3)$ & $\mathbb{P}^3$ & $\mathbb{P}^3$ & $[(0,0)]\oplus[(1,0)]\oplus[(2,0)]\oplus[(3,0)]$\\
$(2,0,2)$ & point & $\mathbb{P}^5$ & $[(\tfrac52,0)]$\\
$(2,1,2)$ & $\mathbb{P}^1$ & $\mathbb{P}^4$ & $[(\tfrac32,0)]\oplus[(\tfrac52,0)]$\\
$(2,2,2)$ & $\mathbb{P}^2$ & $\mathbb{P}^3$ & $[(\tfrac12,0)]\oplus[(\tfrac32,0)]\oplus[(\tfrac52,0)]$\\
$(2,3,2)$ & $\mathbb{P}^3$ & $\mathbb{P}^2$ & $[(\tfrac12,0)]\oplus[(\tfrac32,0)]\oplus[(\tfrac52,0)]$\\
\bottomrule
\end{tabular}
\caption{Seventeen genus-zero charges of $SU(3)_4$.}
\label{tab:SU3-g0}
\end{table}

\paragraph{Genus one curve: elliptic core with attached tails.}
For a genus one curve, contracting rational $f_{1}$ tails preserves the Jacobian of a reduced nodal curve. This suggests that $\h{\mathcal{M}}$ can be constructed by adding a point $p$ on the elliptic core to $\mathcal{M}$. Subtlety arises when the point $p$ collides with one of the attaching point which is a nodal singularity. To resolve this problem we consider the Hilbert scheme. More precisely, consider the flag:
\be
Z\subset W\subset S_{2}\, ,\quad Z\subset G\, , \quad l(Z)=k\, , \quad l(W)=k+1\, .
\ee
Its parameter space is the restricted nested Hilbert scheme
\be
\h{B}_{k}=\{ (Z,W):Z\in \mathrm{Hilb}^kG\cong \mathrm{Sym}^kG\,, \, W\in \mathrm{Hilb}^{k+1}S_{2}\, ,\, Z\subset W  \}\, ,
\ee
which is restricted in the sense that $Z$ is restricted to $\mathrm{Hilb}^kG\subset \mathrm{Hilb}^kS_{2}$. By the proposition 2.2 in \cite{Ellingsrud:1996int}, $\h{B}_{k}$ is isomorphic to $\mathrm{Bl}_{Z_{k}}(S_{2}\times\mathbb{P}^k)$, which is the blow-up of $S_{2}\times\mathrm{Sym}^kG$ along the collision locus
\be
Z_k=\{(p, Z): p\in Z\}\simeq
G\times \mathrm{Sym}^{k-1}G\simeq\mathbb{P}^1\times\mathbb{P}^{k-1},\qquad k\geq1.
\ee
The collision locus has codimension two in $S_{2}\times\mathbb{P}^k$. Away from the collision locus, $B_{k}$ is isomorphic to $S_{2}\times \mathbb{P}^k$ and $\h{\mathcal{M}}$ is a $\mathbb{P}^{N-k-1}$ bundle over $S_{2}\times \mathbb{P}^k$. At each point of the collision locus, the blow-up introduce an exceptional $\mathbb{P}^1$ that records exactly the information of how the point $p$ tends to the point $q$ in $Z$. The moduli space of a curve passing through $(p,Z)$ with $p\in Z$ is $\mathbb{P}^{N-k}$; requiring the curve to be the limit $p\to q\in Z$ of the curve passing through $(p,Z)$ with $p\not\in Z$, the moduli space is $\mathbb{P}^{N-k-1}$. Thus $\h{\mathcal{M}}$ can be described as a $\mathbb{P}^{N-k-1}$ bundle over $\h{B}_{k}$. Without specifying the additional point, the moduli space $\mathcal{M}$ is a $\mathbb{P}^{N-k}$ bundle over $B_{k}=\mathbb{P}^k$. 

For a blow-up along a smooth codimension-two center, the additive cohomology changes by one degree-shifted copy of the center: $H^r(\mathrm{Bl}_{Z}Y)=H^r(Y)\oplus H^{r-2}Z$. Hence we have
\be
\label{eq:g1mhat}
H^*(\h{\mathcal{M}})=\left[ \frac{{N-k-1}}{2} \right]\otimes \left( H^*(S_2)\otimes \left[ \frac{k}{2} \right]+\left[ \frac{1}{2} \right]\otimes \left[ \frac{{k-1}}{2} \right] \right)\, , \, k\geq 1.
\ee
When $k=0$ i.e. there is no attached tails, we have $H^*(\h{\mathcal{M}})=H^*(S_2)\otimes\left[ \frac{{N-1}}{2} \right]$ as usual. Then the BPS spectrum is given by
\be
\label{eq:g1spec}
\begin{aligned}
\mathcal{H}^{\mathrm{BPS}}&=\left[ \left( R_{1 /2},\frac{1}{2} \right) \right]\oplus [(R_{0},0)]\, ,\\
\mathcal{H}^{\mathrm{BPS}}_{\mathrm{diag}}&=H^*(\h{\mathcal{M}})-(\text{reducible corrections})\, , \\
R_{1/ 2}&=H^*(\mathcal{M})=\left[ \frac{{N-k}}{2} \right]\otimes \left[ \frac{k}{2} \right]\, , \\
R_{0}&=\mathcal{H}^{\mathrm{BPS}}_{\mathrm{diag}}-\left[ \frac{1}{2} \right]\otimes R_{1 /2}\, .
\end{aligned}
\ee

\paragraph{The genus-one family $(2,k,3)$.}
For $D=2e_{2}+3f_{2}=-K_{\mathbb{F}_{1}}$, 
\be
h^0(S_{2},\mathcal{O}_{S_{2}}(D))=4+3+2=9\, ,\quad p_a(D)=1\, ,\quad D\cdot G=5\, .
\ee
Thus $N=8$ and $H^*(S_2)=[0]+[1]$ in \eqref{eq:g1mhat}. Using \eqref{eq:g1spec}, the BPS spectrum for $0\leq k\leq 3$ are listed in Table~\ref{tab:planeelliptic}.

\paragraph{The genus-one family $(3,k,3)$.}
Now take $D=3e_2+3f_2$. Equation~\eqref{ex:hirzebruch} gives
\begin{equation}
h^0(S_{2},\mathcal{O}_{S_{2}}(D))=4+3+2+1=10,\qquad p_a(D)=1.
\end{equation}
Since $D\cdot e_2=0$, $p_a(2e_2+3f_2)=1$, there is a possible reducible configuration $(3,k,3)\to(2,k,3)+(1,0,0)$ which needs to be subtracted from the BPS spectrum. As $e_2$ is rigid, the moduli space of this reducible configuration is just the moduli space of $(2,k,3)$ which is given by a $\mathbb{P}^{8-k}$ bundle over $\mathbb{P}^k$. Hence taking $N=9$ in \eqref{eq:g1mhat} and subtracting $[\frac{8-k}{2}]\otimes[\frac{k}{2}]$ in \eqref{eq:g1spec}, we obtain the BPS spectrum for $0\leq k\leq 3$ in Table~\ref{tab:planeelliptic}.

\begin{table}[htbp]
\centering\small
\begin{tabular}{c l l}
\toprule
$\mathbf{d}$ & $R_{1/2}$ & $R_0$\\
\midrule
$(2,0,3)$ & $[4]$ & $[\tfrac52]+[\tfrac72]$\\
$(2,1,3)$ & $[3]+[4]$ & $[\tfrac32]+3[\tfrac52]+2[\tfrac72]$\\
$(2,2,3)$ & $[2]+[3]+[4]$ & $[\tfrac12]+3[\tfrac32]+4[\tfrac52]+2[\tfrac72]$\\
$(2,3,3)$ & $[1]+[2]+[3]+[4]$ & $3[\tfrac12]+4[\tfrac32]+4[\tfrac52]+2[\tfrac72]$\\
$(3,0,3)$ & $[\tfrac92]$ & $[3]$\\
$(3,1,3)$ & $[\tfrac72]+[\tfrac92]$ & $[2]+2[3]+[4]$\\
$(3,2,3)$ & $[\tfrac52]+[\tfrac72]+[\tfrac92]$ & $[1]+2[2]+3[3]+[4]$\\
$(3,3,3)$ & $[\tfrac32]+[\tfrac52]+[\tfrac72]+[\tfrac92]$ & $[0]+2[1]+3[2]+3[3]+[4]$\\
\bottomrule
\end{tabular}
\caption{The eight genus-one $SU(3)_4$ charges $(2,k,3)$ and $(3,k,3)$.
The spectrum is $\mathcal{H}^{\mathrm{BPS}}=\left[ \left( R_{1 /2},\frac{1}{2} \right) \right]\oplus [(R_{0},0)]$. These results agree with
\cite[Table~5]{Kim:2020hhh}}
\label{tab:planeelliptic}
\end{table}

\subsection{Rank-2 example: $SU(3)_5$ theory}
\label{subsec:SU35}

For the 5d $SU(3)_5$ theory, the resolved compact surfaces and their gluing curves are
\be
S_1=\mathbb F_6,\qquad S_2=\mathbb F_0,\qquad
G_1=e_1,\qquad G_2=e_2+2f_2\,.
\ee
Here $e_1^2=-6$, whereas $e_2^2=f_2^2=0$ and $e_2\cdot f_2=1$. Thus
\be
G_1^2=-6,\qquad G_2^2=4,\qquad G_1^2+G_2^2=-2\,,
\ee
as required for a rational gluing curve in a Calabi-Yau threefold. The primitive 2-cycles are $e_{2},f_{1},f_{2}$. We label a curve class with
\be
\mathbf d=(a,k,b),\qquad \beta=ae_2+kf_1+bf_2\,.
\ee
By the same argument as in the $SU(3)_{4}$ geometry, such a curve consists of a core curve $D=ae_{2}+b f_{2}\subset S_{2}$ together with $k$ rational $f_{1}$ tails attached to it. The core curve in $\mathbb{F}_{0}$ has
\be
p_a(D)=(a-1)(b-1),\quad
h^0(S_{2},\mathcal O_{S_{2}}(D))=(a+1)(b+1),\quad
N:=h^0(D)-1=ab+a+b\,.
\ee
When the $k$ attachment conditions are independent, the moduli space of a rational curve with $k$ tails is a $\mathbb{P}^{N-k}$ bundle over $\mathbb{P}^k$.
For $D=e_2$, the three cases $k=0,1,2$ instead all have moduli space $\mathbb P^1$:
for $k=0$ this is the ruling pencil; for $k=1$ an attachment point on $G_2$ determines the ruling;
and for $k=2$ the attachment divisor is the full intersection of the ruling with $G_2$.
The genus-zero results are listed in Table~\ref{tab:SU35-g0}.

\begin{table}[htbp]
\centering\small
\renewcommand{\arraystretch}{1.15}
\begin{tabular}{ccc l}
\toprule
$\mathbf d$ & base & fiber & $R_0$\\
\midrule
$(1,0,0),(1,1,0),(1,2,0)$ & $\mathbb P^1$ & point & $[\tfrac12]$\\
\addlinespace[2pt]
$(1,0,1)$ & point & $\mathbb{P}^{3}$ & $[\tfrac{3}{2}]$\\
$(1,1,1)$ & $\mathbb{P}^{1}$ & $\mathbb{P}^{2}$ & $[\tfrac{1}{2}]+[\tfrac{3}{2}]$\\
$(1,2,1)$ & $\mathbb{P}^{2}$ & $\mathbb{P}^{1}$ & $[\tfrac{1}{2}]+[\tfrac{3}{2}]$\\
\addlinespace[2pt]
$(1,0,2)$ & point & $\mathbb{P}^{5}$ & $[\tfrac{5}{2}]$\\
$(1,1,2)$ & $\mathbb{P}^{1}$ & $\mathbb{P}^{4}$ & $[\tfrac{3}{2}]+[\tfrac{5}{2}]$\\
$(1,2,2)$ & $\mathbb{P}^{2}$ & $\mathbb{P}^{3}$ & $[\tfrac{1}{2}]+[\tfrac{3}{2}]+[\tfrac{5}{2}]$\\
\addlinespace[2pt]
$(2,0,1)$ & point & $\mathbb{P}^{5}$ & $[\tfrac{5}{2}]$\\
$(2,1,1)$ & $\mathbb{P}^{1}$ & $\mathbb{P}^{4}$ & $[\tfrac{3}{2}]+[\tfrac{5}{2}]$\\
$(2,2,1)$ & $\mathbb{P}^{2}$ & $\mathbb{P}^{3}$ & $[\tfrac{1}{2}]+[\tfrac{3}{2}]+[\tfrac{5}{2}]$\\
\addlinespace[2pt]
$(3,0,1)$ & point & $\mathbb{P}^{7}$ & $[\tfrac{7}{2}]$\\
$(3,1,1)$ & $\mathbb{P}^{1}$ & $\mathbb{P}^{6}$ & $[\tfrac{5}{2}]+[\tfrac{7}{2}]$\\
$(3,2,1)$ & $\mathbb{P}^{2}$ & $\mathbb{P}^{5}$ & $[\tfrac{3}{2}]+[\tfrac{5}{2}]+[\tfrac{7}{2}]$\\
\bottomrule
\end{tabular}
\caption{The fifteen genus-zero $SU(3)_5$ charges in the displayed range.
The reduced spectrum is $\mathcal H^{\mathrm{BPS}}=[(R_0,0)]$.
The first row is the exceptional ruling family described in the text;
the remaining rows follow from the projective-bundle formula.
These results agree with \cite[Table~6]{Kim:2020hhh}.}
\label{tab:SU35-g0}
\end{table}

\paragraph{The genus-one family $(2,k,2)$.}
Consider $D=2e_{2}+2f_{2}$ with $0\leq k\leq 2$ tails. It has genus $p_{a}(D)=1$ and $N=8$. Then $\mathcal{M}_{2e_{2}+2f_{2},k}$ is a $\mathbb{P}^{8-k}$ bundle over $\mathbb{P}^k$, leading to $R_{1 /2}=\left[ \frac{k}{2} \right]\otimes\left[ \frac{{8-k}}{2} \right]$ in $\mathcal{H}^{\mathrm{BPS}}=\left[ \left( R_{1 /2},\frac{1}{2} \right) \right]\oplus [(R_{0},0)]$. When $k=0$, $\h{\mathcal{M}}$ is a $\mathbb{P}^{7}$ bundle over $\mathbb{F}_{0}$. When $k\geq 1$, the generic part of $\h{\mathcal{M}}$ is a $\mathbb{P}^{7-k}$ bundle over $\mathbb{F}_{0}\times \mathbb{P}^k$, whereas there is the collision locus $\mathbb{P}^1\times \mathbb{P}^{k-1}$ in the base. Blowing up the collision locus, this moduli space gives rise to 
\be
\mathcal{H}^{\mathrm{BPS}}_{\mathrm{diag}}=H^*(\h{\mathcal{M}})=(H^*(\mathbb{F}_{0})\otimes H^*(\mathbb{P}^k)+H^*(\mathbb{P}^1)\otimes H^*(\mathbb{P}^{k-1}))\otimes H^*(\mathbb{P}^{7-k})\, .
\ee
Hence
\be
\begin{aligned}
R_{0}&=H^*(\h{\mathcal{M}})-\left[ \frac{1}{2} \right]\otimes R_{1 /2}\\
&=\left( ([1]+[0])\otimes \left[ \frac{k}{2} \right]+\left[ \frac{1}{2} \right]\otimes \left[ \frac{{k-1}}{2} \right] \right)\otimes \left[ \frac{{7-k}}{2} \right]-\left[ \frac{1}{2} \right]\otimes \left[ \frac{k}{2} \right]\otimes \left[ \frac{{8-k}}{2} \right]\, .
\end{aligned}
\ee
For $k=0$, the collision term in the preceding expression is omitted.
The resulting genus-one representations are listed in Table~\ref{tab:SU35-g1}.

\begin{table}[htbp]
\centering\small
\renewcommand{\arraystretch}{1.2}
\begin{tabular}{c l l}
\toprule
$\mathbf d$ & $R_{1/2}$ & $R_0$\\
\midrule
$(2,0,2)$ & $[4]$ & $[\tfrac{5}{2}]+[\tfrac{7}{2}]$\\
$(2,1,2)$ & $[3]+[4]$ & $[\tfrac{3}{2}]+3[\tfrac{5}{2}]+2[\tfrac{7}{2}]$\\
$(2,2,2)$ & $[2]+[3]+[4]$ & $[\tfrac{1}{2}]+3[\tfrac{3}{2}]+4[\tfrac{5}{2}]+2[\tfrac{7}{2}]$\\
\bottomrule
\end{tabular}
\caption{The three genus-one $SU(3)_5$ charges $(2,k,2)$, $0\leq k\leq2$.
The reduced spectrum is $\mathcal H^{\mathrm{BPS}}=[(R_{1/2},\tfrac12)]\oplus[(R_0,0)]$.
The representations obtained from the displayed KKV expressions agree with
\cite[Table~6]{Kim:2020hhh}.}
\label{tab:SU35-g1}
\end{table}

\paragraph{The genus-two family $(3,k,2)$.}
Consider $D=3e_{2}+2f_{2}$ with $0\leq k\leq 2$ tails. It has genus $p_{a}(D)=2$ and $N=11$. The moduli space $\mathcal{M}$ is then a $\mathbb{P}^{11-k}$ bundle over $\mathbb{P}^k$, leading to $R_{1}=\left[ \frac{k}{2} \right]\otimes\left[ \frac{{11-k}}{2} \right]$. The BPS spectrum is given by
\be
\mathcal{H}^{\mathrm{BPS}}=[(R_1,1)]\oplus[(R_{1/2},\tfrac12)]\oplus[(R_0,0)]\, .
\ee
To compute the $R_{1 /2}$ which receives contribution from the curve with one node, we use equation \eqref{eq:kkv}
\be
\theta \mathcal{H}^{\mathrm{BPS}}=R_{1 /2}+\left[ \frac{1}{2} \right]\otimes R_{1}=H^*(\tilde{\Sigma}^{[1]})\, .
\ee
For $k=0$, the relative Hilbert scheme $\tilde{\Sigma}^{[1]}$ is a $\mathbb{P}^{10}$ bundle over $\mathbb{F}_{0}$. Hence $R_{1 /2}=[4]+[5]$. For $k=1,2$, $\tilde{\Sigma}^{[1]}$ is a $\mathbb{P}^{10-k}$ bundle over the blow-up of $\mathbb{F}_{0}\times \mathbb{P}^k$ along the collision locus $\mathbb{P}^1\times \mathbb{P}^{k-1}$. Hence we have
\be
R_{1 /2}=\left( ([1]+[0])\otimes \left[ \frac{k}{2} \right]+\left[ \frac{1}{2} \right]\otimes \left[ \frac{{k-1}}{2} \right] \right)\otimes \left[ \frac{{10-k}}{2} \right]-\left[ \frac{1}{2} \right]\otimes \left[ \frac{k}{2} \right]\otimes \left[ \frac{{11-k}}{2} \right]\, , \, k=1,2.
\ee
The $R_{0}$ counts the contribution from curves with two nodes. Using \eqref{eq:kkv} we have
\be
H^*(\widetilde{\Sigma}^{[2]}_{(3,k,2)})=(\mathcal{H}_{(3,k,2)}^{\mathrm{BPS}})_{\mathrm{diag}}\oplus H^*(\mathcal{M}_{(3,k,2)})+\text{correction terms}\, .
\ee
There is a reducible configuration given by $(3,k,2)\to(1,0,0)+(2,k,2)$. The $(2,k,2)$ component has $p_{a}=1$ and intersects the first rational $e_{2}$ at two nodal points, leading to correction that begin at the two-nodes level. The moduli space of the reducible configuration is $\mathcal{M}_{(1,0,0)}\times \mathcal{M}_{(2,k,2)}$, whose cohomology is $H^*(\mathrm{Red})=\left[ \frac{1}{2} \right]\otimes\left( \left[ \frac{k}{2} \right]\otimes\left[ \frac{{8-k}}{2} \right] \right)$. Consider $k=0$, the relative Hilbert scheme $\tilde{\Sigma}^{[2]}$ is a $\mathbb{P}^9$ bundle over $\mathrm{Sym}^2\mathbb{F}_{0}$, whose cohomology is $([2]+2[1]+3[0])\otimes\left[ \frac{9}{2} \right]$. In the $k=1$ case it is generically $\mathbb{P}^8$ bundle over $\mathrm{Sym}^2\mathbb{F}_{0}\times \mathbb{P}^1$ and there is the collision locus $\mathbb{F}_{0}\times \mathbb{P}^1$ in the base. Hence
\be
H^*(\tilde{\Sigma}^{[2]}_{(3,1,2)})=[4]\otimes \left( ([2]+2[1]+3[0])\otimes \left[ \frac{1}{2} \right]+([1]+[0])\otimes \left[ \frac{1}{2} \right] \right)\, .
\ee
For $k=2$, $\tilde{\Sigma}^{[2]}_{(3,2,2)}$ is generically $\mathbb{P}^7$ bundle over $\mathrm{Sym}^2\mathbb{F}_{0}\times \mathbb{P}^2$ with one-point collision locus $\mathbb{F}_{0}\times \mathbb{P}^1\times \mathbb{P}^1$ and two-point collision locus $\mathbb{P}^2$. Blowing up the collision locus, we obtain
\be
H^*(\tilde{\Sigma}^{[2]}_{(3,2,2)})=\left[ \frac{7}{2} \right]\otimes \left( ([2]+2[1]+3[0])\otimes \left[ 1 \right]+([1]+[0])\otimes \left[ \frac{1}{2} \right]\otimes \left[ \frac{1}{2} \right]+[1]  \right)
\ee
Plug in the results of $R_{1},R_{1/2}$ and the cohomologies, we can compute
\be
R_{0}=H^*(\widetilde{\Sigma}^{[2]}_{(3,k,2)})-[1]\otimes R_{1}-\left[ \frac{1}{2} \right]\otimes R_{1 /2}-H^*(\mathcal{M}_{(3,k,2)})-H^*(\mathrm{Red})\, .
\ee
The results are listed in Table~\ref{tab:SU35-g2}. When $k=0$ i.e. there is only the core curve inside $\mathbb{F}_{0}$, the refined GV is computed in \cite{Choi:2012jz} via the KKV approach.

\begin{table}[htbp]
\centering\small
\renewcommand{\arraystretch}{1.2}
\begin{tabular}{c c l}
\toprule
$\mathbf d$ & $j_R$ & $R_{j_R}$\\
\midrule
\multirow{3}{*}{$(3,0,2)$} & $1$ & $[\tfrac{11}{2}]$\\
 & $\tfrac{1}{2}$ & $[4]+[5]$\\
 & $0$ & $[\tfrac{5}{2}]+[\tfrac{7}{2}]+2[\tfrac{9}{2}]$\\
\midrule
\multirow{3}{*}{$(3,1,2)$} & $1$ & $[\tfrac{9}{2}]+[\tfrac{11}{2}]$\\
 & $\tfrac{1}{2}$ & $[3]+3[4]+2[5]$\\
 & $0$ & $[\tfrac{3}{2}]+3[\tfrac{5}{2}]+5[\tfrac{7}{2}]+3[\tfrac{9}{2}]$\\
\midrule
\multirow{3}{*}{$(3,2,2)$} & $1$ & $[\tfrac{7}{2}]+[\tfrac{9}{2}]+[\tfrac{11}{2}]$\\
 & $\tfrac{1}{2}$ & $[2]+3[3]+4[4]+2[5]$\\
 & $0$ & $[\tfrac{1}{2}]+3[\tfrac{3}{2}]+8[\tfrac{5}{2}]+7[\tfrac{7}{2}]+4[\tfrac{9}{2}]$\\
\bottomrule
\end{tabular}
\caption{The three genus-two $SU(3)_5$ charges $(3,k,2)$, $0\leq k\leq2$.
For each charge, the three rows give
$\mathcal H^{\mathrm{BPS}}=[(R_1,1)]\oplus[(R_{1/2},\tfrac12)]\oplus[(R_0,0)]$.
The representations obtained from the displayed KKV expressions agree with
\cite[Table~6]{Kim:2020hhh}.}
\label{tab:SU35-g2}
\end{table}

\subsection{Rank-3 example: $SU(4)_8$ theory}
\label{subsec:SU48}

For the 5d $SU(4)_8$ theory, the resolved compact surfaces and their gluing curves are
\be
\begin{gathered}
S_1=\mathbb F_{10},\qquad S_2=\mathbb F_8,\qquad S_3=\mathbb F_0,\\
e_1\sim h_2=e_2+8f_2,\qquad e_2\sim G=e_3+3f_3.
\end{gathered}
\ee
We use $e_i^2=-n_i$, $e_i\cdot f_i=1$, $f_i^2=0$ on $\mathbb F_{n_i}$.
The two rational gluings satisfy $-10+8=-2$ and $-8+6=-2$.
In the primitive charge convention of \cite[Table~30]{Kim:2020hhh},
\be
\mathbf d=(a,l,k,b),\quad
\beta=ae_3+l f_1+kf_2+bf_3.
\label{eq:SU48-charge}
\ee
We focus on the range displayed there, which is $d_i\leq 2$ for every entry of $\bf d$. Since
$e_2=e_3+3f_3$ and $e_1=e_3+8f_2+3f_3$, a section component
in $S_1$ or $S_2$ would require at least three units of $f_3$.
In this range we therefore consider a core curve in the class
$D=ae_3+bf_3$ on $S_3$, with $k$ rational $f_2$ tails, of which
$l$ points are extended by an $f_1$ tail. All the displayed charges obey
$0\leq l\leq k\leq2$.
The two sections $e_2,h_2$ are disjoint, with an $f_2$ fiber meeting each once. Thus its attachment to the core fixes the point at which an $f_1$ tail can extend it; an extended tail has no additional continuous attachment parameter. Adjunction and the section count on $S_3$ give
\be
p_a(D)=(a-1)(b-1),\quad
N:=h^0(S_3,\mathcal O_{S_3}(D))-1=ab+a+b,\quad
D\cdot G=3a+b.
\label{eq:SU48-core}
\ee
Attaching a rational tree of $\mathbb{P}^1$ at one point leaves the arithmetic genus unchanged. Only genus zero and genus one occur among the range $d_i\leq 2$.

\paragraph{Attachment data and the genus-zero families.}
The $f_2$ tails specify a degree-$k$ divisor $Z_k$ on $G$, and their
$f_1$ extensions specify a sub-divisor $Z_l\subset Z_k$.
Writing $Z_k=A+B$ with $A=Z_l$ gives the attachment space
\be
B_{l,k}=\{Z_l\subset Z_k\subset G\}
\simeq\mathrm{Sym}^{l}G\times\mathrm{Sym}^{k-l}G
\simeq\mathbb P^l\times\mathbb P^{k-l}.
\label{eq:SU48-flag}
\ee
For example, when $(l,k)=(1,2)$, the two points have different tail types: one is an $f_2$, the other is an $f_2$ extended by an $f_1$. So their parameter space is $\mathbb P^1\times\mathbb P^1$.

For $a,b\geq1$, $N=ab+a+b\geq 3$. The condition that the core
contain $Z_k$ therefore imposes $k$ independent conditions for
$k\leq2$. Consequently
$\mathcal M_{D;l,k}$ is a $\mathbb P^{N-k}$ bundle over $B_{l,k}$, and
\be
H^*(\mathcal M_{D;l,k})=
\left[\frac{l}{2}\right]\otimes
\left[\frac{k-l}{2}\right]\otimes
\left[\frac{N-k}{2}\right].
\label{eq:SU48-M}
\ee
For a rational core this is $R_0$, with
$\mathcal H^{\mathrm{BPS}}=[(R_0,0)]$.
For example, $(1,1,2,1)$ gives a $\mathbb P^1$ bundle over
$\mathbb P^1\times\mathbb P^1$, hence
$R_0=[\tfrac12]^{\otimes3}=2[\tfrac12]+[\tfrac32]$.

The exception is the $b=0$ case. The core $D=e_3$ has $N=1$ and $D\cdot G=3$. Hence these three points are not independent. A point of $G\cong\mathbb{P}^1$ fixes an $e_{3}$ which intersects $G$ at three points. These three points give three different choices of the single or the two $f_{2}$ tails. Whereas choosing a point on $G$ to specify $e_{3}$ has a degeneracy: there are three different choices leading to the same $e_{3}$. Thus the total one-tail
family is $G\simeq\mathbb P^1$.
For two tails, the moduli space is also $G\simeq\mathbb P^1$.
Since for every displayed $b=0$ charge, either $l=0$ or $l=k$, so the
sub-divisor adds no choice. The five such families all have
$\mathcal M\simeq\mathbb P^1$ and $R_0=[\tfrac12]$.
The genus-zero results are collected in Table~\ref{tab:SU48-g0}.

\begin{table}[htbp]
\centering\small
\renewcommand{\arraystretch}{1.15}
\begin{tabular}{@{}cccl|cccl@{}}
\toprule
$\mathbf d$ & base & fiber & $R_0$ & $\mathbf d$ & base & fiber & $R_0$\\
\cmidrule(r){1-8}
$(1,0,0,0)$ & $\mathbb P^1$ & point & $[\tfrac12]$ & $(1,1,2,1)$ & $\mathbb P^1\times\mathbb P^1$ & $\mathbb P^1$ & $2[\tfrac12]+[\tfrac32]$\\
$(1,0,0,1)$ & point & $\mathbb P^3$ & $[\tfrac32]$ & $(1,1,2,2)$ & $\mathbb P^1\times\mathbb P^1$ & $\mathbb P^3$ & $[\tfrac12]+2[\tfrac32]+[\tfrac52]$\\
$(1,0,0,2)$ & point & $\mathbb P^5$ & $[\tfrac52]$ & $(1,2,2,0)$ & $\mathbb P^1$ & point & $[\tfrac12]$\\
$(1,0,1,0)$ & $\mathbb P^1$ & point & $[\tfrac12]$ & $(1,2,2,1)$ & $\mathbb P^2$ & $\mathbb P^1$ & $[\tfrac12]+[\tfrac32]$\\
$(1,0,1,1)$ & $\mathbb P^1$ & $\mathbb P^2$ & $[\tfrac12]+[\tfrac32]$ & $(1,2,2,2)$ & $\mathbb P^2$ & $\mathbb P^3$ & $[\tfrac12]+[\tfrac32]+[\tfrac52]$\\
$(1,0,1,2)$ & $\mathbb P^1$ & $\mathbb P^4$ & $[\tfrac32]+[\tfrac52]$ & $(2,0,0,1)$ & point & $\mathbb P^5$ & $[\tfrac52]$\\
$(1,0,2,0)$ & $\mathbb P^1$ & point & $[\tfrac12]$ & $(2,0,1,1)$ & $\mathbb P^1$ & $\mathbb P^4$ & $[\tfrac32]+[\tfrac52]$\\
$(1,0,2,1)$ & $\mathbb P^2$ & $\mathbb P^1$ & $[\tfrac12]+[\tfrac32]$ & $(2,0,2,1)$ & $\mathbb P^2$ & $\mathbb P^3$ & $[\tfrac12]+[\tfrac32]+[\tfrac52]$\\
$(1,0,2,2)$ & $\mathbb P^2$ & $\mathbb P^3$ & $[\tfrac12]+[\tfrac32]+[\tfrac52]$ & $(2,1,1,1)$ & $\mathbb P^1$ & $\mathbb P^4$ & $[\tfrac32]+[\tfrac52]$\\
$(1,1,1,0)$ & $\mathbb P^1$ & point & $[\tfrac12]$ & $(2,1,2,1)$ & $\mathbb P^1\times\mathbb P^1$ & $\mathbb P^3$ & $[\tfrac12]+2[\tfrac32]+[\tfrac52]$\\
$(1,1,1,1)$ & $\mathbb P^1$ & $\mathbb P^2$ & $[\tfrac12]+[\tfrac32]$ & $(2,2,2,1)$ & $\mathbb P^2$ & $\mathbb P^3$ & $[\tfrac12]+[\tfrac32]+[\tfrac52]$\\
$(1,1,1,2)$ & $\mathbb P^1$ & $\mathbb P^4$ & $[\tfrac32]+[\tfrac52]$ &  & & & \\
\bottomrule
\end{tabular}
\caption{The twenty-three genus-zero $SU(4)_8$ charges displayed in
\cite[Table~30]{Kim:2020hhh}. The reduced spectrum is
$\mathcal H^{\mathrm{BPS}}=[(0,R_0)]$. The $b=0$ rows use the exceptional
ruling family; the others follow from \eqref{eq:SU48-M}.}
\label{tab:SU48-g0}
\end{table}

\paragraph{The genus-one family $(2,l,k,2)$.}
The remaining five charges have core $D=2e_3+2f_3$, with $p_a(D)=1$ and
$N=8$. Equation~\eqref{eq:SU48-M} gives
\be
R_{1/2}=H^*(\mathcal M)=
\left[\frac{l}{2}\right]\otimes
\left[\frac{k-l}{2}\right]\otimes
\left[\frac{8-k}{2}\right].
\label{su48r12}
\ee
The genus-one specialization of \eqref{eq:kkv} is
\be
\mathcal H^{\mathrm{BPS}}=[(R_{1/2},\tfrac12)]\oplus[(R_0,0)],\qquad
R_0=H^*(\h{\mathcal M})-[\tfrac12]\otimes R_{1/2}.
\label{eq:SU48-KKV}
\ee
There is no additional reducible-core subtraction at this order.
Indeed, every proper effective split $(2,2)=(u,v)+(2-u,2-v)$ has
intersection $r=2u+2v-2uv\geq2$. Its component-product correction first
occurs at the $r$ nodes level.

To compute $H^*(\h{\mathcal M})$, add a point $p$ on the core and construct the restricted nested Hilbert scheme as in the rank-2 examples. For $l=0$ or $l=k$, all tails have the same type and the base is
\be
\h B_k=\mathrm{Bl}_{\mathcal Z_k}(S_3\times\mathbb P^k),\qquad
\mathcal Z_k=\{(p,Z_k):p\in Z_k\}\simeq G\times\mathbb P^{k-1},
\quad k\geq1.
\ee
For $k=0$ set $\h B_0=S_3$. The center is smooth of codimension two,
so the blow-up formula used in \eqref{eq:g1mhat} gives
\be
H^*(\hat{\mathcal M})=
\left\{([1]+[0])\otimes\left[\frac{k}{2}\right]
+[\tfrac12]\otimes\left[\frac{k-1}{2}\right]\right\}
\otimes\left[\frac{7-k}{2}\right],
\qquad l=0\ \text{or}\ k,
\label{eq:SU48-Mhat-one-type}
\ee
where the collision term is omitted for $k=0$.

\paragraph{Two different tail types: $(2,1,2,2)$.}
Here the attachment points $q_1,q_2\in G$ are distinguished by whether
the tail extends into $S_1$. First the moduli space $\mathcal{M}$ is a $\mathbb{P}^6$ bundle over $\mathbb{P}^1\times\mathbb{P}^1$, whose cohomology is given by \eqref{su48r12}:
\be
R_{1/2}=[\tfrac12]^{\otimes2}\otimes[3]=[2]+2[3]+[4]\, .
\ee
On a generic point, the
base is $S_3\times G\times G$. The collision loci
\be
C_1=\{p=q_1\},\qquad C_2=\{p=q_2\}
\ee
are each isomorphic to $G\times G$ and have codimension two.
Blow-up these two loci gives the resulting smooth base $\widetilde B_{1,2}$ with
\be
H^*(\widetilde B_{1,2})
=([1]+[0])\otimes[\tfrac12]^{\otimes2}
  +2[\tfrac12]^{\otimes2}=[2]+5[1]+4[0].
\label{eq:SU48-two-type-base}
\ee
Thus
\be
H^*(\hat{\mathcal M})=([2]+5[1]+4[0])\otimes[\tfrac52]=[\tfrac12]+6[\tfrac32]+10[\tfrac52]+6[\tfrac72]+[\tfrac92].
\ee
Subtracting
$[\tfrac12]\otimes R_{1/2}=[\tfrac32]+3[\tfrac52]+3[\tfrac72]+[\tfrac92]$
in \eqref{eq:SU48-KKV} yields
\be
R_0=[\tfrac12]+5[\tfrac32]+7[\tfrac52]+3[\tfrac72].
\ee

\begin{table}[htbp]
\centering\small
\renewcommand{\arraystretch}{1.2}
\begin{tabular}{c l l}
\toprule
$\mathbf d$ & $R_{1/2}$ & $R_0$\\
\midrule
$(2,0,0,2)$ & $[4]$ & $[\tfrac52]+[\tfrac72]$\\
$(2,0,1,2)$ & $[3]+[4]$ & $[\tfrac32]+3[\tfrac52]+2[\tfrac72]$\\
$(2,0,2,2)$ & $[2]+[3]+[4]$ & $[\tfrac12]+3[\tfrac32]+4[\tfrac52]+2[\tfrac72]$\\
$(2,1,2,2)$ & $[2]+2[3]+[4]$ & $[\tfrac12]+5[\tfrac32]+7[\tfrac52]+3[\tfrac72]$\\
$(2,2,2,2)$ & $[2]+[3]+[4]$ & $[\tfrac12]+3[\tfrac32]+4[\tfrac52]+2[\tfrac72]$\\
\bottomrule
\end{tabular}
\caption{The five genus-one $SU(4)_8$ charges displayed in
\cite[Table~30]{Kim:2020hhh}. The reduced spectrum is
$\mathcal H^{\mathrm{BPS}}=[(R_{1/2},\tfrac12)]\oplus[(R_0,0)]$.
The two tail types in $(2,1,2,2)$ require the small resolution described
in the text.}
\label{tab:SU48-g1}
\end{table}

The results in Tables~\ref{tab:SU48-g0} and~\ref{tab:SU48-g1} agree with
all twenty-eight entries displayed in \cite[Table~30]{Kim:2020hhh}.

\subsection{Further geometries and singular divisors}
\label{subsec:further}

The KKV approach and its refined version provide a route to all the refined GV invariants of the geometrically engineered 5d SCFTs discussed in \cite{Kim:2020hhh}. By the formula \eqref{eq:kkv}, the relative Hilbert schemes and their reducible-curve corrections reconstruct every $R_{j_R}$.
Hence using surface linear systems, gluing conditions and relative Hilbert schemes, we can perform the same class-by-class computation for the refined GV invariants as above. At higher rank, the resolved Calabi-Yau geometries dual to the 5-brane webs of the $T_N$ theories provide a further family of applications. The refined topological-string computations of \cite{Hayashi:2013qwa} supply useful benchmarks.

We specified our local-surface scope at the beginning of this section. For a smooth surface $S$, the sufficient condition $K_S\cdot\Sigma<0$ in \eqref{eq:negative-degree} guarantees that the deformations within $S$ are unobstructed, hence the moduli space can be obtained as the linear system of the curve. An extension beyond this sufficient condition is the local rational elliptic surface, or half K3, associated with the E-string \cite{Huang:2013yta}. Its elliptic fiber $F$ has $K_S\cdot F=0$, so the absence of transverse deformations cannot be inferred from \eqref{eq:negative-degree}.

Another limitation occurs in \eqref{eq:kkv}, where the smoothness of the relative Hilbert schemes are assumed such that we can compute their Lefschetz cohomology. However this assumption does not always hold. For example in the local $\mathbb{P}^2$ model, the relative Hilbert scheme of degree 6 curves with 8 nodes is singular \cite{Katz:1999xq}. Such singularity imposes corrections to the cohomology in \eqref{eq:kkv}, which correspond roughly to the Lefschetz cohomology of the singular loci on the relative Hilbert scheme. When the degree of the curve is large, the relative Hilbert scheme may often be singular. The stable pair theory is the more proper approach to the curve with singular relative Hilbert scheme. The broader stable pair and sheaf construction also applies to compact examples. In \cite{Huang:2025xkc}, the authors studied elliptic threefolds over $\mathbb P^2$ and Hirzebruch surfaces.

Besides, recall that when the wrapping curve is singular, the M-theory argument is hard to work. In the type IIA description, the moduli space of a D2-brane with singular support requires an extension of the smooth-Jacobian
description. In \cite{Maulik:2016rip} Maulik and Toda provided instructive examples of rigid nodal and cuspidal rational curves: their sheaf moduli spaces in the
local Calabi--Yau threefold are singular, and their GV invariants are computed using perverse sheaves of vanishing cycles.

Obstructions become essential when infinitesimal normal deformations do not extend to actual deformed curves. A nonzero $H^1(\Sigma,N_{\Sigma/X})$ alone does not establish this: in a Calabi-Yau threefold it can be nonzero even when the moduli space is smooth. The relevant datum is the Kuranishi map $\kappa$ of Section~\ref{subsec:obstruction}. The generalized conifold in Section~\ref{subsubsec:singular-divisor} gives a concrete example. For exponent $m\geq2$, the exceptional rational curve has normal bundle $\mathcal O\oplus\mathcal O(-2)$, but its first-order deformation $t$ is obstructed by
\be
\kappa (t)\sim t^m,\qquad \mathcal M=\mathrm{Spec}\,\mathbb C[t]/(t^m).
\ee
The reduced moduli space is one point, whereas the genus-zero GV multiplicity is $m$ \cite{Collinucci:2021wty, Collinucci:2021ofd}. Ordinary cohomology of that point loses this multiplicity. More generally one needs the obstruction-sensitive stable-pair or vanishing-cycle construction \cite{Choi:2012jz, Maulik:2016rip}.

Finally, a smooth resolved threefold can contain a singular compact divisor, as in the hypersurface $x^3+y^3+z^3+w^4=0$ and the sing. $(E_n)$ examples discussed in Section~\ref{subsubsec:singular-divisor} and \ref{sec:3334}\cite{Closset:2020afy, Wang:2026jad}. Curves through the singular locus are not covered by the smooth-surface normal-bundle argument. Their actual deformation moduli space and obstruction, need to be determined in the singular or resolved geometry carefully.

\section{Generalization to the CY4-fold}
\label{sec:CY4}

The above quantization of the M2-brane in the CY3-fold is well developed in the existing literature. In this section, we will have a try to generalize the above discussion to the M2-brane in the CY4-fold, which has two real supercharges instead of four. After quantizing the zero modes, we find a twisted 3d $\mathcal{N}=2$ theory \cite{Ferrari:2019cni} that reduces to an 1d $\mathcal{N}=2$ SQM on the world-line $\mathcal{T}$. The change of the number of the supercharges will change the form of the spectrum significantly.

\subsection{BPS spectrum}

In sec \ref{subsec:Nambu_NLSM}, we have shown that how to transform the Nambu-Goto action with 11d target space spinor into the NLSM with 3d worldvolume spinor in the locally flat frame. Such decomposition is still valid for the CY4-fold compactification. But the target space is $T\times \mb{R}^2\times X$, so the little group on the flat direction is $\op{SO}(2)\simeq \op{U}(1)$. The generator is $Q=\frac{1}{4}[\gamma^{z},\gamma^{\bar{z}}]$, which is induced from the reduction $\op{SO}(4)\rightarrow \op{SO}(2)$. 

We decompose 11d spinors on the M2-brane into 3d+8d components. The $8d$ space has complex coordinates $(m_{1},m_{2},m_{3},m_{4})=(z,w_{1},w_{2},w_{3})$, where $z$ corresponds to the spatial part $\mathbb{R}^2\cong\mathbb{C}$ of $\mathbb{R}^{1,2}$, and $(w_{1},w_{2},w_{3})$ are the normal directions $N_{\Sigma |X}$. The pull-back spinor bundle on the 3d world-volume is
\be
(\mathcal{O}\oplus K_{\Sigma}^{-1})\otimes \bigoplus_{i=1}^4\Lambda^i N_{\Sigma |\mathbb{R}^2\times X} \, .
\ee
Similar to the preceding discussion, we choose the 3d spinor basis to be $v_{+}\in\Gamma(\Sigma,\mathcal{O})$ and $v_{-}\in\Gamma(\Sigma,K_{\Sigma}^{-1})$. The $8d$ spinor basis can also be constructed from the gamma matrices.

The preserved SUSY parameters are covariant constant spinors on the curve $\Sigma$:
\be
\epsilon_{\text{pres}}=\epsilon_{+}v_{+}|0\rangle+\frac{\epsilon_{+}^*}{\mathcal{G}}v_{-}|1234\rangle \, ,
\ee
where $\epsilon_{+}^*=\mathcal{G}\epsilon_{\bar{z}\bar{w}_{1}\bar{w}_{2}\bar{w}_{3},-}$. The SUSY variations of the bosonic fields are
\be
\begin{aligned}
\delta X^z&=-i\sqrt2\,\epsilon_+^*\psi_{\bar{z},+},&
\delta X^{\bar z}&=+i\sqrt2\,\epsilon_+\psi_{\bar{z},+}^*,\\
\delta X^{w_i}&=-i\sqrt2\,G^{w_i\bar w_i}
\epsilon_+^*\psi_{\bar{w}_i,+},&
\delta X^{\bar w_i}&=+i\sqrt2\,G^{w_i\bar w_i}
\epsilon_+\psi_{\bar{w}_i,+}^*,\quad i=1,2,3.
\end{aligned}
\ee
There is no sum over $i$ in this equation. The fermion zero modes are
\be
\begin{aligned}
\psi_{\bar{z},+}&\in H^0(\Sigma,\mathcal{O}),&
\psi_{\bar{z},-}&\in H^1(\Sigma,\mathcal{O}),\\
\psi_{\bar{w}_i,+}&\in H^0(\Sigma,N_{\Sigma |X}),&
\psi_{\bar{w}_i,-}&\in H^1(\Sigma,N_{\Sigma |X}),\\
\psi_{\bar{z}\bar{w}_j\bar{w}_k,+}&\in H^0(\Sigma,\Lambda^2N_{\Sigma |X}),&
\psi_{\bar{z}\bar{w}_j\bar{w}_k,-}&\in H^1(\Sigma,\Lambda^2N_{\Sigma |X}),\\
\psi_{\bar{w}_{1}\bar{w}_{2}\bar{w}_{3},+}&\in H^0(\Sigma,K_\Sigma),&
\psi_{\bar{w}_{1}\bar{w}_{2}\bar{w}_{3},-}&\in H^1(\Sigma,K_\Sigma).
\end{aligned}
\label{CY4_zero_modes}
\ee
The massive little group $SO(2)$ rotating the flat $\mathbb{R}^2$ has generator $Q=\frac{1}{4}[\gamma^z,\gamma^{\bar{z}}]$. An occupied $\bar{z}$ oscillator has $Q=+1/2$; an unoccupied one has $Q=-1/2$, independently of $s=\pm$.

\begin{table}[h]
\centering
\begin{tabular}{c|c|c|c|c}
 Boson & Fermion & $\op{SO}(2)$ of fermion& Zero modes& Multiplet \\\hline
   $X^{z},X^{\bar z}$ & $\psi_{\bar z,+}\, , \, \psi_{\bar{w}_1\bar{w}_2\bar{w}_3,-}$ & $+\frac{1}{2},-\frac{1}{2}$&B: $H^{0}(\Sigma,\mathcal{O})$ ; F: $H^{0}(\Sigma,\mathcal{O})$ &$(2,2,0)$\\
 & $\psi_{\bar z,-}\, , \, \psi_{\bar{w}_1\bar{w}_2\bar{w}_3,+}$ & $+\frac{1}{2},-\frac{1}{2}$&F: $H^{1}(\Sigma,\mathcal{O})$&$(0,2,2)$\\
$X^{w_i},X^{\bar{w}_i}$  & $\psi_{\bar{w}_i,+},\psi_{\bar{z}\bar{w}_j\bar{w}_k,-}$ & $-\frac{1}{2},+\frac{1}{2}$&B: $H^0(\Sigma,N_{\Sigma|X})$ ; F: $H^0(\Sigma,N_{\Sigma|X})$&$(2,2,0)$\\
  & $\psi_{\bar{w}_i,-},\psi_{\bar{z}\bar{w}_j\bar{w}_k,+}$ & $-\frac{1}{2},+\frac{1}{2}$&F: $H^1(\Sigma,N_{\Sigma|X})$& $(0,2,2)$
\end{tabular}
\caption{The bosons and fermions related by the SUSY variation in the $3d$
worldvolume theory for the M2-branes on curves in CY4-fold. Here  $z$ are the complex coordinates of $\op{SO}(2)$ and
$w$ are those of $\op{SO}(6)$. Here for simplicity, we transform the cohomology groups in (\ref{CY4_zero_modes}) into those in this table by Serre duality.}
\label{tab:boson_fermion_zero_cy4}
\end{table}
From the SUSY pairing relation and fermion zero modes, we find the field contents in the world-line 1d $\mathcal{N}=2$ SQM. The multiplets are shown in Table~\ref{tab:boson_fermion_zero_cy4}. In this table, we use the fact in CY4-fold that 
\begin{equation}
    h^0(\Sigma,\Lambda^2 N_{\Sigma|X})=h^1(\Sigma,N_{\Sigma|X})\,,\; h^1(\Sigma,\Lambda^2 N_{\Sigma|X})=h^0(\Sigma,N_{\Sigma|X})\,.
\end{equation}

In the CY4-fold case, $h^0(\Sigma,N_{\Sigma|X})\neq  h^1(\Sigma,N_{\Sigma|X})$ and the $H^{0}(\Sigma,N_{\Sigma|X})$ $(2,4,2)$ chiral multiplets in CY3-fold case will decompose into $H^{0}(\Sigma,N_{\Sigma|X})$ $(2,2,0)$ chiral multiplets and $H^{1}(\Sigma,N_{\Sigma|X})$ $(0,2,2)$ Fermi multiplets.


The $H^{0,0}(\Sigma)$ fermion zero modes acting on the vacuum can generate two states $|0\rangle$ and $\psi_+|0\rangle$. From the discussion in $\mathcal{N}=2B$ NLSM, we can find that the $H^0(\Sigma,N_{\Sigma|X})$ can generate the cohomology related to the moduli space $H^{*}(\mc{M},K_{\mc{M}}^{\frac{1}{2}})$. The $H^{0,1}(\Sigma)$ and $H^1(\Sigma,N_{\Sigma|X})$ will generate the basis of the Jacobian $\mc{J}$ and the obstruction bundle $\mc{E}_{\rm ob}$ over the moduli space $\mc{M}$. So the BPS spectrum should be 
\begin{equation}
    \left((0)+\left(-\frac{1}{2}\right)\right)\otimes \bigoplus_{i,s} H^{i}(\mc{M},(K_{\mc{M}}\otimes \det E)^{\frac{1}{2}}\otimes \Lambda^sE^*)\, , \quad E=\mc{J}\oplus \mc{E}_{\rm ob} \, .
    \label{eq:BPS_cy4}
\end{equation}
When there is no obstruction bundle and Jacobian, i.e. $h^{0,1}(\Sigma)=h^{1}(\Sigma,N_{\Sigma|X})=0$, the BPS spectrum is 
\begin{equation}
    \left((0)+\left(-\frac{1}{2}\right)\right)\otimes \bigoplus_i H^{i}(\mc{M},K_{\mc{M}}^{\frac{1}{2}})\,.
    \label{eq:BPS_cy4_nobundle}
\end{equation}

There exists a tricky point that $|0\rangle$ can have non-zero charge. So the spin of the ground state have an ambiguity up to an overall constant. This ambiguity comes from that the 3d little group is $\op{U}(1)$, all of the irreducible representations are one dimensional so we cannot constrain the spin of $|0\rangle$. In 5d the little group is $\op{SO}(4)$. For its subgroup $\op{SU}(2)\subset\op{SO}(4)$, its representation $(j)$ is composed of states of spin $-j,1-j\dots j-1,j$. In order that the spectrum is composed of the direct sum of the irreducible representation of $\op{SU}(2)$, the spin of $|0\rangle$ is uniquely fixed. In the following discussion, in order to be compatible with the earlier results\cite{Beasley:2008dc,Intriligator:2012ue,Najjar:2023hee}, we will only choose one special choice of $|0\rangle$. 

The choice of the charge of $|0\rangle$ can be transformed to the choice of the charge of the creation operators of the fermion zero modes from $H^{0,0}(\Sigma)$, $H^{1,0}(\Sigma)$, $H^{0}(\Sigma, N_{\Sigma|X})$, $H^{1}(\Sigma, N_{\Sigma|X})$ when $|0\rangle$ is neutral. In (\ref{eq:BPS_cy4}), we choose the creation operator from $H^{0,0}(\Sigma)$ to be of charge $-\frac{1}{2}$

\subsection{Example}

For simplicity, we can discuss genus zero curve $\mb{P}^1$ in CY4-fold, where there is no Jacobian fibration. The vector bundle over $\mb{P}^1$ can be decompose into the direct sum of $\mc{O}(m)$. So the normal bundle is $\mc{O}(m_1)\oplus \mc{O}(m_2)\oplus\mc{O}(m_3)$. From the Calabi-Yau condition, $m_1+m_2+m_3=-2$. To compare with the result from the twisted SYM \cite{Beasley:2008dc,Intriligator:2012ue,Najjar:2023hee}, we will discuss two examples as following. 

We should keep in mind that $m_i$ can only tell us the first order deformation. So we cannot determine the true moduli space from this normal bundle as we have shown in Section (\ref{subsec:obstruction}). For simplicity, we will assume that there is no obstruction that change the dimension of the moduli space, i.e. $\op{dim}\mc{M}=h^0(\Sigma,N_{\Sigma|X})$, which is compatible with example in \cite{Beasley:2008dc,Intriligator:2012ue,Najjar:2023hee}.

\subsubsection{$N_{\mb{P}^1|X}=\mc{O}\oplus\mc{O}(-1)\oplus\mc{O}(-1)$}
In this case, $h^0(\Sigma,N_{\Sigma|X})=1$, $h^{0,1}(\Sigma)=h^{1}(\Sigma,N_{\Sigma|X})=0$. So, the moduli space $\mc{M}$ is a genus $g$ curve and there is no obstruction bundle. Then from (\ref{eq:BPS_cy4_nobundle}), we can obtain the final spectrum as 
\begin{equation}
    \begin{aligned}
        &\left((0)+\left(-\frac{1}{2}\right)\right)\otimes \left(h^0\left(\mc{M},K_{\mc{M}}^{\frac{1}{2}}\right)(0)+h^1\left(\mc{M},K_{\mc{M}}^{\frac{1}{2}}\right)\left(\frac{1}{2}\right)\right)\\
        =&h^0\left(\mc{M},K_{\mc{M}}^{\frac{1}{2}}\right)\left((0)+\left(-\frac{1}{2}\right)\right)+h^1\left(\mc{M},K_{\mc{M}}^{\frac{1}{2}}\right)\left((0)+\left(\frac{1}{2}\right)\right)\,.
    \end{aligned}
\end{equation}
Here we choose the creation operators in $H^{0}(\Sigma,N_{\Sigma})$ to be of charge $\frac{1}{2}$.
There are $h^0\left(\mc{M},K_{\mc{M}}^{\frac{1}{2}}\right)$ chiral multiplets and $h^1\left(\mc{M},K_{\mc{M}}^{\frac{1}{2}}\right)$ anti-chiral multiplets. This is the same spectrum as that of the topologically twisted field theory \cite{Beasley:2008dc,Intriligator:2012ue, Najjar:2023hee}.

\subsubsection{$N_{\mb{P}^1|X}=\mc{O}\oplus\mc{O}\oplus\mc{O}(-2)$}
In this case, $h^0(\Sigma,N_{\Sigma|X})=2$, $h^{0,1}(\Sigma)=0$,$h^{1}(\Sigma,N_{\Sigma|X})=1$. So the moduli space $\mc{M}$ is a complex surface  and the obstruction bundle is a line bundle. The definition of the obstruction bundle is highly dependent on the singular locus on the moduli space, which we have no clean physical interpretation. So in this part, we will use a possible construction of obstruction bundle with a similar result to that in \cite{Cao:2018ddu}, where the obstruction bundle $E=K_{\mc{M}}$. From (\ref{eq:N=2B_spec}), we can find that when we choose the creation operators from $H^0(\Sigma,N_{\Sigma|X})$ to be of charge $\frac{1}{2}$ and the creation operator from $H^1(\Sigma,N_{\Sigma|X})$ to be of charge $-\frac{1}{2}$, there are $h^0(\mc{M},K_{\mc{M}})+h^1(\mc{M},\mc{O})$ anti-chiral multiplets, $h^1(\mc{M},K_{\mc{M}})+h^1(\mc{M},\mc{O})$ chiral multiplets, $H^0(\mc{M},\mc{O})\left((-1)+\left(-\frac{1}{2}\right)\right)$ and $h^2(\mc{M},K_{\mc{M}})$ $\left((1)+\left(\frac{1}{2}\right)\right)$. From Serre duality
\begin{equation}
    h^{i,0}(\mc{M})=h^i(\mc{M},\mc{O})=h^{2-i}(\mc{M},K_{\mc{M}})\,,
\end{equation}
the spectrum is $h^{1,0}(\mc{M})+h^{2,0}(\mc{M})$ vector-like pairs and $h^{0,0}(\mc{M})$ vector multiplets. This is the same spectrum as that of the topologically twisted field theory \cite{Intriligator:2012ue, Najjar:2023hee}.

\paragraph{Construction of the obstruction bundle.} For future reference, here we will show the assumption we need to obtain the obstruction bundle $E=K_{\mc{M}}$. We can construct a space $\mc{C}$ as a fiber bundle over $\mc{M}$ with the fiber as the genus zero curve $\mb{P}^1$ at each point of the moduli space. We have a natural projection
\begin{equation}
    \pi:\mc{C}\rightarrow \mc{M}\,,\;\mc{C}_s=\pi^{-1}(s)\,.
\end{equation}

We can construct the relative normal bundle on $\mc{C}$
\begin{equation}
    \mc{N}:=N_{\mc{C}|X\times \mc{M}}\,,\; \mc{N}|_{\mc{C}_s}=N_{\mc{C}_s|X}\simeq \mc{O}\oplus\mc{O}\oplus\mc{O}(-2)\,.
\end{equation}
We can also construct the relative canonical bundle on $\mc{C}$ as 
\begin{equation}
    \mc{K}:=K_{\mc{C}/\mc{M}}\,,\; \mc{K}|_{\mc{C}_s}=K_{\mc{C}_s}\simeq \mc{O}(-2)\,,\; \mc{K}\simeq K_{\mc{C}}\otimes \pi^* K_{\mc{M}}^{-1}\,.
\end{equation}
A natural definition of the obstruction bundle is 
\begin{equation}
    E=R^1\pi_*\mc{N}\,,\; E|_s\simeq H^1(\mc{C}_s,N_{\mc{C}_s|X})\,.
\end{equation}
Notice that $\pi_*\mc{N}|_s=H^0(\mc{C}_s,N_{\mc{C}_s|X})\simeq T_s\mc{M}$, from the Kodaira-Spencer map, we can make the crucial assumption \footnote{This assumption is highly dependent on how to define the moduli space and $H^0(\mc{C}_s,N_{\mc{C}_s|X})$ at the singular locus.} that
\begin{equation}
    \pi_*\mc{N}=R^0\pi_*\mc{N}\simeq T\mc{M}\,.
\end{equation}
Assuming that $\pi$ is smooth, from the relative Serre duality
\begin{equation}
    (R^i\pi_* E)^*\simeq R^{\op{dim}\mc{C}-\op{dim}\mc{M}-i}\pi_*(E^*\otimes \mc{K})\,,
\end{equation}
We can obtain
\begin{equation}
    R^1\pi_*\mc{N}=(R^0 \pi_*(\mc{N}^*\otimes \mc{K}))^*=(\pi_*(\mc{N}^*\otimes \mc{K}))^*\,.
\end{equation}
Then we only need to determine the relative simple sheaf $\pi_*(\mc{N}^*\otimes \mc{K})$.

There are natural projections $p_X:X\times \mc{M}\rightarrow X$, $p_{\mc{M}}:X\times \mc{M}\rightarrow \mc{M}$. The relative canonical bundle is $K_{X\times\mc{M}|\mc{M}}=p_X^* K_X$. As $K_X\simeq \mc{O}_X$ for CY4-fold $X$,
\begin{equation}
    K_{X\times\mc{M}|\mc{M}}=p_X^* K_X\simeq \mc{O}_{X\times \mc{M}}\,.
\end{equation}
Assume that there exists an embedding
\begin{equation}
    j:\mc{C}\hookrightarrow X\times \mc{M}\,.
\end{equation}
From relative adjunction
\begin{equation}
    \mc{K}\simeq j^* K_{X\times \mc{M}|\mc{M}}\otimes \op{det}\mc{N}\simeq \op{det}\mc{N}\,.
\end{equation}
As $\mc{N}$ is rank three, we can obtain
\begin{equation}
    \mc{N}^*\otimes \mc{K}=\mc{N}^* \otimes \op{det} \mc{N}=\wedge^2 \mc{N}\,.
\end{equation}
Notice that $\wedge^2 \pi_*\mc{N}|_s\simeq \wedge^2 H^0(\mc{C}_s,N_{\mc{C}_s|X})$ is isomorphic to $\pi_*(\wedge^2\mc{N})|_s\simeq H^0(\mc{C}_s,\wedge^2 N_{\mc{C}_s|X})=H^0(\mc{C}_s,\mc{O}\oplus\mc{O}(-2)\oplus\mc{O}(-2))$ fiberwise, we can make an assumption that
\begin{equation}
    \pi_*(\wedge^2\mc{N})\simeq \wedge^2 \pi_*\mc{N}\,.
\end{equation}
Then we can obtain
\begin{equation}
    \pi_*(\mc{N}^*\otimes \mc{K})=\pi_*(\wedge^2\mc{N})=\wedge^2 \pi_*\mc{N}=\wedge^2T\mc{M}=K_{\mc{M}}^{-1}\,.
\end{equation}
So we can obtain the obstruction bundle 
\begin{equation}
    E=R^1\pi_*\mc{N}=(\pi_*(\mc{N}^*\otimes \mc{K}))^*=K_{\mc{M}}\,.
\end{equation}
Notice that this derivation is highly dependent on the example and involve too many assumptions. To generalize to the more complicated case, we will search for more rigorous derivation of the obstruction bundle.

\section{Discussions}

In this note we discussed the supersymmetric quantum mechanics origin of refined GV invariants from M2-branes wrapping curve classes $\Sigma$ with fixed geometric genus on local Calabi-Yau threefolds, which shall be applied to more general examples as long as the moduli space $\mc{M}$ and the fibration $\widehat{\mc{M}}$ of the curve $\Sigma$ are computable from the local geometry. The procedure generalizes the cases of rigid $\Sigma$ in \cite{Dedushenko:2014nya}, and produces results consistent with the results of topological strings and 5d gauge theory. We also extend the discussion to the case of local Calabi-Yau fourfolds, in the simplest cases of $\Sigma\cong\mathbb{P}^1$ with normal bundle $N_{\Sigma|X}$ equals to either $\mc{O}\oplus\mc{O}(-1)\oplus\mc{O}(-1)$ or $\mc{O}\oplus\mc{O}\oplus\mc{O}(-2)$.

In the future, the obvious next steps are to study the higher-genus curves of $g\geq 2$ in Calabi-Yau threefolds, where the Jacobian fibration is much more involved, to discuss the cases of compact Calabi-Yau threefolds, and to obtain a more complete understanding of GV invariants in local Calabi-Yau fourfolds from the physics perspectives.

\section*{Acknowledgement}

We thank Minxin Huang, Go Noshita, Xin Wang for discussions.
YH, QL, YNW are supported by National Natural Science Foundation of China under Grant No. 12422503. JT is supported by National Natural Science Foundation of China under Grant No.~12405085 and by the Natural Science Foundation of Shanghai (Grant No.~24ZR1419300). JT would like to thank Ying Zhang for her love and support.

\bibliographystyle{JHEP}
\bibliography{biblio.bib}

\end{document}